\documentclass[graybox, envcountchap]{svmult}
\usepackage[english]{babel} %%% 'french', 'german', 'spanish', 'danish', etc.
\usepackage{amssymb}
\usepackage{amsmath}
\usepackage{txfonts}
\usepackage{mathdots}
\usepackage{mathptmx}
\usepackage[classicReIm]{kpfonts}
\usepackage{graphicx}
\usepackage{subfigure}
\usepackage{multirow}
\usepackage{aas_macros}
\usepackage{textcomp}
\usepackage[bottom]{footmisc}
\usepackage{booktabs}
\usepackage{color}
\usepackage{helvet}          % selects Helvetica as sans-serif font
\usepackage{courier}         % selects Courier as typewriter font
\usepackage{dirtree}
\usepackage{makeidx}        % allows index generation
\usepackage{multicol}        % used for the two-column index
\usepackage{natbib}

\usepackage[bottom]{footmisc}% places footnotes at page bottom
\usepackage{hyperref}        %for hyperlinks
\usepackage{tabularx}
\usepackage{threeparttable}

\usepackage{hyperref}        %for hyperlinks
\hypersetup{colorlinks=true,urlcolor=blue}
\usepackage[misc]{ifsym}
\makeindex 

\begin{document}

\title{Imaging X-ray Polarimetry Explorer}

\author{Brian Ramsey, Paolo Soffitta and Martin C. Weisskopf}
%\author[0000-0003-1548-1524]{Brian D. Ramsey}
%\affiliation{NASA Marshall Space Flight Center, 
%320 Sparkman Drive \\
%Huntsville, Al 35812, USA}
%\author [0000-0002-5270-4240]{Martin C. Weisskopf}
%\affiliation{NASA Marshall Space Flight Center,
%320 Sparkman Drive \\
%Huntsville, Al 35812, USA}
%\author[0000-0002-7781-4104] {Paolo Soffitta}
%\affiliation{INAF-Instituto di Astrofisica e Planetologia Spaziali,
%Via del Fosso del Cavaliere 100,
%00133 Rome, Italy}
\institute{Brian Ramsey(\Letter) \at NASA-Marshall Space Flight Center (Emeritus),  320 
Sparkman Drive, Huntsville, Al 35812, USA \email{brian.ramsey@nasa.gov}
\and Paolo Soffitta(\Letter) \at INAF-Istituto di Astrofisica e Planetologia Spaziali, Via Fosso Del Cavaliere 100, 00133, Rome (Italy) \email{paolo.soffitta@inaf.it}
\and Martin C. Weisskopf(\Letter) \at NASA-Marshall Space Flight Center (Emeritus),  320 
Sparkman Drive, Huntsville, Al 35812, USA \email{martin.c.weisskopf@nasa.gov}}
\maketitle
\abstract{In 1971 a group from Columbia University  detected polarized X rays from the Crab Nebula  at the $\sim3\sigma$ level during a short sounding rocket flight (\cite{Novick1972}) . This measurement, later confirmed to high precision ( $>20\sigma$) by the OSO-8 satellite in 1976/77 (\cite{weisskopf1978}) marks the real beginning of astrophysical X-ray polarimetry. Since then and before the launch of IXPE, surprisingly few additional measurements had been made. X-ray polarimetry is a demanding field of study. To achieve $\%$-level polarization sensitivity, the collection of more than 10$^{6}$ photons in a polarization-sensitive detector is required. As most cosmic X-ray sources are much fainter than the Crab, to do a systematic study of the X-ray sky demands a highly-sensitive payload and a dedicated mission. Further, as the expected polarization will vary with location in extended sources, such as supernovae and pulsar wind nebula, it is vital that the polarimeter should be imaging. 
First proposed in its current form in 2007, and selected for funding in 2017, the Imaging X-ray Polarimetry Explorer (IXPE) fits this bill. In this chapter we describe: the design and construction of the IXPE payload and spacecraft; its testing and launch; commissioning activities; science operations and; a high-level overview of IXPE science to date.  Finally, we conclude with the current status of the Observatory (as of 2025) and prospects for continued operations.}

\section{IXPE Overview}
\label{sec:intro} 
The Imaging X-ray Polarimetry Explorer (IXPE) is the first mission designed to carry out a systematic study of polarization, over the 2-8 keV band, in a variety of different classes of cosmic X-ray sources. IXPE is a NASA small explorer with a substantial contribution from the Italian Space Agency (ASI), and was launched on December 9, 2021, from complex 39A at Kennedy Space Center, into a 600-km-altitude, equatorial orbit. This orbit was picked for its low background environment and to give passage over the Italian-provided ground station in Malindi, Kenya, every 95 minutes for frequent data downloads. At the time of writing (Summer 2025) IXPE has completed its two-year baseline mission and is now in a general observer phase, providing for the scientific community to competitively select IXPE targets. 

\begin{figure}[ht]
\caption{\textit{The Imaging X-ray Polarimetry Explorer with three
polarization-sensitive detector units at the focus of three X-ray mirror module assemblies}.}
\label{fig:IXPE} \centering
\includegraphics[width=6.5in, height=3.7in, keepaspectratio=true]{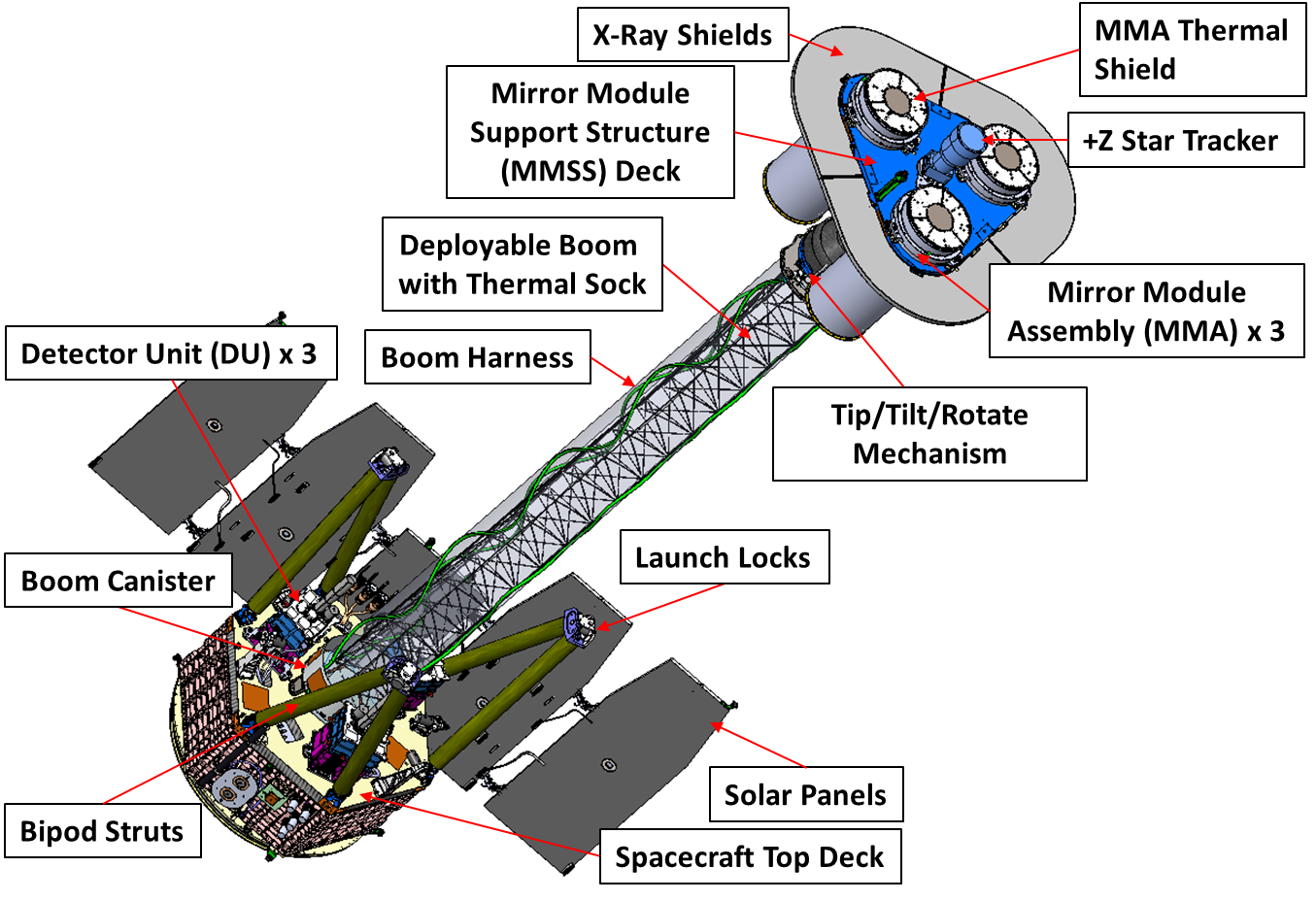}
\end{figure}

The IXPE payload consists of three identical telescopes, each comprised of a mirror module assembly (MMA) with a polarization sensitive X-ray detector unit (DU) at its focus (Figure \ref{fig:IXPE}). Originally designed for launch on a Pegasus rocket the payload features an extensible boom which deploys the mirror modules when on orbit to establish the required 4-m focal length. A tip/tilt/rotate (TTR) mechanism, mounted atop the boom and below the mirror module support structure allows for fine alignment of the optics with respect to the detectors after deployment.  Each mirror module contains 24 concentrically-nested Wolter-1-type mirror shells to provide for imaging and flux concentration, reducing detector background rates to 5 orders of magnitude lower than typical bright IXPE targets. The detectors use photoelectron-track imaging to measure the photoelectron emission direction, allowing a direct measure of the polarization of each absorbed photon. The photoelectron track data also permit the determination of the absorption location in the detector (for imaging) and the energy (through the total charge collected) and time of arrival of each photon.

The IXPE payload rides atop a Ball Aerospace (now BAE Systems) bus that provides for power, telemetry, pointing and thermal control. Typical observation times are quite long ($\sim$ days). Data are downloaded several times per day to a Mission Operations Center (MOC) in Boulder Colorado and from there to a Science Operations Center (SOC) in Huntsville, Alabama. 

IXPE is a collaboration between NASA and ASI, conducted under an implementing arrangement. The Marshall space flight Center manages the mission and provides systems engineering and safety and mission assurance. It was also responsible for the design, fabrication, and testing of the flight MMAs, the MMA and telescope calibrations, and science operations. Italian partners at the Insituto Nazionale di Astrofisica (INAF) and the Insituto di Fisica Nucleare (INFN) provided the flight polarization-sensitive detectors. Ball Aerospace, now BAE systems, built the spacecraft, were responsible for payload integration and Observatory testing, and operate the mission through a collaboration with the Laboratory for Atmospheric and Space Physics at the University of Colorado in Boulder. Finally, the University of Nagoya provided the ultra-thin thermal shields, mounted at each end of the MMAs, which aid in thermal control while permitting X rays in IXPE’s band to pass. Parameters of the IXPE Observatory are given in Table \ref{Tab:Obs}.

%\pagebreak

\begin{table}
\caption{\textit{Performance Parameters of the IXPE Observatory}}
\centering
\begin{tabular}{| l l |}
\hline
\textbf{Parameter} & \textbf{Value} \\
\hline
\text{Polarization Sensitivity} & \text{Minimum detectable polarization
(MDP$_{99}$) $\leq$ 5.5$\%$} \\
& \text {for a point source with an
E$^{-2}$ photon spectrum,} \\ & \text {a 2--8 keV flux of 10$^{-11}$ ergs
cm$^{-2}$ sec$^{-1}$} \\ & \text {and an integration time of 10 days.} \\
\hline
\text{Angular Resolution} & \text{30 arcsec HPD (averaged over 3 telescopes)} \\
\hline
\text{Energy Resolution} & \text{0.6 keV at 2 keV, ( $\propto$ $\sqrt{E}$)} \\
\hline
\text{Timing Resolution} & \text{1-2 $\mu$s } \\
\hline
\text{Rate Capability} & \text{ Set by 1 ms deadtime per event} \\
\hline
\text{Orbit} & \text{Nominal 600 km, 0$^\circ$ inclination} \\
\hline
\text{Slew Rate} & \text{90$^\circ$ in 30 min} \\
\hline
\text{Data Download Rate} & \text{2 Mbps} \\
\hline
\text{On Board Storage } & \text{6 GB} \\
\hline
\text{Ground Stations } & \text{Primary: Malindi, Kenya. Backup: Singapore} \\
\hline
\text{Downlinks per Day } & \text{7-8} \\
\hline
\end{tabular}
\label{Tab:Obs}
\end{table}

\section{IXPE Proposal}

\subsection{Science and Technical Requirements}
The goal for the IXPE mission was to be a pathfinder that would open up the new field of X-ray polarimetry by providing high-sensitivity measurements for a variety of classes of X-ray sources.  A design reference mission was constructed, essentially a 1-year observing program, that showed how a representative sample of sources within 8 different categories could be measured to appropriate levels of polarization sensitivity to confirm or refute current theories. This defined the polarization sensitivity and imaging requirements for the Observatory and in turn defined the lower-level requirement for the payload elements.

The sensitivity of a polarimeter is
given by the equation:

\begin{equation}\label{eq:MDP}
    MDP \sim \frac{4.29 \times 10^{2}\%}{\mu} \times \frac{\sqrt{(R_{s} + R_{b})\times t}}{R_{s}\times t}
\end{equation}

where R$_{s}$ is the source count rate (a product of the source
flux, the mirror module effective area, and the detector quantum
efficiency), R$_{b}$ is the background rate, t is the observation time and $\mu$ is the modulation
factor which describes the response of the polarimeter to 100\%
polarized X rays. The sensitivity, MDP, is the minimum detectable
degree of polarization that can be measured at the 99$\%$
confidence level. The sensitivity to achieve top-level science goals, as exemplified by the design reference mission, is given by the top-level technical requirement:

\emph{IXPE shall provide a minimum detectable polarization
(MDP$_{99}$) not to exceed 5.5$\%$ for a point source with an
E$^{-2}$ photon spectrum and a 2--8 keV flux of 10$^{-11}$ ergs
cm$^{-2}$ sec$^{-1}$ and an integration time of 10 days.}

This describes the level of polarization that can be measured at
the 99$\%$ confidence level for a generic cosmic source spectrum,
flux and integration time. A second top-level requirement concerns
the angular resolution of the observatory, to perform its imaging
role:

\emph{IXPE shall have a system-level angular resolution not to
exceed 30 arcsec half-power diameter (HPD)}.

These two requirements effectively set the requirements for the performance of the mirror assemblies and the focal plane detectors. The source count rate is dependent on the effective area of the optics folded with the quantum efficiency of the detectors. The modulation factor is determined by the detector performance, and the system angular resolution depends both on the angular resolution of the X-ray optics as well as detector spatial resolution effects. Designing a system that meets these IXPE requirements, within the usual flight constraints of mass, geometry, power and available budget, involved a series of trades.

\subsection{Design Trades / Design Philosophy}
\subsubsection{Optics}
The design of the optics was driven by the need to fit within the constraints of the fairing size and lift capability of the Pegasus rocket, the only launch system at that time that had placed Observatories into IXPE’s desired equatorial orbit. Trade studies indicated that a combination of three, 4-m-focal length mirror modules, each with 24 concentrically nested mirror shells, was the best configuration for the Pegasus fairing. Using three mirror modules, rather than one large one, permitted a shorter focal length than a single large optic, given the graze angles needed for operation in the 2-8 keV energy band, and this allowed for a much shorter deployable boom. Also, having three independent telescopes gave a degree or redundancy and permitted clocking of the three at $120^\circ$ intervals to mitigate any systematic polarization effects inherent in the detectors. 
The choice of 24 nested mirror shells was dictated by the required overall effective area (when combined with the detector quantum efficiency), and the substrate thickness of each shell was dictated by the overall observatory mass budget, factoring in the stiffness required to meet the angular resolution budget. It was also found that the bare mirror-shell material, in this case a nickel/cobalt alloy, gave the best reflectivity response over the desired IXPE energy range, so this avoided the additional steps of coating the mirrors with typically-used gold or iridium.

\subsubsection{Detectors}
The principal design trade for the polarization-sensitive detectors was the choice of fill gas, its depth and pressure (\cite{Baldini2021}). Dense, high atomic number fill gases would provide the largest quantum efficiency, but would have the shortest photoelectron tracks. Also, only K shell interactions provide the high modulation required for measuring the polarization of the absorbed photon, and so this limits the choice of gas as the K shell absorption edge needs to be below the desired IXPE energy range. This rules out the typical noble-gas-based fill mixtures, leaving pure dimethyl ether (DME) as the gas filling of choice. This gas has been used in high-energy physics and has a very low diffusion coefficient, necessary for preserving photoelectron track information as the electrons drift from the photon interaction site to the detector readout plane.

The choice of gas pressure and depth is also a trade. Higher pressures for increased quantum efficiency reduce photoelectron track lengths making track measurements more difficult. Lower pressures and increased gas depths (to maintain efficiency) result in more diffusion of the electrons as they drift towards the readout and this blurs the information contained in the photoelectron tracks. As the polarization sensitivity scales linearly with modulation factor and only as the square root of the efficiency (from Equation 1) preserving the track information is vital, and this led to the selection of a 10-mm-deep detector filled with DME to a pressure of 800 mbar. The track lengths at this gas pressure, approximately 1 mm at 8 keV, are well suited to the $50\mu$m readout-pixel-size of a custom Application Specific Integrated Circuit (ASIC) chip which registers the event.

\section{IXPE as built}
\subsection{The Payload}
\subsubsection{Mirror Module Assemblies}
\begin{figure}[ht]
\caption{\textit{MMA mechanical design}.} \label{fig:mirror} \centering
\includegraphics[width=5.1in, height=2.83in, keepaspectratio=true]{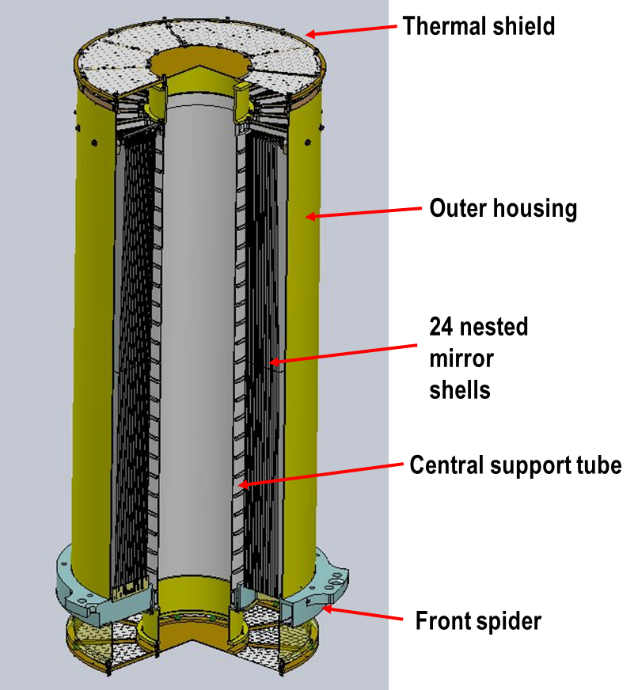}
\end{figure}

The mechanical design  of each mirror module assembly (MMA) is shown in Figure \ref{fig:mirror}. The 24 nested mirror shells forming the heart of the assembly are fabricated from a nickel/cobalt electroforming process and are all identical lengths with each shell being a single structure containing both parabolic and hyperbolic segments of a Wolter-1 configuration. Key parameters of the optical configuration are given in Table \ref{Tab:MirrorChar} and full details
of the IXPE MMA are given in \cite{Ramsey2022}.

\begin{table}
\caption{\textit{GPD Parameters.}}
\label{tab:GPD}
\centering
\begin{threeparttable}
\begin{tabular}{|l|l|}
\hline
\textbf{Parameter} & \textbf{Value} \\
\hline
Sensitive area & 15 $\times$ 15 mm (13 $\times$ 13 arcmin) \\
\hline
Fill gas and fill pressure & DME @ 0.8 atmosphere \\
\hline
Detector window & 50-$\mu$m thick beryllium \\
\hline
Absorption and drift region depth & 10 mm \\
\hline
Spatial resolution (FWHM) & 120 $\mu$m (6.4 arcsec, 2 keV)\tnote{a} \\
\hline
Energy resolution (FWHM) & 0.6 keV @ 2 keV ($\propto \sqrt{E}$) \\
\hline
Useful energy range & 2--8 keV \\
\hline
\end{tabular}

\begin{tablenotes}\footnotesize
\item[a] Measured during flight unit calibration.
\end{tablenotes}
\end{threeparttable}
\end{table}

%\footnote{Measured during flight unit calibration}

 The principal structural element of the MMA is the front spider, to which all 24 nested mirror shells are attached and which provides for attachment to the mirror module support structure, atop the boom. Individual mirror shells are bonded to combs attached to the 9 spokes of the front spider. At the rear end of the shells is a lightweight spider with combs that control shell motion during launch. Completing the design is an outer housing which provides for some mechanical protection and for thermal control of the MMAs via attached heaters. 

The IXPE mirrors are fabricated from an electroforming process in which individual mirror shells are electroformed on figured and super-polished aluminum mandrels from which they are later released by differential thermal contraction. One mandrel is used for each of the 24 shell sizes and is re-used for the four shells needed for the complement of four modules. The shells are fabricated using a nickel/cobalt (80/20) alloy which is more elastic than the pure nickel typically used in electroforming. This makes the shells less susceptible to permanent distortions during fabrication and handling. 

Individual mirror shells are bonded to combs on the spider spokes. This is done by floating the mirror shells in the grooves between the comb teeth and injecting a low-shrinkage epoxy. This process minimizes attachment distortions, vital for maintaining good angular resolution with very thin substrates. Assembly proceeds from the innermost shell outwards. Each shell in turn is suspended from a series of wires, a subset of which is under the control of piezoelectric actuators. By adjusting the actuators, the shape of the shell, monitored by proximity sensors, can be adjusted. Under computer control the shell angular resolution is optimized via this adjustment before the epoxy is injected and the shell fixed in place. Assembly then proceeds to the next shell until the module is complete. For IXPE, four identical MMAs were built, three flight and one spare unit \citep{bongiorno2021}.

Thermal control of the MMAs is vital as the mirror shells are quite susceptible to thermally-induced distortions, particularly for azimuthally varying temperatures. Using thermomechanical models and ray tracing, requirements were placed on operating temperature ($20\pm5 ^\circ$C) and on thermal gradients ($<2 ^\circ$C variation across an individual mirror shell).  To achieve this, the MMA is equipped with six heaters, arranged in two circumferential bands, and is wrapped in multilayer insulation. To control heat loss from the two ends of the mirror shells, without losing low-energy X rays, a pair of ultra-thin thermal shields ($1.4\mu$m thick) are used (see Figure \ref{fig:mirror}).   

When the MMAs were completed, component level environmental tests were carried out. As the hardware is very delicate it was vibration tested at the maximum expected launch loads (from a coupled loads analysis) rather than a generic workmanship level. After vibration testing, the MMAs were thermal vacuum tested to cycle them between their upper and lower survival temperatures (0 to 40 °C) and one unit was acoustically tested. Additional environmental tests were performed later at the Observatory level. Once these components tests were finished the unit were sent for X-ray calibration (see section \ref{Cal}).

\subsubsection{Detector Unit} \label{Detector Unit}

The heart of the IXPE payload are the three detector units (DU). Located at the focus of each MMA, these provide position sensitivity, energy determination, timing information and, uniquely, polarization sensitivity. Inside each DU is a gas pixel detector (GPD), a small proportional counter that can image the photoelectron tracks produced when photons are absorbed in the DME fill gas.   The tracks provide all the necessary information listed above including polarization, as the initial direction of the photoelectron is in the direction of the electric field of the absorbed X-ray photon.

\begin{figure}[ht]
\caption{\textit{Schematic of the GPD}} 
\label{fig:GPD}
\centering
\includegraphics[width=5.1in, height=2.83in, keepaspectratio=true]{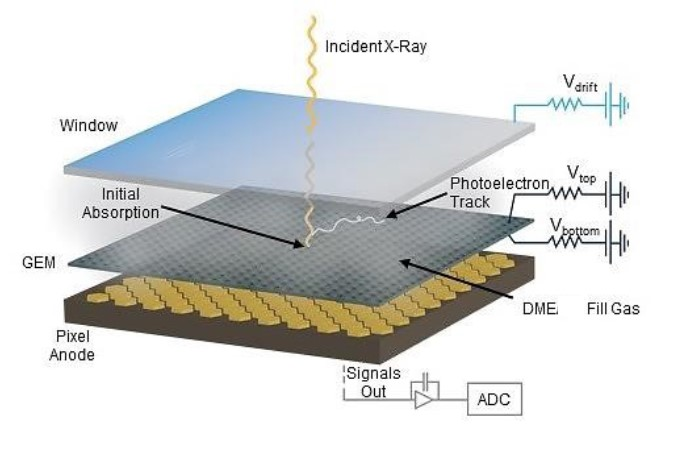}
\end{figure}

Figure \ref{fig:GPD} shows a schematic of the GPD. X rays enter the detector through a beryllium window and interact photoelectrically in the DME fill gas. The resulting photoelectron emitted in the interaction comes to rest in the gas, ionizing gas atoms on the way and forming a track of electrons. This photoelectron track is drifted through a gas electron multiplier (GEM), which boosts the charge in the track, and onto a pixel anode readout, the front end of a custom Application Specific Integrated Circuit (ASIC) which does the initial processing of the event. Table \ref{tab:GPD} gives the parameters of the GPD and more details can be found in \cite{Baldini2021}.

\begin{table}
\caption{\textit{GPD Parameters.}}
\label{tab:GPD}
\centering
\begin{threeparttable}

\begin{tabular}{|l|l|}
\hline
\textbf{Parameter} & \textbf{Value} \\
\hline
Sensitive area & 15 $\times$ 15 mm (13 $\times$ 13 arcmin) \\
\hline
Fill gas and fill pressure & DME @ 0.8 atmosphere \\
\hline
Detector window & 50-$\mu$m thick beryllium \\
\hline
Absorption and drift region depth & 10 mm \\
\hline
Spatial resolution (FWHM) & 120 $\mu$m (6.4 arcsec, 2 keV)\textsuperscript{a} \\
\hline
Energy resolution (FWHM) & 0.6 keV @ 2 keV ($\propto \sqrt{E}$) \\
\hline
Useful energy range & 2--8 keV \\
\hline
\end{tabular}

\begin{tablenotes}[flushleft]
\footnotesize
\item[a] Measured during flight unit calibration.
\end{tablenotes}

\end{threeparttable}
\end{table}

\begin{figure}[ht]
\caption{\textit{Schematic of the DU and photo of flight unit (inset).}} 
\label{fig:DU}
\centering
\includegraphics[width=7.5in, height=4.2in, keepaspectratio=true]{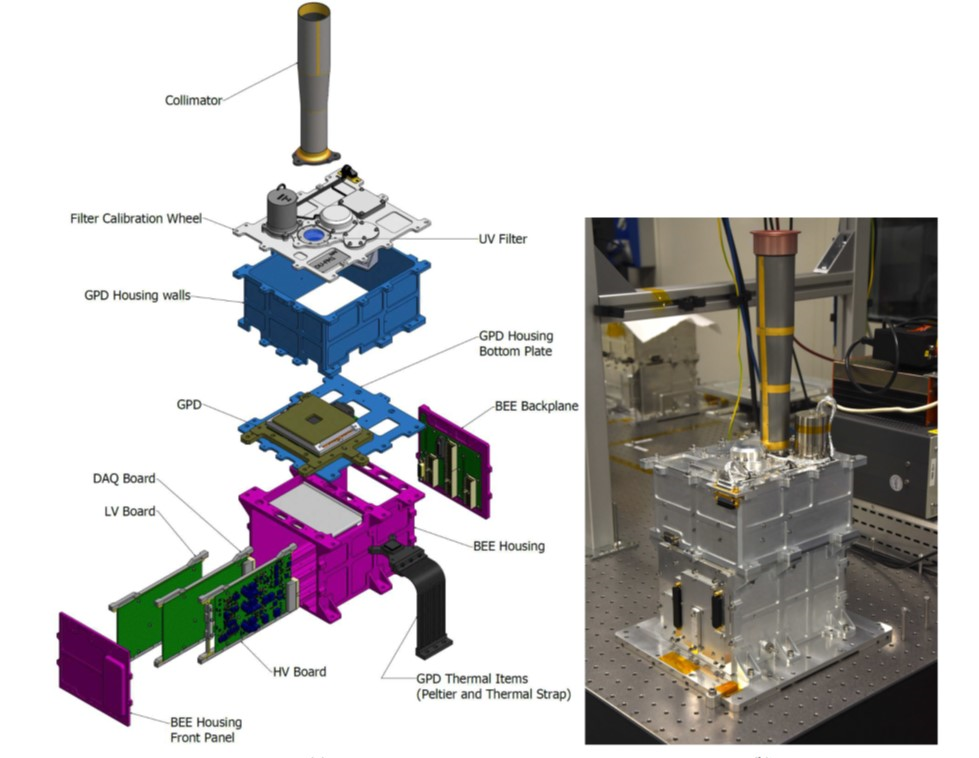}
\end{figure}

\begin{figure}[ht]
\caption{\textit{Filter and calibration wheel assembly, top and bottom view.}} 
\label{fig:FCW}
\centering
\includegraphics[width=5.1in, height=2.83in, keepaspectratio=true]{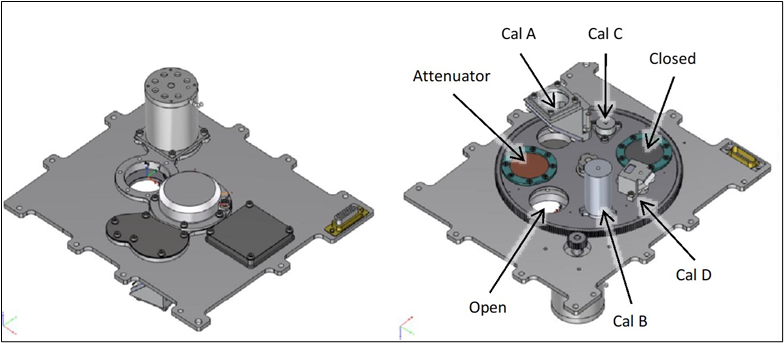}
\end{figure}

A complete detector unit is shown in Figure \ref{fig:DU}. As well as the GPD, the unit houses all the back end electronic to fully process each individual event and the high-voltage power supplies. In addition it has a collimator to restrict background, and a filter and calibration wheel (FCW) assembly for attenuating bright sources as necessary and for on-board monitoring of the performance of the GPD . The FCW (see Figure \ref{fig:FCW}) contains various radioactive sources that can be rotated in front of the GPD to monitor gain, energy resolution, spurious modulation (see section \ref{Sciops}) and modulation factor. The calibration sources are all based on the isotope Fe$^{55}$, which gives Mn lines at 5.9 keV and 6.5 keV. Calibration source A provides near 100$\%$ polarized X rays at 3 keV (via a silver target) and 5.9 keV through Bragg diffraction at $45^\circ$ off a graphite mosaic crystal. Calibration sources B and C give spot ($\sim$3mm) and flood (whole detector) illumination respectively at 5.9 keV and 6.5 keV, while calibration source D gives broad illumination at 1.7 keV via fluorescence from a silicon target. Additional FCW positions include an open position for normal operation and a closed position for internal background monitoring.

Completing the instrument is a single detectors service unit (DSU) which controls all three flight DUs. This unit, effectively a payload computer, acquires and formats the data for transmission to the spacecraft, provides low-voltage power, processes all detector commands, and provides time tagging. In addition, it controls the GPD temperature and the filter and calibration wheel. A complete description of the instrument can be found in \cite{Soffitta2021}.

\subsubsection{Boom and TTR} \label{Boom}
\begin{figure}[ht]
\caption{\textit{IXPE boom deployed during Observatory thermal vacuum testing.}} 
\label{fig:Boom}
\centering
\includegraphics[width=5.1in, height=2.83in, keepaspectratio=true]{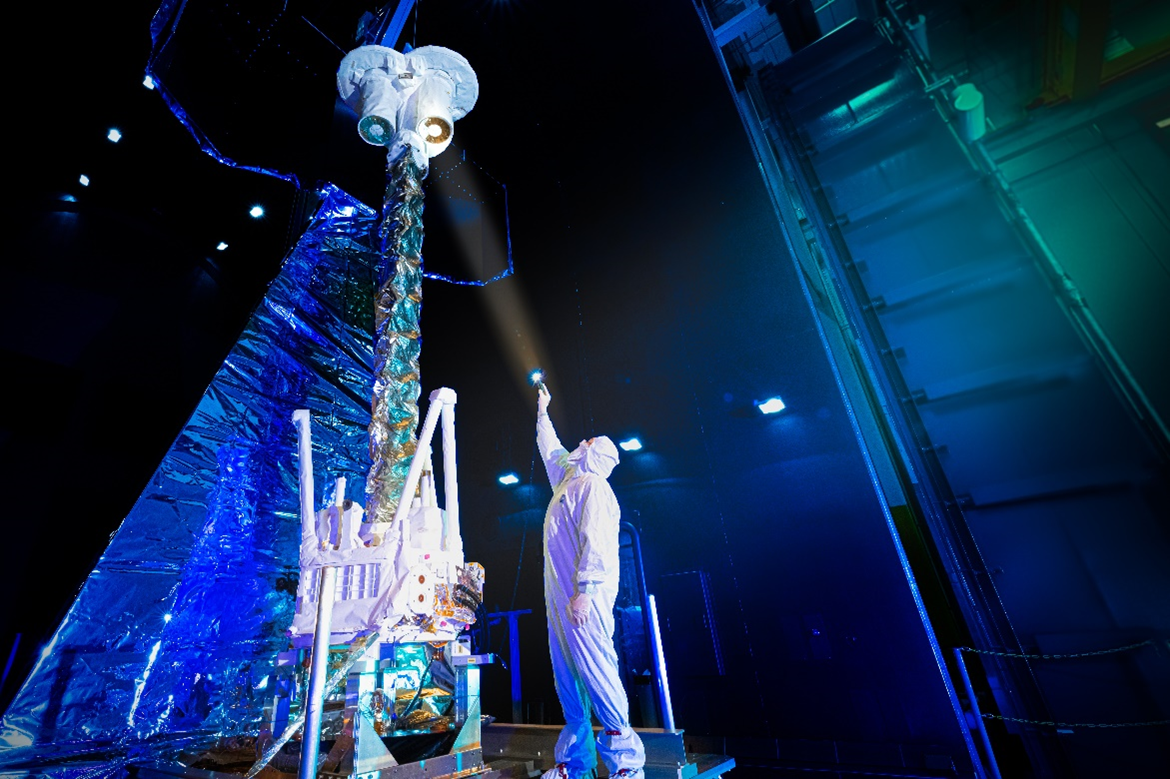}
\end{figure}

To fit within the original Pegasus launch-rocket fairing and yet provide the 4m focal length for the MMAs, IXPE utilizes an extending coilable boom that deploys on orbit. The boom, provided by Orbital ATK (now Northrop Grumman Innovation Systems) is triangular in section and consists of 3 full-length fiber glass longerons with stiffening battens and diagonal members completing the structure. Deployment is via coiled-spring energy, moderated by a speed-controlling damper. A 2-layer multi-layer insulation (MLI) sock covers the boom and provides some thermal isolation to limit boom motion as the Observatory goes in and out of eclipse each orbit. Figure \ref{fig:Boom} shows the sock-covered boom deployed during Observatory thermal vacuum testing.

To facilitate precise alignment of the optics with respect to the detectors once the boom has deployed, a tip/tilt/rotate mechanism was included. Situated immediately below the mirror module support structure (see Figure \ref{fig:IXPE}), this permits small coherent adjustments to be made of the location of the images on the three detectors (via the tip/tilt mechanism) and an adjustment of the relative position of each of the images of the three telescopes via the rotation stage. The range of motions available, $\pm0.15^\circ$ in tip/tilt and $\pm2.6^\circ$ in rotation, are an order of magnitude larger than expected boom deployment uncertainties (see section \ref{Commis} for use of the TTR during commissioning).  

Note that the IXPE optics behave as lenses and so tipping/tilting them would normally not cause the image to move. However, the aspect star tracker is also mounted on the mirror module support structure, so operating the tip/tilt mechanisms effectively repoints the Observatory to move the images on the detectors.

\subsubsection{X-Ray Calibration} \label{Cal}
Detailed and precise X-ray calibrations are vital for an instrument opening a new field of study. MMA calibrations were carried out at the Marshall Space Flight Center, where the optics were built, while detailed detector calibrations were carried out at the Insituto di Astrofisica e Planetologia in Rome. In addition, a single full telescope (spare MMA + spare Detector) was calibrated at MSFC.

Optics calibration took place in the 100-m facility at MSFC. As the name implies, this facility has a $\sim100m$ long vacuum tube, 1.2m in diameter with a series of X-ray sources at one end and a 10m long x 3m diameter test chamber at the other \citep{Thomas2023}. The MMA under test is mounted in the test chamber on a hexapod that enables precise angular and linear control in all axes. Mounted just over 4m away (the effective focal length at $\sim100m$ source distance) is a suite of test detectors.

MMA calibrations consisted of measurements of effective area and angular resolution for a range of energies and off-axis angles covering the bandwidth and the field of view, respectively, of the IXPE telescopes. In addition, measurements were made of ghost rays, the response of the mirror modules to bright sources just outside the telescopes field of view, where singly reflected rays (from just the parabolic or hyperbolic segments) can give rise to a background flux in the detectors.

The DUs including the flight spare underwent a comprehensive calibration in Italy to measure their response to both polarized and unpolarized sources, and to measure their spectral, imaging and timing responses \citep{Muleri2018}. Customized equipment was developed to generate highly-polarized X rays across the IXPE band. These used specific combinations of characteristic X-ray lines and mosaic crystals to produce Bragg diffraction at 45° incident angles, resulting in $>99\%$ polarized X rays. Equally important was to measure the DU response to unpolarized X rays to show that there was either no polarization response or to show that that any systematic effects were stable, low-level, and could be reliably calibrated out. In fact, approximately 60$\%$ of all detector calibration time was spent measuring the polarization response to unpolarized X rays. The IXPE detectors do show a very small amount of position-dependent 'spurious' modulation (typically $\sim0.2\%$ \citep{Baldini2021}), the exact cause of which has never been accurately determined (although it appears to be generated in the GEM), but the effect is very stable and was fully calibrated for later removal from flight data. 

Extensive calibration data were accumulated for the optics in the U.S and for the DUs in Italy. It was therefore important to show that the calibration of a telescope (MMA + DU) could be accurately synthesized from the component calibrations and that the presence of the MMA did not affect the polarization response of the DU. DU calibrations were carried out with essentially parallel beams of X rays, whereas at the focus of a mirror module, the incident X rays are in a cone, at an incident angle of $\sim$2° for the outer shells, focused to the center of the detector’s gas depth. Not only does this effect provide additional blurring to the system angular resolution, but it also potentially changes the spurious modulation that was extensively calibrated. Further, the tipping of the photoelectron emission plane may affect the measured modulation factor. Finally, the reflection of X rays does produce polarization and, although this effect was calculated to be very small for IXPE graze angles, it was important to show that it did not contribute to the source measurement, particularly off axis.

The flight spare telescope was calibrated in the MSFC 100-m facility, as were the MMAs. Measurements were made of telescope effective area and angular resolution and compared with data obtained by combining the respective component calibration data. Then, the modulation factors of the telescope were compared with data taken on the modulation factor during detector calibrations in Italy. Finally, measurements were made of telescope spurious modulations and compared with detector-only data. 

The conclusions of the telescope calibration were that the original stand-alone extensive MMA and DU calibrations could be used to synthesize the flight telescope’s effective area and angular resolution response. This was very important as schedule did not permit an extensive flight telescope calibration campaign. Of equal importance was the finding that the presence of the MMA in no way altered the polarization response of the detectors. That is, the modulation factors and the spurious modulations measured with the flight-spare telescope were statistically identical to those measured (up to nearly 2 years earlier) for the DUs alone. Full details of the telescope calibration are given in \cite{Ramsey2025}. 

\subsection{Spacecraft}
The IXPE spacecraft is based on the Ball configurable platform series, modified for IXPE use. Hexagonal in shape, the spacecraft has six flat aluminum side panels that carry the launch loads through an adapter ring to the launch rocket. The flat panels also provide the real estate for mounting the various satellite and payload related components. The top deck, to which the payload is attached, is made of aluminum honeycomb.

The spacecraft provides power via an array of 5 solar panels capable of generating around 300W of orbit-average power to provide for the payload and spacecraft systems. Batteries provide power during times of eclipse. On launch the solar panels are wrapped around the spacecraft. They deploy automatically shortly after reaching orbit (section \ref{Commis}).

Spacecraft pointing is handled by an attitude control and determination system. Control is via a set of orthogonal reaction wheels, with torquer rods to remove excess wheel angular momentum. For coarse attitude determination, there are a set of 12 sun sensors distributed around the spacecraft, and a 3-axis magnetometer. For fine attitude determination there are a pair of star trackers, one forward looking and mounted on the mirror module support structure, and another rear facing mounted under the satellite bus. The latter can be used to maintain attitude when the primary system is occulted by the earth.

Command and data handling are managed by an integrated avionics unit. This unit runs all the flight software and handles the telemetry, data storage and overall control of the payload via communication with a payload computer contained in a detectors service unit (DSU). Communication is via S band with a data rate of 2 Mbps for downlinking data and a rate of 2 kbps for uplinking commands. Typical ground contact rates are 7-8 times per day. In addition, there are a pair of GPS antennas for essentially full sky coverage. These provide precise location and a highly accurate pulse per second to provide precise timing for analysis of pulsating cosmic sources.

An overview of the spacecraft, ready for launch, is given in \cite{Deininger2022}

\subsection{Integration and Alignment}
Payload integration into the spacecraft took place with delivery of the DUs and MMAs to BAE systems. The DUs were inspected and tested, then mounted onto the top deck of the spacecraft, with the DSU below. The MMAs were then inspected and installed onto the mirror module support structure (MMSS). The final step integrated the mirror assembly (MMAs plus MMSS) onto the stowed boom and bipod structure, but before that the DUs and the MMAs were aligned.

Alignment was a key part of the integration process.  As the field of view of each telescope is quite small (13 x 13 arcminutes) precise alignment is necessary to ensure that images fall near the center of each detector. There are effectively 2 parts to this alignment: ground alignment and on-orbit alignment, the latter correcting for boom deployment uncertainties. Ground alignment makes use of surface mount reflectors (SMR) installed during the builds of the MMAs and the DUs. These SMRs provide precise ($< 100\mu$m) knowledge of the position of the nodes (essentially the centers) of the MMAs and DUs. This knowledge is used, with a laser alignment system, to place the 3 DUs and 3 MMAs on two congruent triangles, the former on the top deck of the spacecraft and the latter on the deployable mirror module support structure. With the MMAs in place, small adjustments were made in their pointing direction, to co-align the modules with each other and with the star tracker. This was accomplished using small, high-quality, alignment cubes whose normal surface was measured with respect to the X-ray axis of each MMA during X-ray calibrations. A similar alignment cube on the star tracker allowed for optical coalignment with the three MMAs. The overall on-ground alignment accuracy between each MMA and DU was about 110$\mu$m in the XY plane (equivalent to $\sim$ 6 arcsec angle) and the optical angular co-alignment between the three telescopes and the star tracker were $<$ 3.6 arcsec. The largest uncertainty in ground alignment was between the X-ray axis and the mirror normal for each MMA, and this was measured to be around 20 arcsec ($3\sigma$).

On orbit alignment is accomplished using the tip$/$tilt$/$rotate stage mounted atop the boom, under the mirror module support structure. Activation of the tip/tilt stages moves the images of the three telescopes in unison, and can be used to correct for lateral offsets in deployed position. As the boom uncoils there is also uncertainty in the final rotation position, and this can be compensated for using the rotate axis of the TTR. As the DUs are clocked on the spacecraft 120$^o$ apart this rotation causes the images on the three telescopes to move independently and can be used to bring the three images into co-alignment as required. See section \ref{Commis}, commissioning, for a description of the actual on-orbit alignment.

\subsection{Ground System}
\begin{figure}[ht]
\caption{\textit{IXPE Ground Network.}} 
\label{fig:Ground}
\centering
\includegraphics[width=5.1in, height=2.83in, keepaspectratio=true]{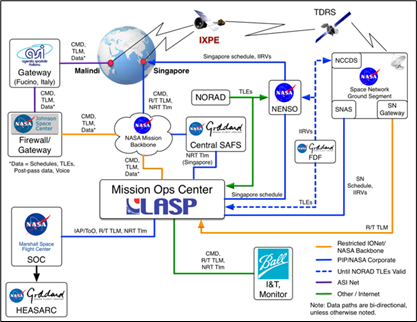}
\end{figure}

The IXPE ground network system is shown in Figure \ref{fig:Ground}. Communication with the spacecraft is via a primary ground station in Malindi, Kenya, with a backup station in Singapore. Data from Malindi go through an Italian gateway then to Johnson Space Center and on to the Mission Operations Center (MOC) in Boulder, Colorado. The MOC communicates with the Science Operations Center in Huntsville, Alabama where the science data are processed. From there the data are sent to the High Energy Astrophysics Science Archive Research Center (HEASARC), where they are archived and made available to the general scientific community.

\section{Launch Vehicle}
At the time of initial selection selection, it was assumed that IXPE would launch on a Pegasus rocket and so the Observatory was designed to fit within the Pegasus fairing. This assumption was based on the fact that IXPE required an equatorial orbit for low background and frequent passages over the Malindi ground station, and Orbital/ATK was the only provider offering this service. Fitting the payload within the small Pegasus fairing was challenging and necessitated a deployable boom to achieve IXPE’s 4-m focal length and deployable shields around the MMAs. The latter were designed to work in conjunction with the detector collimators to block from the detector any X rays that had not passed through the optics.

\begin{figure}[ht]
\caption{\textit{Scale drawing of the IXPE Observatory in the Pegasus and Falcon 9 fairings.}} 
\label{fig:LV}
\centering
\includegraphics[width=5.1in, height=2.83in, keepaspectratio=true]{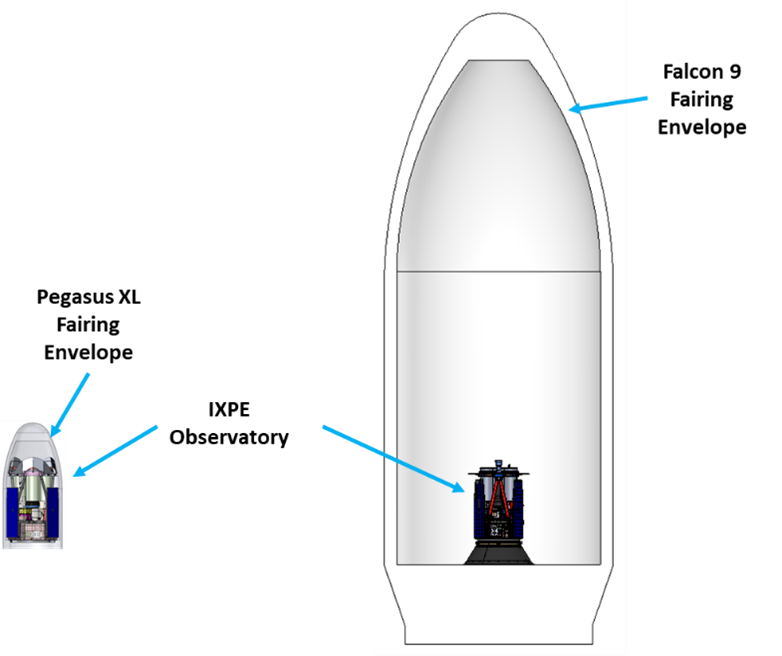}
\end{figure}

Shortly after the project’s critical design review, notification was received from NASA Headquarters that Space X would be launching IXPE on a Falcon 9 rocket. The launch would take place from the Kennedy Space Center and the rocket would carry IXPE down to the equator and place the Observatory into an equatorial orbit. The size of the rocket fairing was considerably larger than that of the Pegasus (see Figure \ref{fig:LV}). However, at this very late stage it was not possible to substantially change the design of IXPE to take advantage of this, so the only design change implemented was to go for fixed X-ray shields around the MMAs rather than the more complex deployable ones. Another benefit that was realized was the capability to launch to a higher orbit (600 km instead of 500 km) which guaranteed a longer lifetime (18 years) before re-entering the atmosphere.

\section{Launch and Commissioning} \label{Commis}

\begin{figure}[ht]
\caption{\textit{IXPE solar panels deploying after Observatory release from Falcon 9 launch vehicle.}} 
\label{fig:Solarpanels}
\centering
\includegraphics[width=2.5in, height=1.4in, keepaspectratio=true]{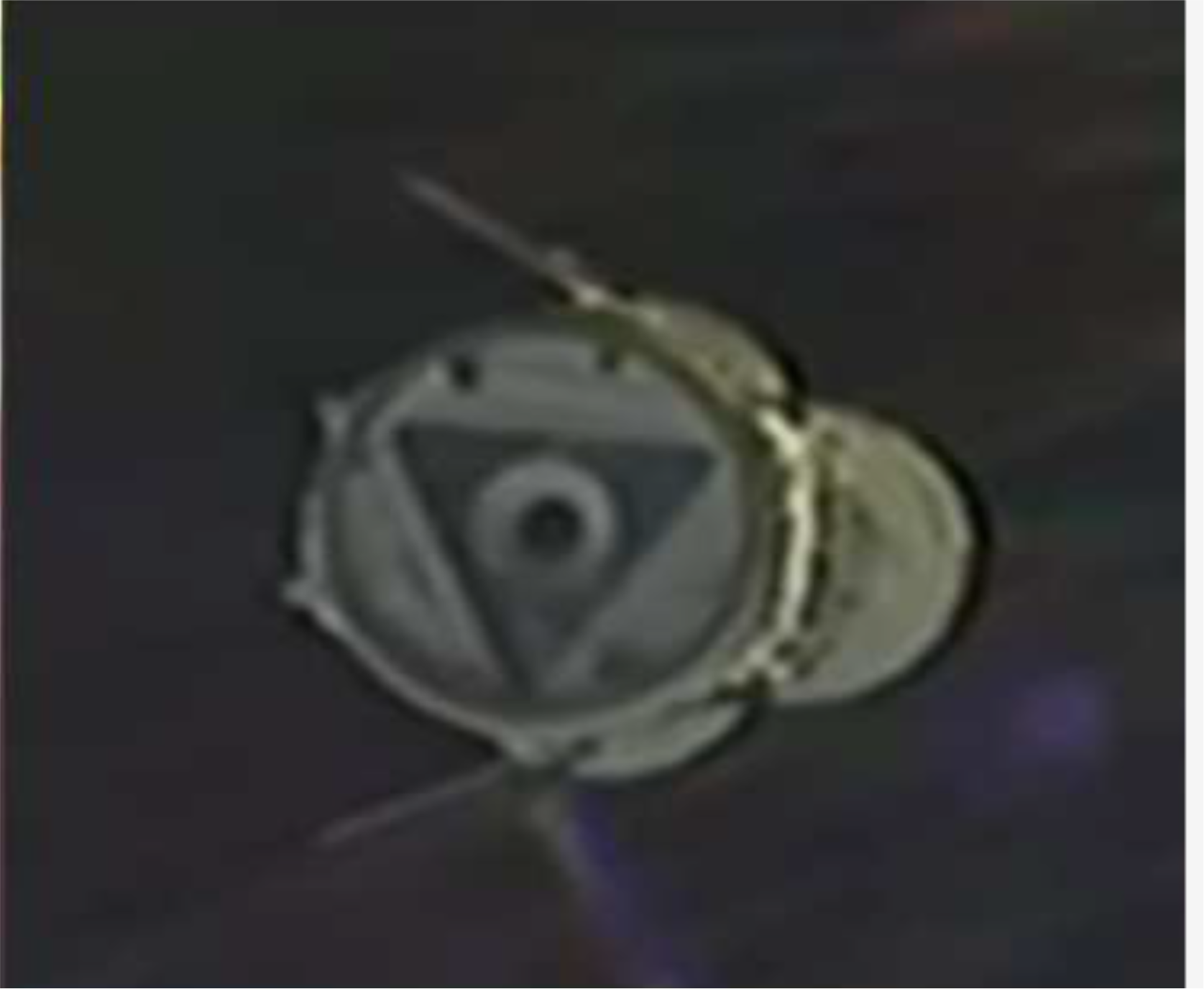}
\end{figure}

The IXPE Observatory was launched on a Falcon-9 rocket from Cape Canaveral at 1 am on 9th December 2021 and released into an equatorial orbit at 600 km altitude about 33 minutes later. Autonomous deployment of the solar panels took place shortly after (still visible from a camera aboard the rocket, see Figure \ref{fig:Solarpanels}) and the Observatory commenced approximately 1 month of commissioning.

After powering up and verifying the operation of the spacecraft the major event in the commissioning was the deployment of the boom. As detailed in section \ref{Boom}, this deployment relies on coiled-spring energy moderated by a lanyard and damper which governs the speed of deployment. Three launch locks attach the MMSS to the spacecraft deck with the compressed boom between them. The three locks must fire simultaneously to allow the boom to unfurl and deploy the optics.  There is no direct monitoring of the booms progress. Rather, data from the magnetometer on the optics bench must be used to verify the correct number of rotations ($\sim3.5$) for full deployment. This was done, and after a few minutes of settling the boom was deemed to be successfully deployed.

\begin{figure}[ht]
\caption{\textit{First composite IXPE image, of 1ES1959+650, showing mis-aligned telescopes.}} 
\label{fig:Firstimage}
\centering
\includegraphics[width=5.1in, height=2.83in, keepaspectratio=true]{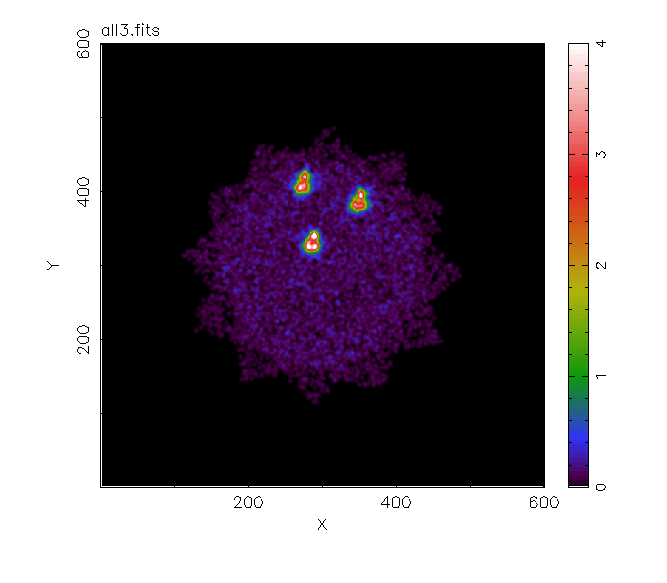}
\end{figure}

The commissioning continued with powering up the detectors, checking the gains with on-board calibration sources and then pointing at test target. These targets were chosen to be at different declinations to provide different thermal environments and assess the boom’s thermal stability.  Moving to the first of these targets, a BL Lac object 1ES1959+650, the composite image showed 3 distinct bright areas indicating that all three telescopes were working but that they were not co-aligned (Figure \ref{fig:Firstimage}). At this point, the tip/tilt/rotate (section \ref{Boom}) stage was used to co-align the 3 images (using the rotate stage) and to position the composite image at the center of the detectors (using the tip/tilt stages).

The first-light image also showed that the telescope images were extended, despite the target being a point source, and splitting the data into time slices showed that this was due to boom motion from the changing thermal environment around the orbit. As these motions repeat around each orbit, the time-resolved X-ray data were used to provide a correction to the boom motion for point sources. Later, a thermal model was incorporated to provide attitude corrections for fainter, extended X-ray sources. 

\section{Operations}
\subsection{Mission Operations}
IXPE is run from a Mission Operations Center (MOC) at the Laboratory for Atmospherics and Space Physics at the University of Colorado. The MOC is responsible for operating both the spacecraft and the payload. It monitors the health and safety of the Observatory, uploads science target information and downloads science and engineering data.

The data are downlinked at 2 Mbps, so that $\sim100$ Mbytes can be downlinked in a single pass. Typically, there will be 7-8 contacts per day with the Malindi ground station, which due to its equatorial location is visible on each of the daily $\sim15$ orbits. More contacts can be requested as required to handle bright targets which will often have their observations times split with fainter targets to allow for the on-board data recorder (6 Gbyte capacity) to unload.

IXPE’s orbit just clips the edge of the South Atlantic Anomaly (SAA), an area of high changed particle background, and so to be conservative the detector high voltage is reduced for passages to avoid any chance of detector damage. This is done based on geographical location and amounts to a maximum of about 10 minutes of time per orbit. This same 'SAA mode’ voltage reduction is also applied automatically if detector count rates hit a pre-determined maximum value, as sometimes happen when solar flares occur. In this case the detectors stay in this reduced high voltage configuration until the next exit from the SAA.

\subsection{Science Operations} \label{Sciops}
The Science Operations Center, in Huntsville, Alabama, is responsible for all science operations. This includes formulating a long-term (9-12 month) observing plan, portions of which are sent to the MOC on a weekly basis for Observatory scheduling and handling requests for targets of opportunity, as they arise.

The SOC also processes all science and payload housekeeping data downloaded from the Observatory. The processed science data are delivered to the High Energy Astrophysics Science Archive Research Center (HEASARC) at NASA’s Goddard Space Flight Center for dissemination to the public, typically within one week of the end of the observation. Standard analysis tools are available for processing the IXPE data, making them easily usable by the community.

The data products provided by the SOC include:
1) Level-1 FITS files containing the individual photoelectron tracks registered in the detectors with information in detector co-ordinates;
2) Level-2 FITS files containing event lists of processed tracks with polarization information in sky co-ordinates;
3) Housekeeping data to allow the Italian Instrument team to monitor the health of the detectors.

The polarization information of a celestial source is retrieved on the ground by subtracting low-energy, percent-level spurious modulation, mapped during calibration and expressed in terms of Stokes parameters, and subsequently deriving the polarization degree and angle. The algorithm used to determine both the emission direction of the photoelectron track and the photon impact point is described in \cite{Bellazzini2003} and more recently in \cite{Fabiani2014b}. This simple algorithm uses the first three moments of the track, and evaluates the emission direction as the one which maximizes the second moment with respect to the impact point. The procedure for extracting polarization parameters follows the method outlined in \cite{Kislat2015}, while the correction for spurious modulation is based on the approach presented in \cite{Rankin2021}. Additional capabilities, such as event time-tagging and pixel charge digitization, enable polarization measurements that are resolved in angle, energy, and time. 

The measurement of the polarization information is performed after some corrections to the data performed automatically in the flight-data pipeline. These corrections take account of:

\begin{enumerate}
    \item A small decrease in quantum efficiency (QE) over time, giving rise to a gradual (and small) increase in the modulation factor and a gradual increase in gas gain.
    \item A rate-dependent gain reduction (due to GEM charging).
    \item An aspect correction for boom thermo-elastic motion.
\end{enumerate}

(1) The decrease in QE is caused mainly by the absorption of the DME gas mixture by the detector construction glue (MASTERBOND SUPREME 10HT), as demonstrated by dedicated post-launch experiments \citep{Tomaiuolo2024}. The Q.E. decreases linearly at a constant rate of approximately 2\% per year, consistent with bulk absorption. The decrease is accompanied by an increase in the length of the photoelectron track and a corresponding increase in the modulation factor, at a rate of approximately 0.3\% per year. Since the sensitivity to polarization depends linearly on the modulation factor but only on the square root of the quantum efficiency, the overall impact of the QE change on the polarization sensitivity is minimal. The decrease in QE is monitored in flight by measuring the rate and event track length of the on-board calibration sources, supported by Monte Carlo simulations, which also provide estimates of the modulation factor at a given flight date. 

(2) The GEM amplifies the photoelectron track, but charge accumulation on the GEM walls or exposed surfaces reduces the electric field responsible for primary charge multiplication. The amount of charge adhering to these surfaces is governed by an equilibrium between the rates of attachment and subsequent release, which, in turn, depends on the local charge density rate (keV/mm$^2$/s). Although pre-flight calibrations and on-board calibration sources were used to model and correct for this behavior, the correction applied in the flight data pipeline is approximate. For very bright sources such as the Crab Nebula, this leads to an inaccuracy in the measured power-law index of up to $\sim$20$\%$ compared to the expected value. This discrepancy is mitigated by allowing the energy-channel calibration to vary freely during spectral fitting. 
%\begin{figure}
%    \centering
%    \includegraphics[width=0.5\linewidth]{figures/ChargingMiscalibration.png}
%    \caption{Despite the modeling of the charging effect and the use of on-board unpolarized calibration sources at 1.7 keV and 5.9 keV, a residual miscalibration of 20–25 eV remains in the data. This can be seen in the figure when the correction is applied to the two on-board polarized sources energies.}
%    \label{fig:Charging Miscalibration}
%\end{figure}

(3) %The original design of IXPE included a metrology system intended to monitor, and thereby facilitate corrections for, potential boom motions, which could arise from thermo-elastic effects when the satellite transitions between Earth's shadow and sunlight. However, this metrology system was eventually removed for programmatic reasons, following a Structural, Thermal, and Optical Performance (STOP) analysis which showed that these thermal effects were small.
After launch, a residual, continuous motion of a point-like X-ray source, with an amplitude of about one arcminute, was observed around each orbit as boom thermal conditions continually changed. For bright point-like sources, this shift could be easily corrected by slicing the data in time and computing time-dependent image centroids. However, for faint or extended sources, this correction was not feasible.
By analyzing the correlation between the Observatory's aspect, as determined independently by its two star trackers (one forward-facing on the end of the boom and the other the other backward-facing on the satellite bus) and data from onboard temperature sensors, it was possible to model with high fidelity the expected boom motion. These data could then be used to predict boom motions for portions of the orbit where only one tracker was available. Boom motions now contribute negligibly to the Observatory angular resolution.

%\begin{figure}
%    \centering
%    \includegraphics[width=0.5\linewidth]{figures/BoomShift.png}
%    \caption{Motion of the satellite's pointing axis during an observation of a typical celestial source, shown for different segments of the orbit. The dotted line represents the modeled shift derived from temperature sensor data. }
%    \label{fig:BoomMotion}
%\end{figure}

The on-board calibration sources (see \S\ref{Detector Unit}) are routinely used to correct the gain and to monitor the detector's performance over time. Each Filter and Calibraton Wheel is rotated one at a time into position during the occultation of celestial sources, allowing the acquisition of events produced by the selected calibration source. Gain calibration with two flood unpolarized sources at 1.7 keV and 5.9 keV also permit checks for any changes in spurious modulation and energy resolution, while the modulation factor is checked less frequently at the two energies (3 keV and 5.9 keV) of the composite polarized source. The energy resolution has been found to be constant over time starting from pre-flight measurements, indicating that the detector fill gas has maintained its purity throughout. 

An additional position on the filter and calibration wheels is equipped with a Kapton film, which reduces the flux by approximately a factor of three \citep{LaMonaca2021,Ferrazzoli2020,Soffitta2021}. So far, this filter has been used in just two observations: one of Sco X-1 \citep{Lamonaca2024} and another of Swift J1727.8–1613 \citep{Veledina2023}. Its purpose is to mitigate the impact of the relatively long dead time (1.1 ms) of the Detector Units (DUs) by reducing the incident flux.

\begin{table}
\centering
\caption{The list of different sources observed by IXPE through the end of June 2025. Some of these sources were re-observed in order to catch possible changes of the polarization properties in different states. PWNe = pulsar wind nebulae, SNR = supernova remnants, BH = black holes, NS = neutron stars, WD = white dwarfs, AGN = active galactic nuclei.}
\begin{tabular}{|p{4cm}|p{6cm}|}\hline  
 & \textbf{Number}\textbf{ of objects} \\\hline
PWNe and isolated pulsars (11 targets)& Crab PWN, Vela PWN, MSH 15-52, PSR B0540-69, G21.5, 3C 58, PSR B1259-63, KES-75, PSR J1723-283.7, Lighthouse N, PSR J1023 +0038\\\hline
SNRs (6 targets,7 pointings)& Cas A, Tycho’s, NE SN 1006, RCW 86, RX J1713.7-3946, Vela Jr, SN1006SW \\\hline
Accreting stellar-BH (17 targets)& Cyg X-1, 4U 1630-472, Cyg X-3, LMC X-1, SS433, 4U 1957-115, SS 433 Lobe East, LMC X-3, SWIFT J1727.8-1613, 4U 1957+115, Swift J0243.6+6124, Swift J1727.8-1613, GX 339-4, SWIFT J151857.0-572, MAXI J1744-294,  SS 433 Lobe West, GRS 1915+105\\\hline
Accreting NS \& WD (38 targets)& Cen X-3, Her X-1, GS1826-67, Vela X-1, Cyg X-2, GX 301-2, Xpersei, GX 9-9, 4U 1820, GRO J1008-57, XTE 1701-46, EXO 2030+375, LS V+44 17, GX 5-1, 4U 1624-49, Sco X-1, Cir X1, GX13+1, SMC X-1, SRGA J144459.2-604207, 4U 1538-52, V395 CAR, PSR J1023+00, GX 340+0, GX 3+1, 4U 1728-34, PSR J1723-2837, 4U 1735-44, GX 9+1, GX 349+2, 4U 1538-52, GX 17+2, 4U 1907+09, EX HYDRAE, 4U 1700-377, IGR J17091-3624,H 1417-624,UW CRB \\\hline
Magnetars (6 targets)& 4U 0142+61, 1RXS J170849, SGR 1806-20, 1E 2259+586, 1E 1841-045, 1E 1547.0–5408 \\\hline
Radio-quiet AGN \& 1 Sgr A* (10 targets)& MCG 5-23-16 (Sey1), Circinus Galaxy (Sey2-CT), NGC 4151 (Sey1), IC 4329A (Sey1), Sgr A* Complex, NGC 1068 (Sey2-CT), NGC 4945 (Sey2-CT), NGC 2110 (Sey2), NGC 3227 (Sey1), NGC 5506 (Sey1), 1ES 1927+654 (Sey2) \\\hline
Blazars \& radio galaxies (20 targets)& Cen A, S5-0716-714, 1ES 1959-650, Mrk 421, BL Lac, 3C 454, 3C 273, 3C 279, Mrk 501,1ES 1959-650, BL-Lac, 1ES 0229-200, PG 1553 -113, S4 0954+65, 1E 2259+586, RGB J0710+591, H 1426+428, 1ES 1101-232, PICTOR A WEST, 1ES 1927+654\\ \hline
\end{tabular}
\label{tab:IXPE Observed Sources}
\end{table}

% Requires: \usepackage{graphicx}
%\begin{table}[h]
%    \centering
%    \begin{tabular}{|l|l|}
%    \hline
%    \textbf{Mission} & \textbf{Allowed Sun Angle (deg)} \\ \hline
%    HST    & 60-180   \\ \hline
%    IXPE   & 56-124 \\ \hline
%    JWST   & 85-135 \\ \hline
%    NICER  & 45-180               \\ \hline
%    NuSTAR & 43-180               \\ \hline
%    SWIFT  & 47-180               \\ \hline
%     XMM& 70-110 (+/-20 deg)   \\ \hline
%    XRISM  & 60-120                \\ \hline
%    \end{tabular}
%    \caption{Allowed Sun Aspect Angles for Various Missions. IXPE during its observational phase increased the allowed Sun Aspect Angles from $\pm$ 25$^\circ$ to $\pm$ 34$^\circ$ for a simultaneous visible sky fraction of about 50$\%$}
 %   \end{table}     
As of the end of June 2025, 13 Targets of Opportunity (ToOs) under Director's Discretionary Time (DDT) were executed out of approximately 54 proposed by the General Users. All accepted ToOs were selected on the basis of scientific merit and feasibility, including compatibility with the requested IXPE repointing time. The mission operations control team requires a  time-to-target  of no less than 72 working hours from the ToO request.

%\textbf{Paolo to add:
%Ground analysis of tracks.
%Monitoring health of detectors. Periodic adjustments for gain/pressure changes
%Correcting data for charging, pressure loss (explain), boom motion, etc.
%On-board calibration source routine, monitoring spurious mod, modulation factor, energy resolution, charging etc. Should we include this under science operations or have a new section ?}

\subsection{General Observer Program}
\label{GOP}
In early 2024, IXPE’s baseline 2-year science mission ended and a general observer (GO) program was begun. In this, the selection of IXPE targets is done by the scientific community via a peer-reviewed proposal process (during the baseline science mission targets were selected by IXPE science team members, although all the data were immediately available to the general public through HEASARC). The IXPE GO program is run by the guest observer facility at Goddard Space Flight Center. Cycles 1 and 2 of the GO program have been very successful with an oversubscription rate of 6 in terms of requested observing time.

\section{Science Highlights}
%\textbf{Paolo to write this section ($\sim4$ pages)}
%While the observing plan of the first two years of IXPE observation (from January 2022 to December 2024) were decided by the IXPE topical working groups, organized in seven subgroups, one for each class of sources for which was, in advance, expected significant polarization, the IXPE targets of the successive phases were decided by the whole interested community, contributing to enlarge the classes of sources beyond those decided for the topical working groups. Hereafter we will maintain the earlier subdivisions in classes highlighting IXPE observations that do not pertain to them. 

IXPE, operating now for over 3.5 years, has looked at $\sim110$ different individual targets spanning a wide range of source types. Table \ref{tab:IXPE Observed Sources} gives a list of these targets, arranged according to target class. In the following sections, we detail some brief science highlights for each of these classes.

\subsection{Pulsar Wind Nebulae and Isolated Pulsars}
Pulsar wind nebulae (PWNe) are structures of magnetized, relativistic plasma produced within supernova remnants (SNRs) or in the interstellar medium (ISM) when the ultrarelativistic wind from the central pulsar interacts with its surroundings. PWNe were among the first astrophysical sources in which nonthermal (synchrotron) emission was identified. To date (see Table~\ref{tab:IXPE Observed Sources}), IXPE has observed 11 objects of this class, all of which yielded significant X-ray polarization measurements. The most intriguing results come from imaging polarimetry of the Vela PWN \citep{Xie2022}. Radio maps show a high linear polarization of $\sim$ 60$\%$ in the outer nebula. IXPE resolved the inner nebula in X-rays, where the polarization fraction at the leading edge exceeds 
$60\%$, approaching the theoretical limit for synchrotron emission. Similarly, large polarization fractions have been measured in the Crab \citep{Bucciantini2023,Wong2024} and in MSH~15$-$52 \citep{Romani2023}. In the latter case, values of $\sim70\%$ are reported in the arcs surrounding the pulsar and toward the end of the jet, i.e., even closer to the synchrotron limit. 

The above measurements show that that the magnetic field is very uniform in these nebulae and that an acceleration mechanism that does not rely on turbulent magnetic fields, such as magnetic reconnection, should be considered. Interesting results were also obtained from phase-resolved analysis of the isolated pulsars in the Crab and in MSH~15$-$52. \cite{Wong2023} use external measurements of the Crab nebular morphology and the pulsar light curve to disentangle their respective contributions to the phase- and spatially-resolved polarization via a single, simultaneous fit. Using six bins over $3\sigma$ in the Main Pulse (MP) and two bins over $3\sigma$ in the Interpulse (IP), \citep{Wong2024} find that the trend of the pulsar’s polarization angle in the MP is broadly consistent with optical measurements \citep{Slowikowska2009}, whereas the polarization degree differs markedly. Similar results are found for the IP, with the polarization angle consistent, but the degree not. \cite{Gonzalez2025} demonstrated that the IXPE Stokes parameters of the pulsar can be derived by applying a linear transformation to the optical Stokes parameters, indicating that both bands may originate from a common synchrotron emission mechanism.
They compared the phase-resolved polarization angle swing of the optical data with that measured by IXPE. In the Main Pulse (MP) they found a phase shift consistent with zero, whereas in the Interpulse (IP) they observed a phase shift that varies across different observations, the origin of which remains unexplained. As already noted by \cite{Slowikowska2009}, existing models of emission sites and mechanisms fail to reproduce optical measurements; the IXPE results provide additional independent constraints for modeling the Crab emission.

%An illustrative case study is the IXPE observation of the pulsar-wind nebula G21.5 \cite{DiLalla2025}. Its angular extent (\(1.3'–1.5'\)) is only slightly larger than the instrument point-spread function (PSF; \(25''–30''\)). To carry out an angularly resolved polarization analysis, we accounted for, and minimized, the “polarization leakage” caused by the previously reported inversion between the impact point and the Bragg-peak location in the track reconstruction \cite{Bucciantini2023,Soffitta2012}. This inversion degrades polarimetry in the presence of steep spatial intensity gradients. The mitigation strategy described in \cite{Cibrario2023,Cibrario2025} was successfully applied to the G21.5 observation. 

The growing set of IXPE measurements now allows the diverse X-ray polarization properties of these systems to be input to improve current models as these currently fail to reproduce the system's full complexity. As noted above, turbulence appears to play a much more limited role than was anticipated before IXPE. 

\subsection{Supernova Remnants}
Supernova remnants are characterized by the shock waves created by the collision of the ejected stellar material with the ambient interstellar gas, leading to X-ray and radio emissions, and sometimes gamma-rays. 
Fast forward shocks and strong magnetic field amplification can accelerate hadrons and electrons to $\ge$ 100 TeV, the latter providing synchrotron emission in X-rays. These shocks are typical of young supernova remnants (not older than 3000-4000 yrs). X-ray polarimetry is capable of determining the orientation and the uniformity (level of turbulence) of the magnetic field, providing insight into the acceleration mechanism and the ambient conditions very close to the acceleration sites.

IXPE observed 6  young SNRs with 7 separate pointings.  The results are reported in Table \ref{tab:SNR results}. 
Before the launch of IXPE, it was known that in the radio band young SNRs, as opposed to older ones, are characterized by a magnetic field oriented radially with respect to the shock front (i.e., approximately along the shock normal) \citep{Dubner2015}. Such patterns were interpreted as either genuine radial stretching driven by hydrodynamic/MHD instabilities or, conversely, as a selection effect in which particle acceleration is most efficient for a radial field.

IXPE observations of Cas~A \citep{Vink2022}, Tycho \citep{Ferrazzoli2023}, and SN~1006 (NE and SW) \citep{Zhou2023,Zhou2025} showed a predominantly radial magnetic field orientation as seen in the radio region. Because X-ray synchrotron emission traces TeV electrons very close to their acceleration sites (within $10^{17}\,\mathrm{cm}$ of the shock, about $2^{\prime\prime}$ at the distance of Cas~A), X-ray polarimetry indicates that the instabilities that produce the radial geometry are already operating very close to the shock fronts.

In contrast, for older/slower remnants such as RX~J1713.7$-$3946 \citep{Ferrazzoli2024} and Vela~Jr.\ \citep{Prokhorov2024}, IXPE found a tangential magnetic field orientation, consistent with shock compression of the upstream field.

As the degree of polarization is expected to be anti-correlated with the level of magnetic field turbulence, the different polarization levels reported in Table \ref{tab:SNR resuts} can be attributed primarily to different turbulence strengths. 
 A hint of positive correlation with the Bohm factor $\eta \equiv \lambda / r_{\mathrm{g}}$, where $\lambda$ is the mean free path of the particle and $r_{\mathrm{g}}$ is the gyroradius, is suggested by the data. By definition $\eta \ge 1$; the case $\eta = 1$ corresponds to the Bohm limit (maximally efficient scattering). The larger $\eta$ implies less efficient scattering and a more ordered effective field, which is consistent with a higher degree of polarization for synchrotron emission. The apparent discrepancy in the behavior of Vela~Jr.\ (NW), and to some degree RX J1713 (W), may be related to differences in the orientation of the magnetic field with respect to the shock-front normal, although a conclusive interpretation is still under debate.

% Requires \usepackage{booktabs}
\begin{table}[t]
\centering
\caption{X-ray polarization and shock properties for young SNRs. R denotes radial magnetic field, T tangential magnetic field and $n_0$ is the ambient density (upstream).}
\resizebox{\textwidth}{!}{
\begin{tabular}{lcccccccc}
\toprule
 & \multicolumn{3}{c}{Polarization Degree (\%)\textsuperscript{a}} 
 & $V_{\mathrm{shock}}$ 
 & $n_{0}$ 
 & Bohm Factor $\eta$ 
 & $B_{\mathrm{low}}$\textsuperscript{b} \\
\cmidrule(lr){2-4}
SNR & Rim & SNR & Peak & (km s$^{-1}$) & (cm$^{-3}$) &  & ($\mu$G) & orient.\\
\midrule
Cas A          & $  4.5 \pm 1.0$  & $2.5 \pm 0.5$ & $\sim 15$     & $\sim 5800$     & $0.9 \pm 0.3$       & $\sim 1$--$6$   & $25$--$40$ & R\\
Tycho          & $12 \pm 2$     & $9 \pm 2$     & $23 \pm 4$      & $\sim 4600$     & $\sim 0.1$--$0.2$   & $\sim 1$--$5$   & $30$--$40$ & R\\
SN 1006 (NE)   & $22.4 \pm 3.5$ & \dots         & $31 \pm 8$      & $\sim 5000$     & $\sim 0.05$--$0.08$ & $\sim 6$--$10$  & $18$--$26$ & R\\
SN 1006 (SW)   & $21.6 \pm 4.5$ & \dots         & $40 \pm 8$      & $\sim 5000$     & $\sim 0.5-16$    &  $\sim 8$--$16$      & $\sim 30$ & R\\
RX~J1713 (W)   & $13.0 \pm 3.5$ & \dots         & $36 \pm 10$     & $1400$--$2900$  & $\sim 0.01$--$0.2$  & $\sim 1.4$      & $ \sim 10$ & T\\
Vela Jr (NW)   & $16.4 \pm 5.2$ & \dots         & $85.2 \pm 30.1$ & $\sim 3000$     & $\le 0.01$          & $0.7 \pm 0.5$   &  $30$--$80$ & T\\
\label{tab:SNR results}
\end{tabular}}

% vshock Vela Jr. Camilloni et al. 2023 (A&A 673, A45), Table 5,
%H.E.S.S. Collaboration (2018), Deeper H.E.S.S. observations of Vela Junior: “…lack of thermal X-ray emission places an upper limit on the ambient density at ~0.01 %cm⁻³ (Slane et al. 2001).” 
%A&A
%Sushch, Brose & Pohl (2018), Modeling of the spatially resolved nonthermal emission from the Vela Jr. SNR: notes an upper limit of ~0.03 cm⁻³ from the same %argument.
\label{tab:SNR resuts}
\raggedright\footnotesize
Notes: \textsuperscript{a}Polarization degree for the rim, whole SNR, and peak region. 
\textsuperscript{b}Estimate of the downstream magnetic field. 
\end{table}

\subsection{Magnetars}
\label{sec:magnetars}
Magnetars are a class of neutron stars that combines both Anomalous X-ray Pulsars (AXPs), i.e., young neutron stars whose steady X-ray luminosity exceeds their spin-down power (unlike “normal” X-ray pulsars), and the Soft Gamma-ray Repeaters (SGRs), isolated neutron stars that emit repeated bursts of relatively soft $\gamma$ rays/hard X-rays. The term “magnetar” reflects their ultrastrong magnetic fields ($\sim10^{14}$–$10^{15}$\,G), proposed by \cite{Thompson2002} to account for both AXP and SGR phenomenology. IXPE has observed six magnetars so far (Table~\ref{tab:IXPE Observed Sources}).

Magnetar polarimetry with IXPE has two goals. The first is astrophysical: to determine the nature of the atmosphere responsible for the transport of the primary emission emitted at its surface and to constrain physical parameters. IXPE helped to determine the locations and geometries of the emitting regions (polar caps, equatorial belts, arc-like loops), and suggested the dominant magnetospheric processes (e.g., resonant Compton scattering, synchrotron emission). The second, more ambitious goal concerns fundamental physics: the search for \emph{vacuum birefringence} \citep{Euler1935}, a quantum electrodynamic effect whereby, above a critical magnetic field, photons polarized parallel and perpendicular to the field propagate with different phase velocities.

The six observed magnetars display markedly different polarimetric behaviors across the 2–8\,keV IXPE band:
\begin{itemize}
  \item \textbf{4U~0142+61} \citep{Taverna2022}: the measured degree of polarization (PD) decreases with energy and then, around the middle of the IXPE energy band, increases again; the polarization angle (PA) rotates by $\sim90^\circ$ near the middle of the energy band. 
  \item \textbf{1RXS~J170849.0$-$400910} \citep{Zane2023}: the measured PD increases steadily with energy, reaching values close to $80\%$, while the PA remains nearly constant.
  \item \textbf{SGR~1806$-$20} \citep{Turolla2023}: the measurements gave only upper limits on polarization, with a possible detection $\sim30\%$ in the central energy bin.
  \item \textbf{1E~1547.0$-$5408} \citep{Heyl2023}: the measurements reveal phase dependent polarization with a PD that reaches $\sim20\%$ at pulse minimum, and a PA shows a clear phase dependence similar to the rotating-vector model (RVM; \citep{Radhakrishnan1969,Poutanen2020}).
  \item \textbf{1E~1841$-$045} \citep{Rigoselli2025,Stewart2025}: the measured PD increases with energy up to $\sim65\%$, correlated with the emergence of a hard, non-thermal tail confirmed at higher energies by simultaneous \textit{NuSTAR} observations. A natural explanation for such high PD in the tail is magnetospheric synchrotron emission from high-energy electron–positron pairs.
\end{itemize}

Taken together, the current results point to a diversity of physical configurations: distinct surface emission mechanisms, different locations and morphologies of the emitting regions, and varying magnetospheric processes. Although most models already incorporate vacuum birefringence, it is not yet \emph{strictly} required by the data. Future detections of sources with a very small pulsed fraction but a very high polarization degree would strengthen the case for vacuum birefringence: in that regime, the observed polarization angle is expected to be set at large radii, where the magnetic field is effectively dipolar. This would constitute a characteristic QED signature, which is already compatible with the observed rotating-vector-model-like PA swings even in the presence of complex near-surface magnetic-field geometries.

\subsection{Radio Quiet Active Galactic Nuclei and Molecular clouds around the SGR A$^{*}$}
\label{sec:AGN&SRA^}

\subsubsection{Radio Quiet Active Galactic Nuclei}
\label{sec:AGN&SRA^}

A primary goal in the development of X-ray polarimetry was the study of Active Galactic Nuclei (AGN). These are intrinsically faint, which challenges photon-starved polarimeters, yet their measurement can be  highly rewarding for constraining the geometry of the inner engine, namely the accretion disk, the hot corona, ionized winds, and the molecular torus.

Unlike black-hole X-ray binaries, the accretion disk in AGN peaks in the optical–UV band. The observed X-rays are produced by inverse Compton scattering (Comptonization) of disk soft photons in a hot corona. A fraction of these primary X-ray photons reaches the observer directly; the remainder is reprocessed and observed after reflection/scattering by the accretion disk, the ionized wind, and the cold molecular torus. The relative contribution of these components strongly depends on the viewing angle. In Seyfert~1s (low inclination), the direct coronal emission and disk reflection dominate the spectrum. In Seyfert~2s (high inclination), reflection from the torus and scattering in the wind can become the only observable components.

IXPE has observed ten radio-quiet AGN and one molecular-cloud region (the Sgr~A$^{\ast}$ complex) near the Galactic Center. Of the ten AGN, only the bright Seyfert~1 galaxy NGC~4151 has been found to have statistically-significant levels of polarization  \citep{Gianolli2023,Gianolli2024}. Combining two observations, IXPE obtained $\mathrm{PD}=4.5\%\pm0.9\%$ and $\mathrm{PA}=81^{\circ}\pm6^{\circ}$. The polarization angle aligns well with the radio emission, supporting a geometry for the X-ray corona in line with what is observed in Galactic black-hole X-ray binaries (parallel to the accretion disk). However, a significant change in polarization angle measured at low energies ($37^{\circ}\pm7^{\circ}$ in the 2--3.5~keV bin) is not yet fully understood, whether due to leakage or the emergence of a different emission component.

These results are compatible with a slab or wedge corona and a moderate inclination angle, as expected for intermediate/changing-look AGN. The observations disfavor a lamppost geometry for the corona and therefore an aborted jet as the coronal origin.

Other Seyfert~1 AGN were observed, but IXPE has provided only upper limits so far. For MCG 5$-$23$-$16, for example, an upper limit of $\mathrm{PD}\le3.2\%$ was measured \citep{Marinucci2022,Tagliacozzo2023,Mondal2024} whereas for IC~4329A, the limit was $\mathrm{PD}\le6.1\%$\citep{Ingram2023,Pal2023}), both at the 99\% confidence level. Both AGNs show a hint that the polarization angle aligns with the ionization cones or radio jet, respectively.

Seyfert~2, and particularly Compton-thick AGN (those with $N_{\rm H}>1/\sigma_T = 1.5\times10^{24}\,\mathrm{cm^{-2}}$) were also observed. Two of these, the Circinus galaxy \citep{Ursini2023} and NGC~1068 \citep{Marin2024}, are found to be significantly polarized, with polarization angles that are (for Circinus) parallel to the H$_2$O maser direction---which traces the accretion disk---and (for NGC~1068) perpendicular to the radio structure. The polarization of the cold reflector (that is, the circumnuclear torus) is $\mathrm{PD}=28\%\pm7\%$ and $\mathrm{PA}=18^{\circ}\pm5^{\circ}$ for Circinus, and $\mathrm{PD}=20\%\pm10\%$ with $\mathrm{PA}=102^{\circ}\pm15^{\circ}$ for NGC~1068. Using Monte Carlo simulations and literature-based inclinations, the IXPE polarization results imply comparable torus semi-apertures ($\sim45^{\circ}$--$55^{\circ}$) for both AGN.

NGC 2110 \citep{Chakraborty2025,Pal2025} is classified as a low-luminosity AGN ($L_{\mathrm{bol}} \sim 10^{38}$--$10^{43}\ \mathrm{erg\ s^{-1}}$) and sub-Eddington, with the Eddington ratio $\lambda_{\mathrm{Edd}} \sim 10^{-2}$ to $10^{-5}$). IXPE measured an upper limit (99\%) to be less than 8.3\% assuming that the corresponding polarization angle (PA) is aligned with the radio jet, and less than 3.6\% if in the perpendicular direction. The PD--PA contour plot shows a low-significance preferential direction of the
polarization angle along the extended radio emission, as for NGC 4151.

\subsubsection{Molecular Clouds in the vicinity of the Galactic center}
\label{sec:MC clouds}

Molecular clouds are among the most demanding sources ever observed by IXPE. Yet, they are especially rewarding: their X-ray diagnostics can trace past activity of the Galactic Center (Sgr A*), which is presently faint in X rays, but may have been up to $10^6$ times more luminous in the past. Reflection of X rays from Sgr A* by nearby dense gas clouds provides a method for determining its past history. Given favorable geometry (and hence scattering angle), such radiation could be significantly polarized. 

IXPE's measurements of the faint, extended reflected emission embedded in hot thermal plasma help determine the true distances to the scattering clouds (without projection effects) and, therefore, the epochs of the illuminating flares. IXPE observed the Sgr~A complex—rich in hotspots of reflected emission (i.e. scattering molecular clouds), with the goal of constraining the three-dimensional geometry via polarimetry and thus dating the flares. The first observation, carried out in 2022 with a total exposure of $1$~Ms \citep{Marin2023}, yielded a polarization degree of $31\%\pm11\%$ and a polarization angle of $-48^\circ\pm11^\circ$ for the integrated Sgr~A complex, an angle consistent with Sgr A* as the source of the emission. If Sgr A* were the source it could have had a Seyfert-galaxy-level luminosity just $\sim$ 200 years ago. A subsequent IXPE observation \citep{Khabibullin2025} added complexity to the interpretation, allowing individual molecular clouds to be isolated and revealing the possibility of multiple candidate sources for the driving emission.

\subsection{Black-hole binaries}
\label{sec:BH binaries}
IXPE observations of black-hole X-ray binaries (BHXRBs) are summarized in~\cite{Dovciak2024}. Of the $\sim$10 persistent and $\sim$73 transient BHXRBs currently known, IXPE has observed 10 sources to date (see Table~\ref{tab:bhxrb_ixpe_all}). Several were targeted in different spectral states: the \emph{soft state} (SS), in which the X-ray emission is dominated by a multi-temperature disk blackbody, and the \emph{hard state} (HS), in which most of the flux arises from Comptonization of soft disk photons in a hot corona. These bracket two transitional states that often occur during outbursts: the \emph{hard intermediate state} (HIMS), a mixed spectrum where the disk is present but not yet dominant, and the \emph{soft intermediate state} (SIMS), a mixed spectrum with a larger disk contribution and markedly reduced fast variability.

During an outburst, sources typically trace a characteristic ‘q’-shaped track in the hardness–intensity diagram. They rise in luminosity from the HS (often by up to two orders of magnitude), pass through HIMS and SIMS into the SS near peak brightness, and then fade back toward the HS at lower luminosity before returning to quiescence.

IXPE observations of black hole sources in these various states has permitted a better understanding of the geometries and physical properties of matter around black-holes. The most significant findings are listed in the subsections below. 

\begin{table*}[ht!]
\centering
\caption{IXPE polarimetric measurements of BHXRB: source name, accretion state, band-integrated polarization degree (PD) and polarization angle (PA), and literature reference. Errors are $1\sigma$; upper limits follow the confidence reported in the cited works. Cyg X-1, GX 339-4 and Swift J1727.8$-$1613 showed a PA in HS parallel to their jet direction (These sources were the only ones in the IXPE sample for which both the PA and jet direction were available).}
\label{tab:bhxrb_ixpe_all}
\resizebox{\textwidth}{!}{
\begin{tabular}{l l c c l}
\toprule
\textbf{Source} & \textbf{State} & \textbf{PD [\%]} & \textbf{PA [deg]} & \textbf{Reference} \\
\midrule
% --- Cygnus X-1 (Kravtsov 2025 weighted where noted)
Cyg X--1 & HS\,1 (sum) & $3.9 \pm 0.2$ & $-23 \pm 2$ & \cite{Krawczynski2022} \\
Cyg X--1 & SS\,1 (sum) & $2.2 \pm 0.1$ & $-21 \pm 2$ & \cite{Steiner2024} \\
Cyg X--1 & HS\,2 (sum; weighted) & $4.0 \pm 0.3$ & $-29 \pm 2$ & \cite{Kravtsov2025} \\
Cyg X--1 & SS\,2 (sum; weighted) & $2.8 \pm 0.2$ & $-22 \pm 2$ & \cite{Kravtsov2025} \\
\addlinespace
% --- Cygnus X-3
Cyg X--3 & HS\,1       & $19.5 \pm 0.4$ & $89.9 \pm 0.5$ & \cite{Veledina2024} \\
Cyg X--3 & HS\,2       & $20.0 \pm 0.5$ & $91.8 \pm 0.7$ & \cite{Veledina2024} \\
Cyg X--3 & IMS         & $9.3 \pm 0.3$  & $92 \pm 1$     & \cite{Veledina2024} \\
Cyg X--3 & SS          & $11.9 \pm 0.5$ & $94 \pm 1$     & \cite{Veledina2024b} \\
\addlinespace
% --- Magellanic sources
LMC X--1 & SS          & $< 2.5$ (99\%) & ---            & \cite{Podgorny2023} \\
LMC X--3 & SS          & $3.2 \pm 0.6$  & $-42 \pm 6$    & \cite{Majumder2024} \\  & & & &\cite{Svoboda2024b} \\
4U 1957+11 & SS        & $1.9 \pm 0.6$  & $-41.8 \pm 7.9$ & \cite{Marra2024} \\
\addlinespace
% --- Swift J1727.8−1613 (values from uploaded papers)
Swift J1727.8$-$1613 & HIMS (sum; weighted) & $3.63 \pm 0.09$ & $2.2 \pm 0.7$ & \cite{Veledina2023},\\ &&&&\cite{Ingram2024} \\
Swift J1727.8$-$1613 & Soft state (combined) & $< 1.2$ (99\%) & --- & \cite{Svoboda2024} \\
Swift J1727.8$-$1613 & Dim hard (reverse transition) & $3.3 \pm 0.4$ & $3 \pm 4$ & \cite{Podgorny2024} \\
\addlinespace
% --- Transients and additional systems
GX 339$-$4 & Soft--intermediate (SIMS) & $1.3 \pm 0.3$ (3--8 keV) & $-74 \pm 7$ & \cite{Mastroserio2025} \\
           & SS                        & $\le 1.2$ ($3\sigma)$         &             &            \cite{Mastroserio2025}\\
4U 1630$-$47 & High/soft (thermal) & $8.32 \pm 0.17$ & $17.8 \pm 0.6$ & \cite{Ratheesh2024} \\
4U 1630$-$47 & Steep power-law (very high) & $6.75 \pm 0.21$ & $21.3 \pm 0.9$ & \cite{RodriguezCavero2023} \\
IGR J17091-3624 & HS                       & $9.1 \pm 1.6$   &  $83 \pm 5$    & \cite{Ewing2025}\\
SWIFT J151857.0 & HSS                      & $ \le 1.1$ (99\%) & --- &  \cite{Ling2024}\\
\bottomrule
\end{tabular}}
\end{table*}
\subsubsection{Corona}
\label{subsubsec: Corona}
Before \textit{IXPE}, the geometry of the X-ray corona was poorly constrained: spectroscopy alone mainly measures its temperature and optical depth and cannot uniquely determine its spatial configuration. The coronal location was inferred instead from X-ray reverberation lags between the hard coronal continuum and the reflected emission (traced by the Fe,K$\alpha$ line) originating in the disk \citep{Fabian2009}. These lags are consistent with a compact source on or near the black-hole spin axis (a so-called “lamppost” model), possibly associated with a failed jet, but degeneracies remained.

For Cygnus X-1, GX 339-4, and Swift J1727.8$-$1613 in HS, the IXPE measured X-ray PA is parallel to the direction of the radio jet. This favors a coronal geometry parallel to the accretion disk plane, i.e., a slab/sandwich-like corona above the disk or an inner hot flow occupying the region left-over by a truncated disk. Remarkably, for Swift J1727.8$-$1613 the IXPE-measured PD is comparable in both dim and bright HS, indicating a similar geometry across large luminosity changes.

A disk-parallel coronal geometry is likewise favored in several Seyfert AGN observed by IXPE: NGC 4151; IC 4329A; and MCG$-$5$-$23$-$16, \citep{Gianolli2024,Ingram2023,Tagliacozzo2023} suggesting that, despite the many-order-of-magnitude difference in black hole mass,
the innermost accretion geometry can be broadly scale invariant.

\subsubsection{Black-hole spin}
\label{subsubsec:BHSpin}
Since the 1970s it has been proposed that X-ray polarimetry—especially the energy dependence of  PA (its rotation with energy) and PD, can be used to measure the spin of a black-hole \citep{Connors1977,Stark1977}. The archetypal target for such a measurement was GRS\,1915+105, owing to its high inclination ($\sim 60^\circ$) and brightness. Unfortunately, shortly before the launch of IXPE the source entered a faint, obscured state \citep{Ratheesh2021} and subsequently returned to quiescence.

The most favorable conditions for spin measurements occur when a BHXRB is in the SS, where the emission is dominated by a geometrically thin, optically thick accretion disk. For a maximally rotating (prograde) Kerr black hole, the innermost stable circular orbit extends to
\[
r_{\rm ISCO} = 1\,r_g = \tfrac{1}{2}\,R_S,
\]
with $r_g \equiv GM/c^2$ and $R_S \equiv 2GM/c^2$.

In practice, measuring black hole spin via X-ray polarimetry proved challenging: in the soft state the PD is often lower than in the hard state, and for several sources \textit{IXPE} obtained only upper limits in their soft state (see Table~\ref{tab:bhxrb_ixpe_all}).

For LMC X-3, IXPE inferred a low spin value ($a_\ast \approx 0.2$), consistent with independent methods. For Cyg~X-1 and 4U\,1957+11, IXPE-derived constraints are compatible with high spin, with results broadly consistent with $a_\ast \gtrsim 0.96$, close to the maximum for the black hole size.

\subsection{Neutron Star binaries}
\label{sec:NeutronStarBinaries}
Neutron-star binaries are valuable laboratories for studying radiative processes in extreme environments. In pulsating systems, where mass transfer from a high-mass companion occurs through stellar winds and is funneled onto the neutron star magnetic poles, radiation propagates through a highly magnetized plasma shaped by the strong magnetic fields of young neutron stars ($\sim 10^{12}\,\mathrm{G}$).

In older low-mass X-ray binaries, mass transfer is thought to occur predominantly via Roche-lobe overflow. The accreting material forms a disk whose emission is intercepted by both the transition layer and the spreading layer on the neutron-star surface, enabling detailed investigations of the system geometry and physical parameters.

\subsubsection{X-ray binary Pulsars}
In accreting binary pulsars, generally young systems harboring a spinning neutron star with a field of $\sim$ 10$^{12}$ gauss, phase-resolved measurements of the position-angle swing further allow one to break the degeneracy between spectral modeling and the underlying geometrical parameters.

IXPE has investigated 14 pulsars so far, measuring for most of them the inclination and magnetic obliquity by applying the RVM~\cite{Radhakrishnan1969}, see Table \ref{tab:PulsarRVMResults}. Despite the often complex phase-dependent behavior of the degree of polarization, the RVM provides a good description in the majority of cases, with only a few exceptions (such as Vela~X$-$1 \cite{Forsblom2023} and 4U 1907+09 \cite{Zhou_M2025}).
Here we report the results in Table \ref{tab:PulsarRVMResults} from \cite{Poutanen2024}, where parameters of the binary pulsars investigated in the first two years by IXPE were measured.

% Requires: \usepackage{booktabs}
\begin{table}[ht!]
    \centering
    \caption{RVM results from IXPE observations of binary pulsars. The table from \cite{Poutanen2024}  has been updated with the most recent measurements.}

    \label{tab:PulsarRVMResults}
    \begin{tabular}{lccc}
        \toprule
        Name & $\iota_{\mathrm{p}}$ [deg] & $\theta_{\mathrm{p}}$ [deg] & $\chi_{\mathrm{p}}$ [deg] \\
        \midrule
        Cen X-3 & 70.2 (fixed) & $16.4 \pm 1.3$ & $49.2 \pm 1.1$ \\
        Her X-1 (main-on) & $56^{+24}_{-20}$ & $3.7^{+2.6}_{-1.9}$ & $42 \pm 2$ \\
        Her X-1 (short-on) & $90 \pm {30}$ & $16.3^{3.5}_{-4.1}$ & $57.9 \pm 2.1$ \\
        GRO J1008$-$57 & $130 \pm 3$ & $74^{+5}_{-4}$ & $75 \pm 4$ \\
        EXO 2030+375 & $128^{+8}_{-6}$ & $60^{+5}_{-6}$ & $-30^{+4}_{-5}$ \\
        X Persei & $162 \pm 12$ & $90 \pm 15$ & $70 \pm 30$ \\
        GX 301$-$2 & $135 \pm 17$ & $43 \pm 12$ & 135 \\
        LS V +44 17/Obs.\ 1 & $56 \pm 12$ & $27 \pm 4$ & $82 \pm 1$ \\
        LS V +44 17/Obs.\ 2 & $102 \pm 2$ & $54 \pm 1$ & $-6.2 \pm 0.4$ \\
        LS V +44 17\textsuperscript{a}     & $108 \pm 2$ & $48 \pm 1$ & $-8.4 \pm 0.6$ \\
        Swift J0243/Obs.\ 1 & $80 \pm 3$ & $87 \pm 2$ & $-70 \pm 4$ \\
        Swift J0243/Obs.\ 2 & $60 \pm 5$ & $88 \pm 3$ & $-87 \pm 7$ \\
        Swift J0243/Obs.\ 3 & $33 \pm 7$ & $75 \pm 5$ & $-66 \pm 7$ \\
        Swift J0243$^{a}$ & $25^{+8}_{+17}$ & $77^{+2}_{-29}$ & $-44^{+12}_{-13}$ \\
        SMC X-1 & $91^{+41}_{-42}$ & $13^{+7}_{-6}$ & $87 \pm 4$ \\
        4U 1538-52 \cite{Loktev2025} & $77^{+50}_{-32}$ & $30^{16}_{-13}$ & $19^{+10}_{-16}$\\ 
        4U 1626-67 \cite{Marshall2022}&  20–80     & Unconstr.   &    50–85   \\
        4U 1907+09 \cite{Zhou_M2025}&               RVM not applicable & &           \\
        Vela X-1 \cite{Forsblom2023}&               RVM not applicable & &           \\
        \bottomrule
\multicolumn{4}{l}{\footnotesize\textsuperscript{a} Including a phase-independent constant polarization component.} \\
    \end{tabular} 
\end{table}

Rather than detail each observation in Table \ref{tab:PulsarRVMResults} we
summarize the most relevant findings using a few representative cases. 

Depending on the measured accretion rate onto the magnetic poles and the 
resulting luminosity, theories predict different emission geometries. The polarization of the emergent X-ray radiation arises from the 
different opacities associated with photons polarized parallel (ordinary, ``O'' mode) 
or perpendicular (extraordinary, ``X'' mode) to the magnetic field. The correlation 
between polarization properties and pulsar intensity, therefore, depends sensitively on 
the accretion regime.

In the low-luminosity, \emph{sub-critical} regime, the accreting plasma forms a hot
spot at the intersection of the magnetic dipole axis with the neutron-star surface.
Radiation is predominantly emitted along the local surface normal, which usually 
coincides with the magnetic dipole direction. In this regime, the degree of polarization is 
expected to decrease with increasing pulse intensity.

In contrast, in the high-luminosity, \emph{super-critical} regime, radiation pressure 
acts on the in-falling plasma and inflates an accretion column whose axis is typically 
aligned with the magnetic dipole. Under these conditions, the degree of polarization 
is expected to increase with increasing pulsar intensity. 

The critical luminosity for the transition between the two regimes is $\sim 10^{37}$ erg s$^{-1}$
\cite{Mushtukov2015}.

Most IXPE observations of accreting X-ray pulsars have so far targeted systems 
operating in the \emph{sub-critical} accretion regime. 

Contrary to early theoretical expectations predicting polarization degrees as high 
as $80$--$100\%$, see for example \cite{Meszaros1988}, the largest values measured by IXPE are at most in the range of $\sim 10$--$15\%$. The origin of this discrepancy is not yet fully understood. 
It is likely related to the much more complex accretion environment than assumed 
in idealized models, which can modify, and in particular reduce, the net 
polarization observed at the detector. The origin of this discrepancy was already recognized after IXPE 
observed Cen X-3, its second accreting pulsar \cite{Tsygankov2022}. A relatively low degree of polarization (PD) is also possible at low mass–accretion rates because of the inverse temperature profile that develops in the accretion atmosphere \cite{Gonzales2019}, similar to what is observed in the atmospheres of magnetars.

Particularly important was the IXPE observation of Her~X--1 at different phases of 
its 35-day superorbital cycle. In addition to the phase-resolved measurements of the 
degree of polarization, the application of the RVM model to the polarization-angle 
variations revealed a change in the inferred inclination along this cycle. This 
behaviour suggests that the neutron-star crust undergoes free precession with the 
same 35-day period, potentially providing the physical origin of the superorbital 
modulation in the source intensity \cite{Heyl2024}. The free precession would arise 
from a slight asymmetry of the stellar moment of inertia, at the level of a few parts 
in ten million.

An interesting finding from IXPE is the nearly orthogonal magnetic-axis obliquity
inferred for GRO~J1808$-$57~\cite{Tsygankov2023} and X~Persei~\cite{Mushtukov2023}, 
in contrast to other X-ray binary pulsars. They also show similar inclinations 
of the magnetic and spin axes. The latter may imply that, notwithstanding the 
large natal kick, accretion torques, even during sporadic accretion episodes 
or at low accretion rates, have had sufficient time to align the spin and 
orbital angular momenta. The first finding, in turn, may provide 
insight into the physical characteristics of the progenitor.

Another peculiar IXPE case is the Be/X-ray binary (BeXRB) LSV+44~17 \cite{Doroshenko2023}. 
LSV+44~17 was likely observed in both subcritical and supercritical accretion regimes, corresponding 
respectively to the absence and presence of an accretion column. In the two observations, the RVM 
formally requires an unphysical, abrupt change in both the inclination and the direction of the 
rotation axis. However, the authors showed that the observed PA--phase dependence in both datasets 
can be reproduced without altering the pulsar geometry if a strongly polarized (PD $=10$--$30\%$, 
PA $\approx 70^\circ$) unpulsed component is included. This component may arise from scattering 
of X-rays emitted at the neutron-star poles by the wind of the Be star or the circumstellar disk.
The same conclusion regarding the presence of a constant, phase-independent polarization component 
may also apply to the ultra-luminous X-ray pulsar Swift~J0243.6+6214 \cite{Poutanen2024b}.

\subsubsection{Low Mass X-ray binaries and millisecond X-ray pulsars}
Low-mass X-ray binaries (LMXBs) are generally old systems harboring neutron stars with magnetic 
field strengths of order $\sim10^{8}$~G, more than four orders of magnitude weaker than those of 
typical X-ray pulsars. This relatively low field results from long-term accretion, which buries 
and screens the magnetic field, as well as from ohmic dissipation over the system’s lifetime. 
Millisecond pulsars are the evolutionary descendants of LMXBs: once sustained accretion ceases, 
the neutron star is left spinning rapidly (P $\approx$ 1--10~ms) with an even weaker magnetic field. 
If accretion resumes sporadically, the source is classified as an accreting millisecond pulsar (AMSP), 
and it may occasionally produce thermonuclear X-ray bursts, as in the archetypal SAX~J1808.4$-$3658.

The spectrum of an LMXB is often decomposed into a thermal component, arising from soft emission 
produced either by the accretion disk or by a spreading layer around the neutron star, and a hard 
component resulting from Comptonization of this thermal radiation in a boundary layer or within 
the spreading layer itself. LMXBs are further classified according to their color--color diagrams 
into Z sources, which are generally more luminous, and Atoll sources, which are typically less 
luminous. Occasionally, Comptonized radiation may, in turn, illuminate the disk, producing a 
reflection spectrum characterized by a fluorescent iron line and a high-energy hump in the 
20--30~keV range.

Unlike black-hole binaries, in systems hosting a neutron star the dominant radiative processes occur 
at distances of at least a few Schwarzschild radii. Consequently, any energy-dependent behavior of the 
polarization angle is difficult to attribute to general- or special-relativistic effects, which become 
significant only much closer to the compact object. Instead, such behavior must arise from geometrical 
configurations in which the emitting components are not strictly orthogonal to each other.

X-ray polarimetry with IXPE provides a powerful means to recover key geometrical parameters of these 
bright yet still poorly understood celestial sources: we summarize here the most important results obtained for both classes of sources, as discussed in  \cite{Ursini2024}. 

Atoll sources are generally weakly polarized systems (below 1-2\%, with few exceptions), 
with some evidence of an increasing degree of polarization at higher energies. This trend is typically 
interpreted as arising from a combination of Comptonization of radiation from the boundary layer and a 
reflected component from the disk, the latter being chiefly responsible for the rise with energy. 
Possible alternative explanations include a slab-like corona or scattering in an out-flowing wind.

In Z sources, IXPE observations have enabled polarization measurements along the full Z track, 
typically consisting of a Horizontal Branch (HB), a Normal Branch (NB), and a Flaring Branch (FB). 
In general, the HB spectrum is harder than in the other two branches, and correspondingly the 
polarization degree (PD) is higher---up to 4--5\% in the HB and only 1--2\% in the NB and FB. 
Thus, Z sources observed in different branches tend to exhibit larger polarization in the HB.

IXPE measured relatively low polarization degrees (1--2\%) in Cyg~X-2 \cite{Farinelli2023} and 
Sco~X-1 \cite{Lamonaca2024}, both observed only in their soft branches (NB and FB, respectively). 
These two objects are among the few NS--LMXBs known to host resolved radio jets. Remarkably, in 
Cyg~X-2 the polarization angle is aligned with the direction of the radio jet, while in Sco~X-1 the 
polarization is significantly rotated westward by $46^\circ \pm 9^\circ$ relative to the jet axis.
The latter case is particularly intriguing, as it may reflect the specific orientation and geometry 
of the various emitting and scattering regions in this very bright neutron-star LMXB.

A particularly peculiar source is GX~13+1. During the two-day IXPE observation, \cite{Bobrikova2024} 
reported a rotation of the polarization angle (PA) by approximately $70^\circ$, accompanied by variations in 
the polarization degree (PD), which changed from $\sim$2\% to non-detectable levels and then increased 
again to $\sim$5\%. Notably, these polarization changes were not associated with any visible variations 
in the spectral state of the source. A rotation of the PA consistent with a misalignment between the 
neutron-star spin axis and the orbital axis—again without significant spectral changes—was also found . 
As discussed by \cite{Dimarco2025}, these effects appear to also be related to transitions between the in-dip 
and out-of-dip intervals. The observed variations in PA and PD were therefore ascribed to a scattering 
component in an accretion-disk corona or a disk wind.

X-ray polarimetry of AMSPs can provide valuable constraints on their geometrical parameters, which is 
of paramount importance for determining the neutron-star equation of state, for example through the 
effects of surface gravity on the beam profile. Accreting millisecond pulsars are usually faint, but 
occasionally exhibit outbursts and thermonuclear bursts. The thermal radiation emerging from the surface 
is thought to be scattered at the shock front formed where the infalling plasma, producing a measurable 
degree of polarization.

SRGA \,J144459.2 is the first AMSP observed by IXPE \cite{Papitto2025}, and for the first time a significant polarization 
signal was detected: a degree of polarization of $2.3\% \pm 0.4\%$ at a polarization angle of 
$59^\circ \pm 6^\circ$. An upper limit of $<2\%$ (99\% LC) was measured below 3~keV, while a polarization 
of $4\% \pm 0.5\%$ was detected above this energy, consistent with the scattering of the infalling plasma. 
Unfortunately, the nearly flat phase-resolved light curves of the Stokes parameters prevented the 
measurement of the geometrical parameters required to model the pulse profile for mass-to-radius 
constraints.
 
\subsection{Blazars and Radio Loud AGNs}
\label{sec:Blazars&RLAGN}   
Blazars are active galactic nuclei (AGNs) whose relativistic jets are oriented close to the line of sight of the observer. The large bulk Lorentz factor of the jet and the associated Doppler boosting strongly enhance the non-thermal emission, effectively masking the radiation from the accretion disk and from the broad- and narrow-line regions. As a result, the spectral energy distribution (SED) is characterized by two major components (or ``bumps'').  

The low-energy bump is attributed to synchrotron emission from relativistic particles in the jet, extending from radio frequencies up to the optical--UV band and, in some cases, into the X-ray regime. The high-energy bump is generally interpreted as inverse Compton (IC) emission, produced when relativistic electrons (or possibly hadrons) upscatter soft photons to high energies. The origin of these seed photons is still debated: they may arise from the synchrotron emission itself (internal origin; synchrotron self-Compton), or from external photon fields associated with the accretion disk, the broad- and narrow-line regions, or even from the cosmic microwave background.  

The location of the X-ray emission within the SED provides an important diagnostic of the dominant radiative process. In synchrotron-dominated blazars, the optical--UV portion of the SED can be smoothly connected to the X-ray spectrum, indicating that the X-rays are produced by synchrotron radiation. In contrast, in inverse-Compton-dominated blazars, the extrapolation of the synchrotron component falls below the observed X-ray emission, which is typically characterized by a flatter spectrum.

IXPE and X-ray polarimetry provide a powerful tool for investigating both synchrotron-dominated and inverse-Compton-dominated blazars. In the former case, the measurements confirm the synchrotron origin of the emission, offering valuable insight into the particle-acceleration mechanisms, while in the latter they shed light on the leptonic origin of the observed emission.

Results fom synchrotron-dominated blazars are reported in Tab. \ref{tab:HSP_pol}.

\begin{table}
\centering
\caption{Contemporaneous X-ray and optical polarization of HSP blazars\cite{Marscher2024}}
\label{tab:HSP_pol}
\small
\begin{threeparttable}
\begin{tabularx}{\linewidth}{l c c c c c}
\toprule
\multirow{2}{*}{Object (Date)} 
& \multicolumn{2}{c}{X-ray} 
& \multicolumn{2}{c}{Optical R-band\tnote{a}} 
& \multirow{2}{*}{Jet PA ($^\circ$) at 43 GHz\tnote{b}} \\
\cmidrule(lr){2-3}\cmidrule(lr){4-5}
& $\Pi_X$ (\%) & $\psi_X$ ($^\circ$) 
& $\Pi_O$ (\%) & $\psi_O$ ($^\circ$) & \\
\midrule
1ES0229$+$200 (23 Jan 2023) 
& $18 \pm 3$ & $25 \pm 5$ 
& $2.4 \pm 0.7$ & $2 \pm 8$ 
& $163 \pm 8$ \\

\midrule
Mrk~421 (5 May 2022) 
& $15 \pm 2$ & $35 \pm 4$ 
& $2.9 \pm 0.5$ & $32 \pm 5$ 
& $-29 \pm 18$ \\
Mrk~421 (5 Jun 2022) 
& $10 \pm 1$ & Rotation 
& $4.4 \pm 0.4$ & $-40 \pm 6$ 
& $-29 \pm 18$ \\
Mrk~421 (8 Jun 2022) 
& $10 \pm 1$ & Rotation 
& $5.4 \pm 0.4$ & $-35 \pm 1$ 
& $-29 \pm 18$ \\
Mrk~421 (17 Dec 2022) 
& $14 \pm 1$ & $-73 \pm 3$ 
& $4.6 \pm 1.3$ & $26 \pm 9$ 
& $-29 \pm 18$ \\

\midrule
PG~1553$+$113 (2 Feb 2023) 
& $10 \pm 2$ & $86 \pm 8$ 
& $4.2 \pm 0.5$ & Rotation 
& $50 \pm 10$ \\

\midrule
Mrk~501 (7 Mar 2022) 
& $9.8 \pm 1.7$ & $136 \pm 5$ 
& $6.6 \pm 0.4$ & $110 \pm 5$ 
& $120 \pm 12$ \\
Mrk~501 (27 Mar 2022) 
& $10.3 \pm 1.4$ & $115 \pm 4$ 
& $4.7 \pm 0.3$ & $120 \pm 3$ 
& $120 \pm 12$ \\
Mrk~501 (9 Jul 2022) 
& $6.9 \pm 1.8$ & $134 \pm 8$ 
& $2.7 \pm 0.5$ & $109 \pm 5$ 
& $120 \pm 12$ \\
Mrk~501 (12 Feb 2023) 
& $9.0 \pm 2.4$ & $110 \pm 8$ 
& $6.6 \pm 0.9$ & $150 \pm 4$ 
& $120 \pm 12$ \\
Mrk~501 (19 Mar 2023) 
& $6.0 \pm 2.1$ & $107 \pm 11$ 
& $6.1 \pm 0.7$ & $125 \pm 3$ 
& $120 \pm 12$ \\
Mrk~501 (16 Apr 2023) 
& $18.5 \pm 2.2$ & $103 \pm 3$ 
& $5.9 \pm 1.5$ & $108 \pm 6$ 
& $120 \pm 12$ \\

\midrule
1ES~1959$+$650 (3 May 2022) 
& $8.0 \pm 2.3$ & $123 \pm 8$ 
& $4.5 \pm 0.2$ & $159 \pm 1$ 
& $120$--$150$ \\
1ES~1959$+$650 (10 Jun 2022) 
& $<5.1$\tnote{d} & --- 
& $4.7 \pm 0.6$ & $151 \pm 19$ 
& $120$--$150$ \\

\midrule
1ES~2155$-$304 (30 Oct 2023) 
& $31 \pm 2$ & $129 \pm 2$ 
& $4.3 \pm 0.7$ & $116 \pm 8$ 
& $135 \pm 45$ \\
1ES~2155$-$304 (4 Nov 2023) 
& $15 \pm 2$ & $125 \pm 4$ 
& $3.8 \pm 0.9$ & $116 \pm 8$ 
& $135 \pm 45$ \\
\bottomrule
\end{tabularx}
\begin{tablenotes}\footnotesize
\item Dates correspond to the middle of IXPE exposures.
\item[\textit{a}] Median optical polarization during the IXPE observation; corrected for host-galaxy dilution.
\item[\textit{b}] Determined from VLBA images.
\item[\textit{d}] When the 200 ks IXPE exposure is divided into four equal time bins, the third bin yields a significant detection.
\end{tablenotes}
\end{threeparttable}
\end{table}

IXPE reported only upper limits for low-synchrotron-peaked (LSP) blazars (see Table~\ref{tab:ixpe_lbl_ul}). A relatively low polarization degree is expected if the X-ray emission is produced via synchrotron self-Compton (SSC), and an even lower polarization degree is predicted if the emission originates from inverse Compton scattering of external photon fields, such as those from the accretion disk, the broad-line region, or the cosmic microwave background.

The leptonic (Compton scattering) versus hadronic origin of the jet emission is still under debate. This issue can be investigated by comparing the optical and X-ray polarization degrees: similar values would favor a hadronic origin, whereas significantly different values would support a leptonic scenario.

An important step forward in addressing this question was presented by \cite{Agudo2025}, who reported an optical polarization flare in BL~Lac with a polarization degree reaching up to 45\%, while simultaneous IXPE observations provided a stringent upper limit of 7.4\% in the X-ray band. As discussed by \citep{Zhang2013}, such a large discrepancy between optical and X-ray polarization favors a leptonic origin of the emission, whereas comparable polarization degrees would instead indicate a hadronic origin.

Indeed, IXPE has played a key role in constraining the inverse Compton origin of the high-energy component in low-synchrotron-peaked blazars. However, its current sensitivity is not yet sufficient to achieve a firm polarization detection in these sources. Most likely, future instruments with a significantly larger effective area will be required to obtain a robust measurement.

\begin{table}
\centering
\footnotesize
\setlength{\tabcolsep}{3pt}
\caption{IXPE upper limits on X-ray polarization for LBL/LSP blazars.}
\label{tab:ixpe_lbl_ul}
%\begin{tabular}{p{2.1cm} p{1.6cm} p{2.1cm} p{1.6cm}}
%\begin{tabular}{l c c l}
\begin{tabular}{p{2.1cm} p{1.6cm} >{\centering\arraybackslash}p{2.5cm} p{1.6cm}}
\toprule
Source & Epoch & $\Pi_{\rm X}$ UL & Ref. \\
\midrule
BL Lacertae 
& 2022 
& $<12$--$14\%$(99\%) 
& \citep{Middei2023} \\

BL Lacertae 
& 2023   
& $<7.4\%$(3$\sigma$) 
& \citep{Agudo2025} \\

S4 0954+65 
& 2023  
&$<$$14\%$(3$\sigma$) 
& \citep{Kouch2025} \\
&
&$<$$8.8\%$\textsuperscript{a}
&\\

3C 273 
& 2022& $\lesssim 9\%$ (99\%)& \citep{Marshall2024} (Tab.2) \\

3C 279 
& 2022& $\lesssim 12.7\%$ (99\%)& \citep{Marshall2024} (Tab.2) \\

3C 454.3& 2022-2023 &$\lesssim10\text{--}28\%$ (99\%) & \citep{Marshall2024} (Tab.2)\\

S5 0716+714 
& 2022& $\lesssim$ 26\% (99\%) 
& \citep{Marshall2024} (Tab.2)\\
\bottomrule
\multicolumn{4}{l}{\footnotesize\textsuperscript{a} Pol. Angle fixed} \\
\end{tabular}
\end{table}
 
We recall that IXPE and X-ray polarimetry can open new perspectives in the emerging field of multi-messenger astrophysics, particularly for blazars. The detection of neutrino emission from the blazar TXS~0506+056 \citep{ICECube2018}, together with IXPE results indicating a probable leptonic origin of Bl-Lac's  high-energy emission, still needs to be fully explored and consistently incorporated into models addressing the leptonic and hadronic scenarios for the production of high-energy radiation.

Neutrino emission has also been reported from the Seyfert~2 galaxy NGC~1068 \citep{Icecube2022}. In this source, IXPE revealed the geometrical structure of the molecular torus. However, for such a radio-quiet AGN, the connection between the X-ray polarization — arising from reflection of the inner nuclear emission — and the mechanisms responsible for neutrino production is far from straightforward.

\section{Current Status of Observatory}
At the time of writing, Summer 2025, IXPE is midway through its 4th year of operation. The Observatory has no consumables and so far no hardware failures. Originally designed for a 2-year mission, but with a 18-year projected orbit lifetime before re-entry, the current limiting factor appears to be battery life, specifically depth of discharge during eclipses. To address this, small changes have been made in operations to reduce power consumption during these times, and currently BAE predicts that IXPE can support science operations at least through 2030.

Future observations will include not only deeper studies of current targets, but also science cases beyond the primary mission, some in coordination with other missions. IXPE is changing our understanding of the X-ray sky in a fundamental way in many areas of astrophysics. The large oversubscription in the general observer cycles to date indicates that we can expect IXPE to provide continued scientific advances for the foreseeable future.

\textbf{A final note:} The groundbreaking nature of IXPE was recognized by the awarding of the 2024 Bruno Rossi Prize of the High Energy Astrophysics Division of the American Astronomical Society to Martin Weisskopf, Paolo Soffitta and the IXPE team for the development of IXPE “whose novel measurements advance our understanding of particle acceleration and emission from astrophysical shocks, black holes and neutron stars.” The Rossi Prize is awarded annually for significant contributions to high-energy astrophysics. 

In June 2025, Martin Weisskopf, Enrico Costa and Ronaldo Bellazzini were awarded the prestigious Antonio Feltrinelli prize for the “realization and success of the Imaging X-ray Polarimetry Explorer, this mission representing a historic milestone for contemporary astrophysics  …”. The award citation further detailed the scientific breakthroughs that IXPE had enabled.

In November 2025 the Ernst Mach medal for Physical Science of the Czech Academy of Science was awarded to Enrico Costa for his contributions to the development of X-ray polarimetry in Astrophysics.

\bibliography{References}{}

@ARTICLE{Agudo2025,
       author = {{Agudo}, Iv{\'a}n and {Liodakis}, Ioannis and {Otero-Santos}, Jorge and {Middei}, Riccardo and {Marscher}, Alan and {Jorstad}, Svetlana and {Zhang}, Haocheng and {Li}, Hui and {Di Gesu}, Laura and {Romani}, Roger W. and {Kim}, Dawoon E. and {Fenu}, Francesco and {Marshall}, Herman L. and {Pacciani}, Luigi and {Escudero Pedrosa}, Juan and {Aceituno}, Francisco Jos{\'e} and {Ag{\'\i}s-Gonz{\'a}lez}, Beatriz and {Bonnoli}, Giacomo and {Casanova}, V{\'\i}ctor and {Morcuende}, Daniel and {Piirola}, Vilppu and {Sota}, Alfredo and {Kouch}, Pouya M. and {Lindfors}, Elina and {McCall}, Callum and {Jermak}, Helen E. and {Steele}, Iain A. and {Borman}, George A. and {Grishina}, Tatiana S. and {Hagen-Thorn}, Vladimir A. and {Kopatskaya}, Evgenia N. and {Larionova}, Elena G. and {Morozova}, Daria A. and {Savchenko}, Sergey S. and {Shishkina}, Ekaterina V. and {Troitskiy}, Ivan S. and {Troitskaya}, Yulia V. and {Vasilyev}, Andrey A. and {Zhovtan}, Alexey V. and {Myserlis}, Ioannis and {Gurwell}, Mark and {Keating}, Garrett and {Rao}, Ramprasad and {Kang}, Sincheol and {Lee}, Sang-Sung and {Kim}, Sanghyun and {Cheong}, Whee Yeon and {Jeong}, Hyeon-Woo and {Angelakis}, Emmanouil and {Kraus}, Alexander and {Blinov}, Dmitry and {Maharana}, Siddharth and {Bachev}, Rumen and {Jormanainen}, Jenni and {Nilsson}, Kari and {Fallah Ramazani}, Vandad and {Casadio}, Carolina and {Fuentes}, Antonio and {Traianou}, Efthalia and {Thum}, Clemens and {G{\'o}mez}, Jos{\'e} L. and {Antonelli}, Lucio Angelo and {Bachetti}, Matteo and {Baldini}, Luca and {Baumgartner}, Wayne H. and {Bellazzini}, Ronaldo and {Bianchi}, Stefano and {Bongiorno}, Stephen D. and {Bonino}, Raffaella and {Brez}, Alessandro and {Bucciantini}, Niccol{\`o} and {Capitanio}, Fiamma and {Castellano}, Simone and {Cavazzuti}, Elisabetta and {Chen}, Chien-Ting and {Ciprini}, Stefano and {Costa}, Enrico and {De Rosa}, Alessandra and {Del Monte}, Ettore and {Di Lalla}, Niccol{\`o} and {Di Marco}, Alessandro and {Donnarumma}, Immacolata and {Doroshenko}, Victor and {Dov{\v{c}}iak}, Michal and {Ehlert}, Steven R. and {Enoto}, Teruaki and {Evangelista}, Yuri and {Fabiani}, Sergio and {Ferrazzoli}, Riccardo and {Garc{\'\i}a}, Javier A. and {Gunji}, Shuichi and {Hayashida}, Kiyoshi and {Heyl}, Jeremy and {Iwakiri}, Wataru and {Kaaret}, Philip and {Karas}, Vladimir and {Kislat}, Fabian and {Kitaguchi}, Takao and {Kolodziejczak}, Jeffery J. and {Krawczynski}, Henric and {La Monaca}, Fabio and {Latronico}, Luca and {Maldera}, Simone and {Manfreda}, Alberto and {Marin}, Fr{\'e}d{\'e}ric and {Marinucci}, Andrea and {Massaro}, Francesco and {Matt}, Giorgio and {Mitsuishi}, Ikuyuki and {Mizuno}, Tsunefumi and {Muleri}, Fabio and {Negro}, Michela and {Ng}, Chi-Yung and {O'Dell}, Stephen L. and {Omodei}, Nicola and {Oppedisano}, Chiara and {Papitto}, Alessandro and {Pavlov}, George G. and {Peirson}, Abel L. and {Perri}, Matteo and {Pesce-Rollins}, Melissa and {Petrucci}, Pierre-Olivier and {Pilia}, Maura and {Possenti}, Andrea and {Poutanen}, Juri and {Puccetti}, Simonetta and {Ramsey}, Brian D. and {Rankin}, John and {Ratheesh}, Ajay and {Roberts}, Oliver J. and {Sgr{\`o}}, Carmelo and {Slane}, Patrick and {Soffitta}, Paolo and {Spandre}, Gloria and {Swartz}, Douglas A. and {Tamagawa}, Toru and {Tavecchio}, Fabrizio and {Taverna}, Roberto and {Tawara}, Yuzuru and {Tennant}, Allyn F. and {Thomas}, Nicholas E. and {Tombesi}, Francesco and {Trois}, Alessio and {Tsygankov}, Sergey S. and {Turolla}, Roberto and {Vink}, Jacco and {Weisskopf}, Martin C. and {Wu}, Kinwah and {Xie}, Fei and {Zane}, Silvia},
        title = "{High Optical-to-X-Ray Polarization Ratio Reveals Compton Scattering in BL Lacertae's Jet}",
      journal = {\apjl},
         year = 2025,
        month = may,
       volume = {985},
       number = {1},
          eid = {L15},
        pages = {L15},
          doi = {10.3847/2041-8213/adc572},
archivePrefix = {arXiv},
       eprint = {2505.01832},
 primaryClass = {astro-ph.HE},
       adsurl = {https://ui.adsabs.harvard.edu/abs/2025ApJ...985L..15A}
}

@ARTICLE{Baldini2021,
title = {Design, construction, and test of the Gas Pixel Detectors for the IXPE mission},
journal = {Astroparticle Physics},
volume = {133},
pages = {102628},
year = {2021},
issn = {0927-6505},
doi = {https://doi.org/10.1016/j.astropartphys.2021.102628},
url = {https://www.sciencedirect.com/science/article/pii/S0927650521000670},
author = {L. Baldini and M. Barbanera and R. Bellazzini and R. Bonino and F. Borotto and A. Brez and C. Caporale and C. Cardelli and S. Castellano and M. Ceccanti and S. Citraro and N. {Di Lalla} and L. Latronico and L. Lucchesi and C. Magazzu' and G. Magazzu' and S. Maldera and A. Manfreda and M. Marengo and A. Marrocchesi and P. Mereu and M. Minuti and F. Mosti and H. Nasimi and A. Nuti and C. Oppedisano and L. Orsini and M. Pesce-Rollins and M. Pinchera and A. Profeti and C. Sgro' and G. Spandre and M. Tardiola and D. Zanetti and F. Amici and H. Andersson and P. Attina' and M. Bachetti and W. Baumgartner and D. Brienza and R. Carpentiero and M. Castronuovo and L. Cavalli and E. Cavazzuti and M. Centrone and E. Costa and E. DAlba and F. D'Amico and E. {Del Monte} and S. {Di Cosimo} and A. {Di Marco} and G. {Di Persio} and I. Donnarumma and Y. Evangelista and S. Fabiani and R. Ferrazzoli and T. Kitaguchi and F. {La Monaca} and C. Lefevre and P. Loffredo and P. Lorenzi and E. Mangraviti and G. Matt and T. Meilahti and A. Morbidini and F. Muleri and T. Nakano and B. Negri and S. Nenonen and S.L. O'Dell and M. Perri and R. Piazzolla and S. Pieraccini and M. Pilia and S. Puccetti and B.D. Ramsey and J. Rankin and A. Ratheesh and A. Rubini and F. Santoli and P. Sarra and E. Scalise and A. Sciortino and P. Soffitta and T. Tamagawa and A.F. Tennant and A. Tobia and A. Trois and K. Uchiyama and M. Vimercati and M.C. Weisskopf and F. Xie and F. Zanetti and Y. Zhou}
}

@ARTICLE{Bucciantini2023,
       author = {{Bucciantini}, Niccol{\`o} and {Ferrazzoli}, Riccardo and {Bachetti}, Matteo and {Rankin}, John and {Di Lalla}, Niccol{\`o} and {Sgr{\`o}}, Carmelo and {Omodei}, Nicola and {Kitaguchi}, Takao and {Mizuno}, Tsunefumi and {Gunji}, Shuichi and {Watanabe}, Eri and {Baldini}, Luca and {Slane}, Patrick and {Weisskopf}, Martin C. and {Romani}, Roger W. and {Possenti}, Andrea and {Marshall}, Herman L. and {Silvestri}, Stefano and {Pacciani}, Luigi and {Negro}, Michela and {Muleri}, Fabio and {de O{\~n}a Wilhelmi}, Emma and {Xie}, Fei and {Heyl}, Jeremy and {Pesce-Rollins}, Melissa and {Wong}, Josephine and {Pilia}, Maura and {Agudo}, Iv{\'a}n and {Antonelli}, Lucio A. and {Baumgartner}, Wayne H. and {Bellazzini}, Ronaldo and {Bianchi}, Stefano and {Bongiorno}, Stephen D. and {Bonino}, Raffaella and {Brez}, Alessandro and {Capitanio}, Fiamma and {Castellano}, Simone and {Cavazzuti}, Elisabetta and {Chen}, Chien-Ting and {Ciprini}, Stefano and {Costa}, Enrico and {De Rosa}, Alessandra and {Del Monte}, Ettore and {Di Gesu}, Laura and {Di Marco}, Alessandro and {Donnarumma}, Immacolata and {Doroshenko}, Victor and {Dov{\v{c}}iak}, Michal and {Ehlert}, Steven R. and {Enoto}, Teruaki and {Evangelista}, Yuri and {Fabiani}, Sergio and {Garcia}, Javier A. and {Hayashida}, Kiyoshi and {Iwakiri}, Wataru and {Jorstad}, Svetlana G. and {Kaaret}, Philip and {Karas}, Vladimir and {Kislat}, Fabian and {Kolodziejczak}, Jeffery J. and {Krawczynski}, Henric and {La Monaca}, Fabio and {Latronico}, Luca and {Liodakis}, Ioannis and {Maldera}, Simone and {Manfreda}, Alberto and {Marin}, Fr{\'e}d{\'e}ric and {Marinucci}, Andrea and {Marscher}, Alan P. and {Massaro}, Francesco and {Matt}, Giorgio and {Mitsuishi}, Ikuyuki and {Ng}, C. -Y. and {O'Dell}, Stephen L. and {Oppedisano}, Chiara and {Papitto}, Alessandro and {Pavlov}, George G. and {Peirson}, Abel L. and {Perri}, Matteo and {Petrucci}, Pierre-Olivier and {Poutanen}, Juri and {Puccetti}, Simonetta and {Ramsey}, Brian D. and {Ratheesh}, Ajay and {Roberts}, Oliver J. and {Soffitta}, Paolo and {Spandre}, Gloria and {Swartz}, Doug and {Tamagawa}, Toru and {Tavecchio}, Fabrizio and {Taverna}, Roberto and {Tawara}, Yuzuru and {Tennant}, Allyn F. and {Thomas}, Nicolas E. and {Tombesi}, Francesco and {Trois}, Alessio and {Tsygankov}, Sergey and {Turolla}, Roberto and {Vink}, Jacco and {Wu}, Kinwah and {Zane}, Silvia},
        title = "{Simultaneous space and phase resolved X-ray polarimetry of the Crab pulsar and nebula}",
      journal = {Nature Astronomy},
         year = 2023,
        month = may,
       volume = {7},
        pages = {602-610},
          doi = {10.1038/s41550-023-01936-8},
archivePrefix = {arXiv},
       eprint = {2207.05573},
 primaryClass = {astro-ph.HE},
       adsurl = {https://ui.adsabs.harvard.edu/abs/2023NatAs...7..602B}
}

@INPROCEEDINGS{Bellazzini2003,
   author = {{Bellazzini}, R. and {Angelini}, F. and {Baldini}, L. and {Brez}, A. and
{Costa}, E. and {Di Persio}, G. and {Latronico}, L. and {Massai}, M.~M. and
{Omodei}, N. and {Pacciani}, L. and {Soffitta}, P. and {Spandre}, G.
},
    title = "{Novel gaseus X-ray polarimeter: data analysis and simulation}",
booktitle = {Proc. SPIE},
     year = 2003,
   volume = 4843,
    pages = {383},
   adsurl = {http://adsabs.harvard.edu/abs/2003SPIE.4843..383B}
}

@inproceedings{bongiorno2021,
author = {Stephen D. Bongiorno and Jeffery J. Kolodziejczak and Kiranmayee Kilaru and Ron Eng and Mark Stahl and Wayne H. Baumgartner and Nicholas Thomas and Jaganathan Ranganathan and Brian D. Ramsey and John Tucker},
title = {{Assembly of the IXPE mirror modules}},
volume = {11822},
booktitle = {Optics for EUV, X-Ray, and Gamma-Ray Astronomy X},
editor = {Stephen L. O'Dell and Jessica A. Gaskin and Giovanni Pareschi},
organization = {International Society for Optics and Photonics},
publisher = {SPIE},
pages = {189 -- 200},
year = {2021},
doi = {10.1117/12.2594316},
URL = {https://doi.org/10.1117/12.2594316}
}

@ARTICLE{Bobrikova2024,
       author = {{Bobrikova}, Anna and {Forsblom}, Sofia V. and {Di Marco}, Alessandro and {La Monaca}, Fabio and {Poutanen}, Juri and {Ng}, Mason and {Ravi}, Swati and {Loktev}, Vladislav and {Kajava}, Jari J.~E. and {Ursini}, Francesco and {Veledina}, Alexandra and {Rogantini}, Daniele and {Salmi}, Tuomo and {Bianchi}, Stefano and {Capitanio}, Fiamma and {Done}, Chris and {Fabiani}, Sergio and {Gnarini}, Andrea and {Heyl}, Jeremy and {Kaaret}, Philip and {Matt}, Giorgio and {Muleri}, Fabio and {Nitindala}, Anagha P. and {Rankin}, John and {Weisskopf}, Martin C. and {Agudo}, Iv{\'a}n and {Antonelli}, Lucio A. and {Bachetti}, Matteo and {Baldini}, Luca and {Baumgartner}, Wayne H. and {Bellazzini}, Ronaldo and {Bongiorno}, Stephen D. and {Bonino}, Raffaella and {Brez}, Alessandro and {Bucciantini}, Niccol{\`o} and {Castellano}, Simone and {Cavazzuti}, Elisabetta and {Chen}, Chien-Ting and {Ciprini}, Stefano and {Costa}, Enrico and {De Rosa}, Alessandra and {Del Monte}, Ettore and {Di Gesu}, Laura and {Di Lalla}, Niccol{\`o} and {Donnarumma}, Immacolata and {Doroshenko}, Victor and {Dov{\v{c}}iak}, Michal and {Ehlert}, Steven R. and {Enoto}, Teruaki and {Evangelista}, Yuri and {Ferrazzoli}, Riccardo and {Garc{\'\i}a}, Javier A. and {Gunji}, Shuichi and {Hayashida}, Kiyoshi and {Iwakiri}, Wataru and {Jorstad}, Svetlana G. and {Karas}, Vladimir and {Kislat}, Fabian and {Kitaguchi}, Takao and {Kolodziejczak}, Jeffery J. and {Krawczynski}, Henric and {Latronico}, Luca and {Liodakis}, Ioannis and {Maldera}, Simone and {Manfreda}, Alberto and {Marin}, Fr{\'e}d{\'e}ric and {Marinucci}, Andrea and {Marscher}, Alan P. and {Marshall}, Herman L. and {Massaro}, Francesco and {Mitsuishi}, Ikuyuki and {Mizuno}, Tsunefumi and {Negro}, Michela and {Ng}, Chi-Yung and {O'Dell}, Stephen L. and {Omodei}, Nicola and {Oppedisano}, Chiara and {Papitto}, Alessandro and {Pavlov}, George G. and {Peirson}, Abel L. and {Perri}, Matteo and {Pesce-Rollins}, Melissa and {Petrucci}, Pierre-Olivier and {Pilia}, Maura and {Possenti}, Andrea and {Puccetti}, Simonetta and {Ramsey}, Brian D. and {Ratheesh}, Ajay and {Roberts}, Oliver J. and {Romani}, Roger W. and {Sgr{\`o}}, Carmelo and {Slane}, Patrick and {Soffitta}, Paolo and {Spandre}, Gloria and {Swartz}, Douglas A. and {Tamagawa}, Toru and {Tavecchio}, Fabrizio and {Taverna}, Roberto and {Tawara}, Yuzuru and {Tennant}, Allyn F. and {Thomas}, Nicholas E. and {Tombesi}, Francesco and {Trois}, Alessio and {Tsygankov}, Sergey S. and {Turolla}, Roberto and {Vink}, Jacco and {Wu}, Kinwah and {Xie}, Fei and {Zane}, Silvia},
        title = "{Discovery of a strong rotation of the X-ray polarization angle in the galactic burster GX 13+1}",
      journal = {\aap},
         year = 2024,
        month = aug,
       volume = {688},
          eid = {A170},
        pages = {A170},
          doi = {10.1051/0004-6361/202449318},
archivePrefix = {arXiv},
       eprint = {2401.13058},
 primaryClass = {astro-ph.HE},
       adsurl = {https://ui.adsabs.harvard.edu/abs/2024A&A...688A.170B}
}

@ARTICLE{Chakraborty2025,
       author = {{Chakraborty}, Sudip and {Ratheesh}, Ajay and {Tagliacozzo}, Daniele and {Kaaret}, Philip and {Podgorn{\'y}}, Jakub and {Marin}, Fr{\'e}d{\'e}ric and {Tombesi}, Francesco and {Ehlert}, Steven R. and {Chen}, Chien-Ting J. and {Kim}, Dawoon E. and {Liodakis}, Ioannis and {Ursini}, Francesco and {Middei}, Riccardo and {Di Marco}, Alessandro and {La Monaca}, Fabio and {Banerjee}, Srimanta and {Fukumura}, Keigo and {Maksym}, W. Peter and {Miku{\v{s}}incov{\'a}}, Romana and {Nemmen}, Rodrigo and {Petrucci}, Pierre-Olivier and {Soffitta}, Paolo and {Svoboda}, Ji{\v{r}}{\'\i} and {Zhang}, Wenda},
        title = "{First X-Ray Polarimetric View of a Low-luminosity Active Galactic Nucleus: The Case of NGC 2110}",
      journal = {\apj},
         year = 2025,
        month = sep,
       volume = {990},
       number = {1},
          eid = {89},
        pages = {89},
          doi = {10.3847/1538-4357/ade87d},
archivePrefix = {arXiv},
       eprint = {2503.01071},
 primaryClass = {astro-ph.HE},
       adsurl = {https://ui.adsabs.harvard.edu/abs/2025ApJ...990...89C}
}

@ARTICLE{Connors1977,
       author = {{Connors}, P.~A. and {Stark}, R.~F.},
        title = "{Observable gravitational effects on polarised radiation coming from near a black hole}",
      journal = {\nat},
         year = 1977,
        month = sep,
       volume = {269},
       number = {5624},
        pages = {128-129},
          doi = {10.1038/269128a0},
       adsurl = {https://ui.adsabs.harvard.edu/abs/1977Natur.269..128C}
}

@INPROCEEDINGS{Deininger2022,
  author={Deininger, William D. and Kalinowski, William and Head, Michael and Peterson, Colin and Antoniak, Spencer and Mize, Rondal and Weisskopf, Martin C. and Ramsey, Brian and O'Dell, Stephen L. and Tennant, Allyn and Soffitta, Paolo and Santoli, Francesco and Del Monte, Ettore and Pinchera, Michele and Trois, Alessio and Osborne, Darren},
  booktitle={2022 IEEE Aerospace Conference (AERO)}, 
  title={Imaging X-Ray Polarimetry Explorer (IXPE) - Ready for Flight}, 
  year={2022},
  volume={},
  number={},
  pages={1-16},
  doi={10.1109/AERO53065.2022.9843233}}

@ARTICLE{Dimarco2025,
       author = {{Di Marco}, Alessandro and {La Monaca}, Fabio and {Bobrikova}, Anna and {Stella}, Luigi and {Papitto}, Alessandro and {Poutanen}, Juri and {Baglio}, Maria Cristina and {Bachetti}, Matteo and {Loktev}, Vladislav and {Pilia}, Maura and {Rogantini}, Daniele},
        title = "{X-Ray Dips and Polarization Angle Swings in GX 13+1}",
      journal = {\apjl},
         year = 2025,
        month = feb,
       volume = {979},
       number = {2},
          eid = {L47},
        pages = {L47},
          doi = {10.3847/2041-8213/ada7f8},
archivePrefix = {arXiv},
       eprint = {2501.05511},
 primaryClass = {astro-ph.HE},
       adsurl = {https://ui.adsabs.harvard.edu/abs/2025ApJ...979L..47D}
}

@ARTICLE{Doroshenko2023,
       author = {{Doroshenko}, Victor and {Poutanen}, Juri and {Heyl}, Jeremy and {Tsygankov}, Sergey S. and {Caiazzo}, Ilaria and {Turolla}, Roberto and {Veledina}, Alexandra and {Weisskopf}, Martin C. and {Forsblom}, Sofia V. and {Gonz{\'a}lez-Caniulef}, Denis and {Loktev}, Vladislav and {Malacaria}, Christian and {Mushtukov}, Alexander A. and {Suleimanov}, Valery F. and {Lutovinov}, Alexander A. and {Mereminskiy}, Ilya A. and {Molkov}, Sergey V. and {Salganik}, Alexander and {Santangelo}, Andrea and {Berdyugin}, Andrei V. and {Kravtsov}, Vadim and {Nitindala}, Anagha P. and {Agudo}, Iv{\'a}n and {Antonelli}, Lucio A. and {Bachetti}, Matteo and {Baldini}, Luca and {Baumgartner}, Wayne H. and {Bellazzini}, Ronaldo and {Bianchi}, Stefano and {Bongiorno}, Stephen D. and {Bonino}, Raffaella and {Brez}, Alessandro and {Bucciantini}, Niccol{\`o} and {Capitanio}, Fiamma and {Castellano}, Simone and {Cavazzuti}, Elisabetta and {Chen}, Chien-Ting and {Ciprini}, Stefano and {Costa}, Enrico and {De Rosa}, Alessandra and {Del Monte}, Ettore and {Di Gesu}, Laura and {Di Lalla}, Niccol{\`o} and {Di Marco}, Alessandro and {Donnarumma}, Immacolata and {Dov{\v{c}}iak}, Michal and {Ehlert}, Steven R. and {Enoto}, Teruaki and {Evangelista}, Yuri and {Fabiani}, Sergio and {Ferrazzoli}, Riccardo and {Garc{\'\i}a}, Javier A. and {Gunji}, Shuichi and {Hayashida}, Kiyoshi and {Iwakiri}, Wataru and {Jorstad}, Svetlana G. and {Kaaret}, Philip and {Karas}, Vladimir and {Kislat}, Fabian and {Kitaguchi}, Takao and {Kolodziejczak}, Jeffery J. and {Krawczynski}, Henric and {La Monaca}, Fabio and {Latronico}, Luca and {Liodakis}, Ioannis and {Maldera}, Simone and {Manfreda}, Alberto and {Marin}, Fr{\'e}d{\'e}ric and {Marinucci}, Andrea and {Marscher}, Alan P. and {Marshall}, Herman L. and {Massaro}, Francesco and {Matt}, Giorgio and {Mitsuishi}, Ikuyuki and {Mizuno}, Tsunefumi and {Muleri}, Fabio and {Negro}, Michela and {Ng}, Chi-Yung and {O'Dell}, Stephen L. and {Omodei}, Nicola and {Oppedisano}, Chiara and {Papitto}, Alessandro and {Pavlov}, George G. and {Peirson}, Abel L. and {Perri}, Matteo and {Pesce-Rollins}, Melissa and {Petrucci}, Pierre-Olivier and {Pilia}, Maura and {Possenti}, Andrea and {Puccetti}, Simonetta and {Ramsey}, Brian D. and {Rankin}, John and {Ratheesh}, Ajay and {Roberts}, Oliver J. and {Romani}, Roger W. and {Sgr{\`o}}, Carmelo and {Slane}, Patrick and {Soffitta}, Paolo and {Spandre}, Gloria and {Swartz}, Douglas A. and {Tamagawa}, Toru and {Tavecchio}, Fabrizio and {Taverna}, Roberto and {Tawara}, Yuzuru and {Tennant}, Allyn F. and {Thomas}, Nicholas E. and {Tombesi}, Francesco and {Trois}, Alessio and {Vink}, Jacco and {Wu}, Kinwah and {Xie}, Fei and {Zane}, Silvia},
        title = "{Complex variations in X-ray polarization in the X-ray pulsar LS V +44 17/RX J0440.9+4431}",
      journal = {\aap},
         year = 2023,
        month = sep,
       volume = {677},
          eid = {A57},
        pages = {A57},
          doi = {10.1051/0004-6361/202347088},
archivePrefix = {arXiv},
       eprint = {2306.02116},
 primaryClass = {astro-ph.HE},
       adsurl = {https://ui.adsabs.harvard.edu/abs/2023A&A...677A..57D}
}

@ARTICLE{Dovciak2024,
       author = {{Dov{\v{c}}iak}, Michal and {Podgorn{\'y}}, Jakub and {Svoboda}, Ji{\v{r}}{\'\i} and {Steiner}, James F. and {Kaaret}, Philip and {Krawczynski}, Henric and {Ingram}, Adam and {Kravtsov}, Vadim and {Marra}, Lorenzo and {Muleri}, Fabio and {Garc{\'\i}a}, Javier A. and {Mastroserio}, Guglielmo and {Miku{\v{s}}incov{\'a}}, Romana and {Ratheesh}, Ajay and {Cavero}, Nicole Rodriguez},
        title = "{IXPE View of BH XRBs during the First 2.5 Years of the Mission}",
      journal = {Galaxies},
         year = 2024,
        month = sep,
       volume = {12},
       number = {5},
          eid = {54},
        pages = {54},
          doi = {10.3390/galaxies12050054},
       adsurl = {https://ui.adsabs.harvard.edu/abs/2024Galax..12...54D}
}

@ARTICLE{Dubner2015,
       author = {{Dubner}, Gloria and {Giacani}, Elsa},
        title = "{Radio emission from supernova remnants}",
      journal = {\aapr},
         year = 2015,
        month = sep,
       volume = {23},
          eid = {3},
        pages = {3},
          doi = {10.1007/s00159-015-0083-5},
archivePrefix = {arXiv},
       eprint = {1508.07294},
 primaryClass = {astro-ph.HE},
       adsurl = {https://ui.adsabs.harvard.edu/abs/2015A&ARv..23....3D}
}

@ARTICLE{Euler1935,
   author = {{Euler}, H. and {Kockel}, B.},
    title = "{{\"U}ber die Streuung von Licht an Licht nach der Diracschen Theorie}",
  journal = {Naturwissenschaften},
     year = 1935,
    month = apr,
   volume = 23,
    pages = {246-247},
      doi = {10.1007/BF01493898},
   adsurl = {http://adsabs.harvard.edu/abs/1935NW.....23..246E}
}

@ARTICLE{Ewing2025,
       author = {{Ewing}, Melissa and {Parra}, Maxime and {Mastroserio}, Guglielmo and {Veledina}, Alexandra and {Ingram}, Adam and {Dov{\v{c}}iak}, Michal and {Garc{\'\i}a}, Javier A. and {Russell}, Thomas D. and {Baglio}, Maria C. and {Poutanen}, Juri and {Adegoke}, Oluwashina and {Bianchi}, Stefano and {Capitanio}, Fiamma and {Connors}, Riley and {Del Santo}, Melania and {De Marco}, Barbara and {Trigo}, Mar{\'\i}a D{\'\i}az and {Gandhi}, Poshak and {Gupta}, Maitrayee and {Kang}, Chulsoo and {Kammoun}, Elias and {Loktev}, Vladislav and {Marra}, Lorenzo and {Matt}, Giorgio and {Nathan}, Edward and {Petrucci}, Pierre-Olivier and {Shidatsu}, Megumi and {Steiner}, James F. and {Tombesi}, Francesco and {Vincentelli}, Federico M.},
        title = "{The very high X-ray polarization of accreting black hole IGR J17091‑3624 in the hard state}",
      journal = {\mnras},
         year = 2025,
        month = aug,
       volume = {541},
       number = {2},
        pages = {1774-1781},
          doi = {10.1093/mnras/staf859},
archivePrefix = {arXiv},
       eprint = {2503.22665},
 primaryClass = {astro-ph.HE},
       adsurl = {https://ui.adsabs.harvard.edu/abs/2025MNRAS.541.1774E}
}

@ARTICLE{Fabian2009,
       author = {{Fabian}, A.~C. and {Zoghbi}, A. and {Ross}, R.~R. and {Uttley}, P. and {Gallo}, L.~C. and {Brandt}, W.~N. and {Blustin}, A.~J. and {Boller}, T. and {Caballero-Garcia}, M.~D. and {Larsson}, J. and {Miller}, J.~M. and {Miniutti}, G. and {Ponti}, G. and {Reis}, R.~C. and {Reynolds}, C.~S. and {Tanaka}, Y. and {Young}, A.~J.},
        title = "{Broad line emission from iron K- and L-shell transitions in the active galaxy 1H0707-495}",
      journal = {\nat},
         year = 2009,
        month = may,
       volume = {459},
       number = {7246},
        pages = {540-542},
          doi = {10.1038/nature08007},
       adsurl = {https://ui.adsabs.harvard.edu/abs/2009Natur.459..540F}
}

@BOOK{Fabiani2014b,
       author = {{Fabiani}, Sergio and {Muleri}, Fabio},
        title = "{Astronomical X-Ray Polarimetry}",
         year = 2014,
          PUBLISHER = "Aracne",
       adsurl = {https://ui.adsabs.harvard.edu/abs/2014axp..book.....F}
}

@ARTICLE{Farinelli2023,
       author = {{Farinelli}, R. and {Fabiani}, S. and {Poutanen}, J. and {Ursini}, F. and {Ferrigno}, C. and {Bianchi}, S. and {Cocchi}, M. and {Capitanio}, F. and {De Rosa}, A. and {Gnarini}, A. and {Kislat}, F. and {Matt}, G. and {Mikusincova}, R. and {Muleri}, F. and {Agudo}, I. and {Antonelli}, L.~A. and {Bachetti}, M. and {Baldini}, L. and {Baumgartner}, W.~H. and {Bellazzini}, R. and {Bongiorno}, S.~D. and {Bonino}, R. and {Brez}, A. and {Bucciantini}, N. and {Castellano}, S. and {Cavazzuti}, E. and {Ciprini}, S. and {Costa}, E. and {Del Monte}, E. and {Di Gesu}, L. and {Di Lalla}, N. and {Di Marco}, A. and {Donnarumma}, I. and {Doroshenko}, V. and {Dov{\v{c}}iak}, M. and {Ehlert}, S.~R. and {Enoto}, T. and {Evangelista}, Y. and {Ferrazzoli}, R. and {Garcia}, J.~A. and {Gunji}, S. and {Hayashida}, K. and {Heyl}, J. and {Iwakiri}, W. and {Jorstad}, S.~G. and {Karas}, V. and {Kitaguchi}, T. and {Kolodziejczak}, J.~J. and {Krawczynski}, H. and {La Monaca}, F. and {Latronico}, L. and {Liodakis}, I. and {Maldera}, S. and {Manfreda}, A. and {Marin}, F. and {Marscher}, A.~P. and {Marshall}, H.~L. and {Mitsuishi}, I. and {Mizuno}, T. and {Ng}, C. -Y. and {O'Dell}, S.~L. and {Omodei}, N. and {Oppedisano}, C. and {Papitto}, A. and {Pavlov}, G.~G. and {Peirson}, A.~L. and {Perri}, M. and {Pesce-Rollins}, M. and {Petrucci}, P.~O. and {Pilia}, M. and {Possenti}, A. and {Puccetti}, S. and {Ramsey}, B.~D. and {Rankin}, J. and {Ratheesh}, A. and {Romani}, R.~W. and {Sgr{\`o}}, C. and {Slane}, P. and {Soffitta}, P. and {Spandre}, G. and {Tamagawa}, T. and {Tavecchio}, F. and {Taverna}, R. and {Tawara}, Y. and {Tennant}, A.~F. and {Thomas}, N.~E. and {Tombesi}, F. and {Trois}, A. and {Tsygankov}, S.~S. and {Turolla}, R. and {Vink}, J. and {Weisskopf}, M.~C. and {Wu}, K. and {Xie}, F. and {Zane}, S.},
        title = "{Accretion geometry of the neutron star low mass X-ray binary Cyg X-2 from X-ray polarization measurements}",
      journal = {\mnras},
         year = 2023,
        month = mar,
       volume = {519},
       number = {3},
        pages = {3681-3690},
          doi = {10.1093/mnras/stac3726},
archivePrefix = {arXiv},
       eprint = {2212.13119},
 primaryClass = {astro-ph.HE},
       adsurl = {https://ui.adsabs.harvard.edu/abs/2023MNRAS.519.3681F}
}

@ARTICLE{Ferrazzoli2020,
       author = {{Ferrazzoli}, Riccardo and {Muleri}, Fabio and {Lefevre}, Carlo and {Morbidini}, Alfredo and {Amici}, Fabrizio and {Brienza}, Daniele and {Costa}, Enrico and {Del Monte}, Ettore and {Di Marco}, Alessandro and {Di Persio}, Giuseppe and {Donnarumma}, Immacolata and {Fabiani}, Sergio and {La Monaca}, Fabio and {Loffredo}, Pasqualino and {Maiolo}, Luca and {Maita}, Francesco and {Piazzolla}, Raffaele and {Ramsey}, Brian and {Rankin}, John and {Ratheesh}, Ajay and {Rubini}, Alda and {Sarra}, Paolo and {Soffitta}, Paolo and {Tobia}, Antonino and {Xie}, Fei},
        title = "{In-flight calibration system of imaging x-ray polarimetry explorer}",
      journal = {Journal of Astronomical Telescopes, Instruments, and Systems},
         year = 2020,
        month = oct,
       volume = {6},
          eid = {048002},
        pages = {048002},
          doi = {10.1117/1.JATIS.6.4.048002},
archivePrefix = {arXiv},
       eprint = {2010.14185},
 primaryClass = {astro-ph.IM},
       adsurl = {https://ui.adsabs.harvard.edu/abs/2020JATIS...6d8002F}
}

@ARTICLE{Ferrazzoli2023,
       author = {{Ferrazzoli}, Riccardo and {Slane}, Patrick and {Prokhorov}, Dmitry and {Zhou}, Ping and {Vink}, Jacco and {Bucciantini}, Niccol{\`o} and {Costa}, Enrico and {Di Lalla}, Niccol{\`o} and {Di Marco}, Alessandro and {Soffitta}, Paolo and {Weisskopf}, Martin C. and {Asakura}, Kazunori and {Baldini}, Luca and {Heyl}, Jeremy and {Kaaret}, Philip E. and {Marin}, Fr{\'e}d{\'e}ric and {Mizuno}, Tsunefumi and {Ng}, C. -Y. and {Pesce-Rollins}, Melissa and {Silvestri}, Stefano and {Sgr{\`o}}, Carmelo and {Swartz}, Douglas A. and {Tamagawa}, Toru and {Yang}, Yi-Jung and {Agudo}, Iv{\'a}n and {Antonelli}, Lucio A. and {Bachetti}, Matteo and {Baumgartner}, Wayne H. and {Bellazzini}, Ronaldo and {Bianchi}, Stefano and {Bongiorno}, Stephen D. and {Bonino}, Raffaella and {Brez}, Alessandro and {Capitanio}, Fiamma and {Castellano}, Simone and {Cavazzuti}, Elisabetta and {Chen}, Chien-Ting and {Ciprini}, Stefano and {De Rosa}, Alessandra and {Del Monte}, Ettore and {Di Gesu}, Laura and {Donnarumma}, Immacolata and {Doroshenko}, Victor and {Dov{\v{c}}iak}, Michal and {Ehlert}, Steven R. and {Enoto}, Teruaki and {Evangelista}, Yuri and {Fabiani}, Sergio and {Garcia}, Javier A. and {Gunji}, Shuichi and {Hayashida}, Kiyoshi and {Iwakiri}, Wataru and {Jorstad}, Svetlana G. and {Kislat}, Fabian and {Karas}, Vladimir and {Kitaguchi}, Takao and {Kolodziejczak}, Jeffery J. and {Krawczynski}, Henric and {La Monaca}, Fabio and {Latronico}, Luca and {Liodakis}, Ioannis and {Maldera}, Simone and {Manfreda}, Alberto and {Marinucci}, Andrea and {Marscher}, Alan P. and {Marshall}, Herman L. and {Matt}, Giorgio and {Mitsuishi}, Ikuyuki and {Muleri}, Fabio and {Negro}, Michela and {O'Dell}, Stephen L. and {Omodei}, Nicola and {Oppedisano}, Chiara and {Papitto}, Alessandro and {Pavlov}, George G. and {Peirson}, Abel L. and {Perri}, Matteo and {Petrucci}, Pierre-Olivier and {Pilia}, Maura and {Possenti}, Andrea and {Poutanen}, Juri and {Puccetti}, Simonetta and {Ramsey}, Brian D. and {Rankin}, John and {Ratheesh}, Ajay and {Roberts}, Oliver and {Romani}, Roger W. and {Spandre}, Gloria and {Tavecchio}, Fabrizio and {Taverna}, Roberto and {Tawara}, Yuzuru and {Tennant}, Allyn F. and {Thomas}, Nicholas E. and {Tombesi}, Francesco and {Trois}, Alessio and {Tsygankov}, Sergey S. and {Turolla}, Roberto and {Wu}, Kinwah and {Xie}, Fei and {Zane}, Silvia},
        title = "{X-Ray Polarimetry Reveals the Magnetic-field Topology on Sub-parsec Scales in Tycho's Supernova Remnant}",
      journal = {\apj},
         year = 2023,
        month = mar,
       volume = {945},
       number = {1},
          eid = {52},
        pages = {52},
          doi = {10.3847/1538-4357/acb496},
archivePrefix = {arXiv},
       eprint = {2301.07397},
 primaryClass = {astro-ph.HE},
       adsurl = {https://ui.adsabs.harvard.edu/abs/2023ApJ...945...52F}
}

@ARTICLE{Ferrazzoli2024,
       author = {{Ferrazzoli}, Riccardo and {Prokhorov}, Dmitry and {Bucciantini}, Niccol{\`o} and {Slane}, Patrick and {Vink}, Jacco and {Cardillo}, Martina and {Yang}, Yi-Jung and {Silvestri}, Stefano and {Zhou}, Ping and {Costa}, Enrico and {Omodei}, Nicola and {Ng}, C. -Y. and {Soffitta}, Paolo and {Weisskopf}, Martin C. and {Baldini}, Luca and {Di Marco}, Alessandro and {Doroshenko}, Victor and {Heyl}, Jeremy and {Kaaret}, Philip and {Kim}, Dawoon E. and {Marin}, Fr{\'e}d{\'e}ric and {Mizuno}, Tsunefumi and {Pesce-Rollins}, Melissa and {Sgr{\`o}}, Carmelo and {Swartz}, Douglas A. and {Tamagawa}, Toru and {Xie}, Fei and {Agudo}, Iv{\'a}n and {Antonelli}, Lucio A. and {Bachetti}, Matteo and {Baumgartner}, Wayne H. and {Bellazzini}, Ronaldo and {Bianchi}, Stefano and {Bongiorno}, Stephen D. and {Bonino}, Raffaella and {Brez}, Alessandro and {Capitanio}, Fiamma and {Castellano}, Simone and {Cavazzuti}, Elisabetta and {Chen}, Chien-Ting and {Ciprini}, Stefano and {De Rosa}, Alessandra and {Del Monte}, Ettore and {Di Gesu}, Laura and {Di Lalla}, Niccol{\`o} and {Donnarumma}, Immacolata and {Dov{\v{c}}iak}, Michal and {Ehlert}, Steven R. and {Enoto}, Teruaki and {Evangelista}, Yuri and {Fabiani}, Sergio and {Garcia}, Javier A. and {Gunji}, Shuichi and {Hayashida}, Kiyoshi and {Iwakiri}, Wataru and {Jorstad}, Svetlana G. and {Karas}, Vladimir and {Kislat}, Fabian and {Kitaguchi}, Takao and {Kolodziejczak}, Jeffery J. and {Krawczynski}, Henric and {La Monaca}, Fabio and {Latronico}, Luca and {Liodakis}, Ioannis and {Maldera}, Simone and {Manfreda}, Alberto and {Marinucci}, Andrea and {Marscher}, Alan P. and {Marshall}, Herman L. and {Massaro}, Francesco and {Matt}, Giorgio and {Mitsuishi}, Ikuyuki and {Muleri}, Fabio and {Negro}, Michela and {O'Dell}, Stephen L. and {Oppedisano}, Chiara and {Papitto}, Alessandro and {Pavlov}, George G. and {Peirson}, Abel L. and {Perri}, Matteo and {Petrucci}, Pierre-Olivier and {Pilia}, Maura and {Possenti}, Andrea and {Poutanen}, Juri and {Puccetti}, Simonetta and {Ramsey}, Brian D. and {Rankin}, John and {Ratheesh}, Ajay and {Roberts}, Oliver J. and {Romani}, Roger W. and {Spandre}, Gloria and {Tavecchio}, Fabrizio and {Taverna}, Roberto and {Tawara}, Yuzuru and {Tennant}, Allyn F. and {Thomas}, Nicholas E. and {Tombesi}, Francesco and {Trois}, Alessio and {Tsygankov}, Sergey S. and {Turolla}, Roberto and {Wu}, Kinwah and {Zane}, Silvia},
        title = "{Discovery of a Shock-compressed Magnetic Field in the Northwestern Rim of the Young Supernova Remnant RX J1713.7{\textendash}3946 with X-Ray Polarimetry}",
      journal = {\apjl},
         year = 2024,
        month = jun,
       volume = {967},
       number = {2},
          eid = {L38},
        pages = {L38},
          doi = {10.3847/2041-8213/ad4a68},
archivePrefix = {arXiv},
       eprint = {2405.07577},
 primaryClass = {astro-ph.HE},
       adsurl = {https://ui.adsabs.harvard.edu/abs/2024ApJ...967L..38F}
}

@ARTICLE{Forsblom2023,
       author = {{Forsblom}, Sofia V. and {Poutanen}, Juri and {Tsygankov}, Sergey S. and {Bachetti}, Matteo and {Di Marco}, Alessandro and {Doroshenko}, Victor and {Heyl}, Jeremy and {La Monaca}, Fabio and {Malacaria}, Christian and {Marshall}, Herman L. and {Muleri}, Fabio and {Mushtukov}, Alexander A. and {Pilia}, Maura and {Rogantini}, Daniele and {Suleimanov}, Valery F. and {Taverna}, Roberto and {Xie}, Fei and {Agudo}, Iv{\'a}n and {Antonelli}, Lucio A. and {Baldini}, Luca and {Baumgartner}, Wayne H. and {Bellazzini}, Ronaldo and {Bianchi}, Stefano and {Bongiorno}, Stephen D. and {Bonino}, Raffaella and {Brez}, Alessandro and {Bucciantini}, Niccol{\`o} and {Capitanio}, Fiamma and {Castellano}, Simone and {Cavazzuti}, Elisabetta and {Chen}, Chien-Ting and {Ciprini}, Stefano and {Costa}, Enrico and {De Rosa}, Alessandra and {Del Monte}, Ettore and {Di Gesu}, Laura and {Di Lalla}, Niccol{\`o} and {Donnarumma}, Immacolata and {Dov{\v{c}}iak}, Michal and {Ehlert}, Steven R. and {Enoto}, Teruaki and {Evangelista}, Yuri and {Fabiani}, Sergio and {Ferrazzoli}, Riccardo and {Garcia}, Javier A. and {Gunji}, Shuichi and {Hayashida}, Kiyoshi and {Iwakiri}, Wataru and {Jorstad}, Svetlana G. and {Kaaret}, Philip and {Karas}, Vladimir and {Kitaguchi}, Takao and {Kolodziejczak}, Jeffery J. and {Krawczynski}, Henric and {Latronico}, Luca and {Liodakis}, Ioannis and {Maldera}, Simone and {Manfreda}, Alberto and {Marin}, Fr{\'e}d{\'e}ric and {Marinucci}, Andrea and {Marscher}, Alan P. and {Matt}, Giorgio and {Mitsuishi}, Ikuyuki and {Mizuno}, Tsunefumi and {Negro}, Michela and {Ng}, Chi-Yung and {O'Dell}, Stephen L. and {Omodei}, Nicola and {Oppedisano}, Chiara and {Papitto}, Alessandro and {Pavlov}, George G. and {Peirson}, Abel L. and {Perri}, Matteo and {Pesce-Rollins}, Melissa and {Petrucci}, Pierre-Olivier and {Possenti}, Andrea and {Puccetti}, Simonetta and {Ramsey}, Brian D. and {Rankin}, John and {Ratheesh}, Ajay and {Roberts}, Oliver J. and {Romani}, Roger W. and {Sgr{\`o}}, Carmelo and {Slane}, Patrick and {Soffitta}, Paolo and {Spandre}, Gloria and {Sunyaev}, Rashid A. and {Swartz}, Douglas A. and {Tamagawa}, Toru and {Tavecchio}, Fabrizio and {Tawara}, Yuzuru and {Tennant}, Allyn F. and {Thomas}, Nicholas E. and {Tombesi}, Francesco and {Trois}, Alessio and {Turolla}, Roberto and {Vink}, Jacco and {Weisskopf}, Martin C. and {Wu}, Kinwah and {Zane}, Silvia and {IXPE Collaboration}},
        title = "{IXPE Observations of the Quintessential Wind-accreting X-Ray Pulsar Vela X-1}",
      journal = {\apjl},
         year = 2023,
        month = apr,
       volume = {947},
       number = {2},
          eid = {L20},
        pages = {L20},
          doi = {10.3847/2041-8213/acc391},
archivePrefix = {arXiv},
       eprint = {2303.01800},
 primaryClass = {astro-ph.HE},
       adsurl = {https://ui.adsabs.harvard.edu/abs/2023ApJ...947L..20F}
}

@ARTICLE{Gianolli2023,
       author = {{Gianolli}, V.~E. and {Kim}, D.~E. and {Bianchi}, S. and {Ag{\'\i}s-Gonz{\'a}lez}, B. and {Madejski}, G. and {Marin}, F. and {Marinucci}, A. and {Matt}, G. and {Middei}, R. and {Petrucci}, P. -O. and {Soffitta}, P. and {Tagliacozzo}, D. and {Tombesi}, F. and {Ursini}, F. and {Barnouin}, T. and {De Rosa}, A. and {Di Gesu}, L. and {Ingram}, A. and {Loktev}, V. and {Panagiotou}, C. and {Podgorny}, J. and {Poutanen}, J. and {Puccetti}, S. and {Ratheesh}, A. and {Veledina}, A. and {Zhang}, W. and {Agudo}, I. and {Antonelli}, L.~A. and {Bachetti}, M. and {Baldini}, L. and {Baumgartner}, W.~H. and {Bellazzini}, R. and {Bongiorno}, S.~D. and {Bonino}, R. and {Brez}, A. and {Bucciantini}, N. and {Capitanio}, F. and {Castellano}, S. and {Cavazzuti}, E. and {Chen}, C. -T. and {Ciprini}, S. and {Costa}, E. and {Del Monte}, E. and {Di Lalla}, N. and {Di Marco}, A. and {Donnarumma}, I. and {Doroshenko}, V. and {Dov{\v{c}}iak}, M. and {Ehlert}, S.~R. and {Enoto}, T. and {Evangelista}, Y. and {Fabiani}, S. and {Ferrazzoli}, R. and {Garc{\'\i}a}, J.~A. and {Gunji}, S. and {Heyl}, J. and {Iwakiri}, W. and {Jorstad}, S.~G. and {Kaaret}, P. and {Karas}, V. and {Kislat}, F. and {Kitaguchi}, T. and {Kolodziejczak}, J.~J. and {Krawczynski}, H. and {La Monaca}, F. and {Latronico}, L. and {Liodakis}, I. and {Maldera}, S. and {Manfreda}, A. and {Marscher}, A.~P. and {Marshall}, H.~L. and {Massaro}, F. and {Mitsuishi}, I. and {Mizuno}, T. and {Muleri}, F. and {Negro}, M. and {Ng}, C. -Y. and {O'Dell}, S.~L. and {Omodei}, N. and {Oppedisano}, C. and {Papitto}, A. and {Pavlov}, G.~G. and {Peirson}, A.~L. and {Perri}, M. and {Pesce-Rollins}, M. and {Pilia}, M. and {Possenti}, A. and {Ramsey}, B.~D. and {Rankin}, J. and {Roberts}, O.~J. and {Romani}, R.~W. and {Sgr{\`o}}, C. and {Slane}, P. and {Spandre}, G. and {Swartz}, D.~A. and {Tamagawa}, T. and {Tavecchio}, F. and {Taverna}, R. and {Tawara}, Y. and {Tennant}, A.~F. and {Thomas}, N.~E. and {Trois}, A. and {Tsygankov}, S.~S. and {Turolla}, R. and {Vink}, J. and {Weisskopf}, M.~C. and {Wu}, K. and {Xie}, F. and {Zane}, S.},
        title = "{Uncovering the geometry of the hot X-ray corona in the Seyfert galaxy NGC 4151 with IXPE}",
      journal = {\mnras},
         year = 2023,
        month = aug,
       volume = {523},
       number = {3},
        pages = {4468-4476},
          doi = {10.1093/mnras/stad1697},
archivePrefix = {arXiv},
       eprint = {2303.12541},
 primaryClass = {astro-ph.GA},
       adsurl = {https://ui.adsabs.harvard.edu/abs/2023MNRAS.523.4468G}
}

@ARTICLE{Gianolli2024,
       author = {{Gianolli}, V.~E. and {Bianchi}, S. and {Kammoun}, E. and {Gnarini}, A. and {Marinucci}, A. and {Ursini}, F. and {Parra}, M. and {Tortosa}, A. and {De Rosa}, A. and {Kim}, D.~E. and {Marin}, F. and {Matt}, G. and {Serafinelli}, R. and {Soffitta}, P. and {Tagliacozzo}, D. and {Di Gesu}, L. and {Done}, C. and {Marshall}, H.~L. and {Middei}, R. and {Mikusincova}, R. and {Petrucci}, P. -O. and {Ravi}, S. and {Svoboda}, J. and {Tombesi}, F.},
        title = "{A second view on the X-ray polarization of NGC 4151 with IXPE}",
      journal = {\aap},
         year = 2024,
        month = nov,
       volume = {691},
          eid = {A29},
        pages = {A29},
          doi = {10.1051/0004-6361/202451645},
archivePrefix = {arXiv},
       eprint = {2407.17243},
 primaryClass = {astro-ph.HE},
       adsurl = {https://ui.adsabs.harvard.edu/abs/2024A&A...691A..29G}
}

@ARTICLE{Gonzales2019,
       author = {{Gonz{\'a}lez-Caniulef}, Denis and {Zane}, Silvia and {Turolla}, Roberto and {Wu}, Kinwah},
        title = "{Atmosphere of strongly magnetized neutron stars heated by particle bombardment}",
      journal = {\mnras},
         year = 2019,
        month = feb,
       volume = {483},
       number = {1},
        pages = {599-613},
          doi = {10.1093/mnras/sty3159},
archivePrefix = {arXiv},
       eprint = {1811.08526},
 primaryClass = {astro-ph.HE},
       adsurl = {https://ui.adsabs.harvard.edu/abs/2019MNRAS.483..599G}
}

@ARTICLE{Heyl2024,
       author = {{Heyl}, Jeremy and {Doroshenko}, Victor and {Gonz{\'a}lez-Caniulef}, Denis and {Caiazzo}, Ilaria and {Poutanen}, Juri and {Mushtukov}, Alexander and {Tsygankov}, Sergey S. and {Kirmizibayrak}, Demet and {Bachetti}, Matteo and {Pavlov}, George G. and {Forsblom}, Sofia V. and {Malacaria}, Christian and {Suleimanov}, Valery F. and {Agudo}, Iv{\'a}n and {Antonelli}, Lucio Angelo and {Baldini}, Luca and {Baumgartner}, Wayne H. and {Bellazzini}, Ronaldo and {Bianchi}, Stefano and {Bongiorno}, Stephen D. and {Bonino}, Raffaella and {Brez}, Alessandro and {Bucciantini}, Niccol{\`o} and {Capitanio}, Fiamma and {Castellano}, Simone and {Cavazzuti}, Elisabetta and {Chen}, Chien-Ting and {Ciprini}, Stefano and {Costa}, Enrico and {De Rosa}, Alessandra and {Del Monte}, Ettore and {Di Gesu}, Laura and {Di Lalla}, Niccol{\`o} and {Di Marco}, Alessandro and {Donnarumma}, Immacolata and {Dov{\v{c}}iak}, Michal and {Ehlert}, Steven R. and {Enoto}, Teruaki and {Evangelista}, Yuri and {Fabiani}, Sergio and {Ferrazzoli}, Riccardo and {Garcia}, Javier A. and {Gunji}, Shuichi and {Hayashida}, Kiyoshi and {Iwakiri}, Wataru and {Jorstad}, Svetlana G. and {Kaaret}, Philip and {Karas}, Vladimir and {Kislat}, Fabian and {Kitaguchi}, Takao and {Kolodziejczak}, Jeffery J. and {Krawczynski}, Henric and {La Monaca}, Fabio and {Latronico}, Luca and {Liodakis}, Ioannis and {Maldera}, Simone and {Manfreda}, Alberto and {Marin}, Fr{\'e}d{\'e}ric and {Marinucci}, Andrea and {Marscher}, Alan P. and {Marshall}, Herman L. and {Massaro}, Francesco and {Matt}, Giorgio and {Mitsuishi}, Ikuyuki and {Mizuno}, Tsunefumi and {Muleri}, Fabio and {Negro}, Michela and {Ng}, C.-Y. and {O'Dell}, Stephen L. and {Omodei}, Nicola and {Oppedisano}, Chiara and {Papitto}, Alessandro and {Peirson}, Abel Lawrence and {Perri}, Matteo and {Pesce-Rollins}, Melissa and {Petrucci}, Pierre-Olivier and {Pilia}, Maura and {Possenti}, Andrea and {Puccetti}, Simonetta and {Ramsey}, Brian D. and {Rankin}, John and {Ratheesh}, Ajay and {Roberts}, Oliver J. and {Romani}, Roger W. and {Sgr{\`o}}, Carmelo and {Slane}, Patrick and {Soffitta}, Paolo and {Spandre}, Gloria and {Swartz}, Douglas A. and {Tamagawa}, Toru and {Tavecchio}, Fabrizio and {Taverna}, Roberto and {Tawara}, Yuzuru and {Tennant}, Allyn F. and {Thomas}, Nicholas E. and {Tombesi}, Francesco and {Trois}, Alessio and {Turolla}, Roberto and {Vink}, Jacco and {Weisskopf}, Martin C. and {Wu}, Kinwah and {Xie}, Fei and {Zane}, Silvia},
        title = "{Complex rotational dynamics of the neutron star in Hercules X-1 revealed by X-ray polarization}",
      journal = {Nature Astronomy},
         year = 2024,
        month = aug,
       volume = {8},
        pages = {1047-1053},
          doi = {10.1038/s41550-024-02295-8},
       adsurl = {https://ui.adsabs.harvard.edu/abs/2024NatAs...8.1047H}
}

@ARTICLE{Heyl2023,
       author = {{Heyl}, Jeremy and {Taverna}, Roberto and {Turolla}, Roberto and {Israel}, Gian Luca and {Ng}, Mason and {K{\i}rm{\i}z{\i}bayrak}, Demet and {Gonz{\'a}lez-Caniulef}, Denis and {Caiazzo}, Ilaria and {Zane}, Silvia and {Ehlert}, Steven R. and {Negro}, Michela and {Agudo}, Iv{\'a}n and {Antonelli}, Lucio Angelo and {Bachetti}, Matteo and {Baldini}, Luca and {Baumgartner}, Wayne H. and {Bellazzini}, Ronaldo and {Bianchi}, Stefano and {Bongiorno}, Stephen D. and {Bonino}, Raffaella and {Brez}, Alessandro and {Bucciantini}, Niccol{\`o} and {Capitanio}, Fiamma and {Castellano}, Simone and {Cavazzuti}, Elisabetta and {Chen}, Chien-Ting and {Ciprini}, Stefano and {Costa}, Enrico and {De Rosa}, Alessandra and {Del Monte}, Ettore and {Di Gesu}, Laura and {Di Lalla}, Niccol{\`o} and {Di Marco}, Alessandro and {Donnarumma}, Immacolata and {Doroshenko}, Victor and {Dov{\v{c}}iak}, Michal and {Enoto}, Teruaki and {Evangelista}, Yuri and {Fabiani}, Sergio and {Ferrazzoli}, Riccardo and {Garcia}, Javier A. and {Gunji}, Shuichi and {Hayashida}, Kiyoshi and {Iwakiri}, Wataru and {Jorstad}, Svetlana G. and {Kaaret}, Philip and {Karas}, Vladimir and {Kislat}, Fabian and {Kitaguchi}, Takao and {Kolodziejczak}, Jeffery J. and {Krawczynski}, Henric and {La Monaca}, Fabio and {Latronico}, Luca and {Liodakis}, Ioannis and {Maldera}, Simone and {Manfreda}, Alberto and {Marin}, Fr{\'e}d{\'e}ric and {Marinucci}, Andrea and {Marscher}, Alan P. and {Marshall}, Herman L. and {Massaro}, Francesco and {Matt}, Giorgio and {Mitsuishi}, Ikuyuki and {Mizuno}, Tsunefumi and {Muleri}, Fabio and {Ng}, C. -Y. and {O'Dell}, Stephen L. and {Omodei}, Nicola and {Oppedisano}, Chiara and {Papitto}, Alessandro and {Pavlov}, George G. and {Peirson}, Abel Lawrence and {Perri}, Matteo and {Pesce-Rollins}, Melissa and {Petrucci}, Pierre-Olivier and {Pilia}, Maura and {Possenti}, Andrea and {Poutanen}, Juri and {Puccetti}, Simonetta and {Ramsey}, Brian D. and {Rankin}, John and {Ratheesh}, Ajay and {Roberts}, Oliver J. and {Romani}, Roger W. and {Sgr{\`o}}, Carmelo and {Slane}, Patrick and {Soffitta}, Paolo and {Spandre}, Gloria and {Swartz}, Douglas A. and {Tamagawa}, Toru and {Tavecchio}, Fabrizio and {Tawara}, Yuzuru and {Tennant}, Allyn F. and {Thomas}, Nicholas E. and {Tombesi}, Francesco and {Trois}, Alessio and {Tsygankov}, Sergey S. and {Vink}, Jacco and {Weisskopf}, Martin C. and {Wu}, Kinwah and {Xie}, Fei},
        title = "{The detection of polarized x-ray emission from the magnetar 1E 2259+586}",
      journal = {\mnras},
         year = 2023,
        month = dec,
          doi = {10.1093/mnras/stad3680},
archivePrefix = {arXiv},
       eprint = {2311.03637},
 primaryClass = {astro-ph.HE},
       adsurl = {https://ui.adsabs.harvard.edu/abs/2023MNRAS.tmp.3565H}
}

@ARTICLE{ICECube2018,
       author = {{IceCube Collaboration} and {Aartsen}, M.~G. and {Ackermann}, M. and {Adams}, J. and {Aguilar}, J.~A. and {Ahlers}, M. and {Ahrens}, M. and {Samarai}, I. Al and {Altmann}, D. and {Andeen}, K. and {Anderson}, T. and {Ansseau}, I. and {Anton}, G. and {Arg{\"u}elles}, C. and {Arsioli}, B. and {Auffenberg}, J. and {Axani}, S. and {Bagherpour}, H. and {Bai}, X. and {Barron}, J.~P. and {Barwick}, S.~W. and {Baum}, V. and {Bay}, R. and {Beatty}, J.~J. and {Becker Tjus}, J. and {Becker}, K.-H. and {BenZvi}, S. and {Berley}, D. and {Bernardini}, E. and {Besson}, D.~Z. and {Binder}, G. and {Bindig}, D. and {Blaufuss}, E. and {Blot}, S. and {Bohm}, C. and {B{\"o}rner}, M. and {Bos}, F. and {B{\"o}ser}, S. and {Botner}, O. and {Bourbeau}, E. and {Bourbeau}, J. and {Bradascio}, F. and {Braun}, J. and {Brenzke}, M. and {Bretz}, H.-P. and {Bron}, S. and {Brostean-Kaiser}, J. and {Burgman}, A. and {Busse}, R.~S. and {Carver}, T. and {Cheung}, E. and {Chirkin}, D. and {Christov}, A. and {Clark}, K. and {Classen}, L. and {Coenders}, S. and {Collin}, G.~H. and {Conrad}, J.~M. and {Coppin}, P. and {Correa}, P. and {Cowen}, D.~F. and {Cross}, R. and {Dave}, P. and {Day}, M. and {de Andr{\'e}}, J.~P.~A.~M. and {De Clercq}, C. and {DeLaunay}, J.~J. and {Dembinski}, H. and {DeRidder}, S. and {Desiati}, P. and {de Vries}, K.~D. and {de Wasseige}, G. and {de With}, M. and {DeYoung}, T. and {D{\'\i}az-V{\'e}lez}, J.~C. and {di Lorenzo}, V. and {Dujmovic}, H. and {Dumm}, J.~P. and {Dunkman}, M. and {Dvorak}, E. and {Eberhardt}, B. and {Ehrhardt}, T. and {Eichmann}, B. and {Eller}, P. and {Evenson}, P.~A. and {Fahey}, S. and {Fazely}, A.~R. and {Felde}, J. and {Filimonov}, K. and {Finley}, C. and {Flis}, S. and {Franckowiak}, A. and {Friedman}, E. and {Fritz}, A. and {Gaisser}, T.~K. and {Gallagher}, J. and {Gerhardt}, L. and {Ghorbani}, K. and {Giommi}, P. and {Glauch}, T. and {Gl{\"u}senkamp}, T. and {Goldschmidt}, A. and {Gonzalez}, J.~G. and {Grant}, D. and {Griffith}, Z. and {Haack}, C. and {Hallgren}, A. and {Halzen}, F. and {Hanson}, K. and {Hebecker}, D. and {Heereman}, D. and {Helbing}, K. and {Hellauer}, R. and {Hickford}, S. and {Hignight}, J. and {Hill}, G.~C. and {Hoffman}, K.~D. and {Hoffmann}, R. and {Hoinka}, T. and {Hokanson-Fasig}, B. and {Hoshina}, K. and {Huang}, F. and {Huber}, M. and {Hultqvist}, K. and {H{\"u}nnefeld}, M. and {Hussain}, R. and {In}, S. and {Iovine}, N. and {Ishihara}, A. and {Jacobi}, E. and {Japaridze}, G.~S. and {Jeong}, M. and {Jero}, K. and {Jones}, B.~J.~P. and {Kalaczynski}, P. and {Kang}, W. and {Kappes}, A. and {Kappesser}, D. and {Karg}, T. and {Karle}, A. and {Katz}, U. and {Kauer}, M. and {Keivani}, A. and {Kelley}, J.~L. and {Kheirandish}, A. and {Kim}, J. and {Kim}, M. and {Kintscher}, T. and {Kiryluk}, J. and {Kittler}, T. and {Klein}, S.~R. and {Koirala}, R. and {Kolanoski}, H. and {K{\"o}pke}, L. and {Kopper}, C. and {Kopper}, S. and {Koschinsky}, J.~P. and {Koskinen}, D.~J. and {Kowalski}, M. and {Krammer}, B. and {Krings}, K. and {Kroll}, M. and {Kr{\"u}ckl}, G. and {Kunwar}, S. and {Kurahashi}, N. and {Kuwabara}, T. and {Kyriacou}, A. and {Labare}, M. and {Lanfranchi}, J.~L. and {Larson}, M.~J. and {Lauber}, F. and {Leonard}, K. and {Lesiak-Bzdak}, M. and {Leuermann}, M. and {Liu}, Q.~R. and {Lozano Mariscal}, C.~J. and {Lu}, L. and {L{\"u}nemann}, J. and {Luszczak}, W. and {Madsen}, J. and {Maggi}, G. and {Mahn}, K.~B.~M. and {Mancina}, S. and {Maruyama}, R. and {Mase}, K. and {Maunu}, R. and {Meagher}, K. and {Medici}, M. and {Meier}, M. and {Menne}, T. and {Merino}, G. and {Meures}, T. and {Miarecki}, S. and {Micallef}, J. and {Moment{\'e}}, G. and {Montaruli}, T. and {Moore}, R.~W. and {Morse}, R. and {Moulai}, M. and {Nahnhauer}, R.},
        title = "{Neutrino emission from the direction of the blazar TXS 0506+056 prior to the IceCube-170922A alert}",
      journal = {Science},
         year = 2018,
        month = jul,
       volume = {361},
       number = {6398},
        pages = {147-151},
          doi = {10.1126/science.aat2890},
archivePrefix = {arXiv},
       eprint = {1807.08794},
 primaryClass = {astro-ph.HE},
       adsurl = {https://ui.adsabs.harvard.edu/abs/2018Sci...361..147I}
}

@ARTICLE{Icecube2022,
       author = {{IceCube Collaboration} and {Abbasi}, R. and {Ackermann}, M. and {Adams}, J. and {Aguilar}, J.~A. and {Ahlers}, M. and {Ahrens}, M. and {Alameddine}, J.~M. and {Alispach}, C. and {Alves}, Jr., A.~A. and {Amin}, N.~M. and {Andeen}, K. and {Anderson}, T. and {Anton}, G. and {Arg{\"u}elles}, C. and {Ashida}, Y. and {Axani}, S. and {Bai}, X. and {Balagopal}, A.~V. and {Barbano}, V.~A. and {Barwick}, S.~W. and {Bastian}, B. and {Basu}, V. and {Baur}, S. and {Bay}, R. and {Beatty}, J.~J. and {Becker}, K.-H. and {Becker Tjus}, J. and {Bellenghi}, C. and {Benzvi}, S. and {Berley}, D. and {Bernardini}, E. and {Besson}, D.~Z. and {Binder}, G. and {Bindig}, D. and {Blaufuss}, E. and {Blot}, S. and {Boddenberg}, M. and {Bontempo}, F. and {Borowka}, J. and {B{\"o}ser}, S. and {Botner}, O. and {B{\"o}ttcher}, J. and {Bourbeau}, E. and {Bradascio}, F. and {Braun}, J. and {Brinson}, B. and {Bron}, S. and {Brostean-Kaiser}, J. and {Browne}, S. and {Burgman}, A. and {Burley}, R.~T. and {Busse}, R.~S. and {Campana}, M.~A. and {Carnie-Bronca}, E.~G. and {Chen}, C. and {Chen}, Z. and {Chirkin}, D. and {Choi}, K. and {Clark}, B.~A. and {Clark}, K. and {Classen}, L. and {Coleman}, A. and {Collin}, G.~H. and {Conrad}, J.~M. and {Coppin}, P. and {Correa}, P. and {Cowen}, D.~F. and {Cross}, R. and {Dappen}, C. and {Dave}, P. and {de Clercq}, C. and {Delaunay}, J.~J. and {Delgado L{\'o}pez}, D. and {Dembinski}, H. and {Deoskar}, K. and {Desai}, A. and {Desiati}, P. and {de Vries}, K.~D. and {de Wasseige}, G. and {de With}, M. and {Deyoung}, T. and {Diaz}, A. and {D{\'\i}az-V{\'e}lez}, J.~C. and {Dittmer}, M. and {Dujmovic}, H. and {Dunkman}, M. and {Duvernois}, M.~A. and {Dvorak}, E. and {Ehrhardt}, T. and {Eller}, P. and {Engel}, R. and {Erpenbeck}, H. and {Evans}, J. and {Evenson}, P.~A. and {Fan}, K.~L. and {Fazely}, A.~R. and {Fedynitch}, A. and {Feigl}, N. and {Fiedlschuster}, S. and {Fienberg}, A.~T. and {Filimonov}, K. and {Finley}, C. and {Fischer}, L. and {Fox}, D. and {Franckowiak}, A. and {Friedman}, E. and {Fritz}, A. and {F{\"u}rst}, P. and {Gaisser}, T.~K. and {Gallagher}, J. and {Ganster}, E. and {Garcia}, A. and {Garrappa}, S. and {Gerhardt}, L. and {Ghadimi}, A. and {Glaser}, C. and {Glauch}, T. and {Gl{\"u}senkamp}, T. and {Goldschmidt}, A. and {Gonzalez}, J.~G. and {Goswami}, S. and {Grant}, D. and {Gr{\'e}goire}, T. and {Griswold}, S. and {G{\"u}nther}, C. and {Gutjahr}, P. and {Haack}, C. and {Hallgren}, A. and {Halliday}, R. and {Halve}, L. and {Halzen}, F. and {Hanson}, M. Ha Minh K. and {Hardin}, J. and {Harnisch}, A.~A. and {Haungs}, A. and {Hebecker}, D. and {Helbing}, K. and {Henningsen}, F. and {Hettinger}, E.~C. and {Hickford}, S. and {Hignight}, J. and {Hill}, C. and {Hill}, G.~C. and {Hoffman}, K.~D. and {Hoffmann}, R. and {Hokanson-Fasig}, B. and {Hoshina}, K. and {Huang}, F. and {Huber}, M. and {Huber}, T. and {Hultqvist}, K. and {H{\"u}nnefeld}, M. and {Hussain}, R. and {Hymon}, K. and {in}, S. and {Iovine}, N. and {Ishihara}, A. and {Jansson}, M. and {Japaridze}, G.~S. and {Jeong}, M. and {Jin}, M. and {Jones}, B.~J.~P. and {Kang}, D. and {Kang}, W. and {Kang}, X. and {Kappes}, A. and {Kappesser}, D. and {Kardum}, L. and {Karg}, T. and {Karl}, M. and {Karle}, A. and {Katz}, U. and {Kauer}, M. and {Kellermann}, M. and {Kelley}, J.~L. and {Kheirandish}, A. and {Kin}, K. and {Kintscher}, T. and {Kiryluk}, J. and {Klein}, S.~R. and {Koirala}, R. and {Kolanoski}, H. and {Kontrimas}, T. and {K{\"o}pke}, L. and {Kopper}, C. and {Kopper}, S. and {Koskinen}, D.~J. and {Koundal}, P. and {Kovacevich}, M. and {Kowalski}, M. and {Kozynets}, T. and {Kun}, E. and {Kurahashi}, N. and {Lad}, N. and {Lagunas Gualda}, C. and {Lanfranchi}, J.~L. and {Larson}, M.~J. and {Lauber}, F. and {Lazar}, J.~P.},
        title = "{Evidence for neutrino emission from the nearby active galaxy NGC 1068}",
      journal = {Science},
         year = 2022,
        month = nov,
       volume = {378},
       number = {6619},
        pages = {538-543},
          doi = {10.1126/science.abg3395},
archivePrefix = {arXiv},
       eprint = {2211.09972},
 primaryClass = {astro-ph.HE},
       adsurl = {https://ui.adsabs.harvard.edu/abs/2022Sci...378..538I}
}

@ARTICLE{Ingram2023,
       author = {{Ingram}, A. and {Ewing}, M. and {Marinucci}, A. and {Tagliacozzo}, D. and {Rosario}, D.~J. and {Veledina}, A. and {Kim}, D.~E. and {Marin}, F. and {Bianchi}, S. and {Poutanen}, J. and {Matt}, G. and {Marshall}, H.~L. and {Ursini}, F. and {De Rosa}, A. and {Petrucci}, P. -O. and {Madejski}, G. and {Barnouin}, T. and {Gesu}, L. Di and {Dov{\v{c}}iak}, M. and {Gianolli}, V.~E. and {Krawczynski}, H. and {Loktev}, V. and {Middei}, R. and {Podgorny}, J. and {Puccetti}, S. and {Ratheesh}, A. and {Soffitta}, P. and {Tombesi}, F. and {Ehlert}, S.~R. and {Massaro}, F. and {Agudo}, I. and {Antonelli}, L.~A. and {Bachetti}, M. and {Baldini}, L. and {Baumgartner}, W.~H. and {Bellazzini}, R. and {Bongiorno}, S.~D. and {Bonino}, R. and {Brez}, A. and {Bucciantini}, N. and {Capitanio}, F. and {Castellano}, S. and {Cavazzuti}, E. and {Chen}, C. -T. and {Ciprini}, S. and {Costa}, E. and {Del Monte}, E. and {Lalla}, N. Di and {Marco}, A. Di and {Donnarumma}, I. and {Doroshenko}, V. and {Enoto}, T. and {Evangelista}, Y. and {Fabiani}, S. and {Ferrazzoli}, R. and {Garc{\'\i}a}, J.~A. and {Gunji}, S. and {Heyl}, J. and {Iwakiri}, W. and {Jorstad}, S.~G. and {Kaaret}, P. and {Karas}, V. and {Kislat}, F. and {Kitaguchi}, T. and {Kolodziejczak}, J.~J. and {Monaca}, F. La and {Latronico}, L. and {Liodakis}, I. and {Maldera}, S. and {Manfreda}, A. and {Marscher}, A.~P. and {Mitsuishi}, I. and {Mizuno}, T. and {Muleri}, F. and {Negro}, M. and {Ng}, C. -Y. and {O'Dell}, S.~L. and {Omodei}, N. and {Oppedisano}, C. and {Papitto}, A. and {Pavlov}, G.~G. and {Peirson}, A.~L. and {Perri}, M. and {Pesce-Rollins}, M. and {Pilia}, M. and {Possenti}, A. and {Ramsey}, B.~D. and {Rankin}, J. and {Roberts}, O.~J. and {Romani}, R.~W. and {Sgr{\`o}}, C. and {Slane}, P. and {Spandre}, G. and {Swartz}, D.~A. and {Tamagawa}, T. and {Tavecchio}, F. and {Taverna}, R. and {Tawara}, Y. and {Tennant}, A.~F. and {Thomas}, N.~E. and {Trois}, A. and {Tsygankov}, S.~S. and {Turolla}, R. and {Vink}, J. and {Weisskopf}, M.~C. and {Wu}, K. and {Xie}, F. and {Zane}, S.},
        title = "{The X-ray polarization of the Seyfert 1 galaxy IC 4329A}",
      journal = {\mnras},
         year = 2023,
        month = nov,
       volume = {525},
       number = {4},
        pages = {5437-5449},
          doi = {10.1093/mnras/stad2625},
archivePrefix = {arXiv},
       eprint = {2305.13028},
 primaryClass = {astro-ph.HE},
       adsurl = {https://ui.adsabs.harvard.edu/abs/2023MNRAS.525.5437I}
}

@ARTICLE{Ingram2024,
       author = {{Ingram}, Adam and {Bollemeijer}, Niek and {Veledina}, Alexandra and {Dov{\v{c}}iak}, Michal and {Poutanen}, Juri and {Egron}, Elise and {Russell}, Thomas D. and {Trushkin}, Sergei A. and {Negro}, Michela and {Ratheesh}, Ajay and {Capitanio}, Fiamma and {Connors}, Riley and {Neilsen}, Joseph and {Kraus}, Alexander and {Iacolina}, Maria Noemi and {Pellizzoni}, Alberto and {Pilia}, Maura and {Carotenuto}, Francesco and {Matt}, Giorgio and {Mastroserio}, Guglielmo and {Kaaret}, Philip and {Bianchi}, Stefano and {Garc{\'\i}a}, Javier A. and {Bachetti}, Matteo and {Wu}, Kinwah and {Costa}, Enrico and {Ewing}, Melissa and {Kravtsov}, Vadim and {Krawczynski}, Henric and {Loktev}, Vladislav and {Marinucci}, Andrea and {Marra}, Lorenzo and {Miku{\v{s}}incov{\'a}}, Romana and {Nathan}, Edward and {Parra}, Maxime and {Petrucci}, Pierre-Olivier and {Righini}, Simona and {Soffitta}, Paolo and {Steiner}, James F. and {Svoboda}, Ji{\v{r}}{\'\i} and {Tombesi}, Francesco and {Tugliani}, Stefano and {Ursini}, Francesco and {Yang}, Yi-Jung and {Zane}, Silvia and {Zhang}, Wenda and {Agudo}, Iv{\'a}n and {Antonelli}, Lucio A. and {Baldini}, Luca and {Baumgartner}, Wayne H. and {Bellazzini}, Ronaldo and {Bongiorno}, Stephen D. and {Bonino}, Raffaella and {Brez}, Alessandro and {Bucciantini}, Niccol{\`o} and {Castellano}, Simone and {Cavazzuti}, Elisabetta and {Chen}, Chien-Ting and {Ciprini}, Stefano and {De Rosa}, Alessandra and {Del Monte}, Ettore and {Di Gesu}, Laura and {Di Lalla}, Niccol{\`o} and {Di Marco}, Alessandro and {Donnarumma}, Immacolata and {Doroshenko}, Victor and {Ehlert}, Steven R. and {Enoto}, Teruaki and {Evangelista}, Yuri and {Fabiani}, Sergio and {Ferrazzoli}, Riccardo and {Gunji}, Shuichi and {Hayashida}, Kiyoshi and {Heyl}, Jeremy and {Iwakiri}, Wataru and {Jorstad}, Svetlana G. and {Karas}, Vladimir and {Kislat}, Fabian and {Kitaguchi}, Takao and {Kolodziejczak}, Jeffery J. and {La Monaca}, Fabio and {Latronico}, Luca and {Liodakis}, Ioannis and {Maldera}, Simone and {Manfreda}, Alberto and {Marin}, Fr{\'e}d{\'e}ric and {Marscher}, Alan P. and {Marshall}, Herman L. and {Massaro}, Francesco and {Mitsuishi}, Ikuyuki and {Mizuno}, Tsunefumi and {Muleri}, Fabio and {Ng}, Chi-Yung and {O'Dell}, Stephen L. and {Omodei}, Nicola and {Oppedisano}, Chiara and {Papitto}, Alessandro and {Pavlov}, George G. and {Peirson}, Abel L. and {Perri}, Matteo and {Pesce-Rollins}, Melissa and {Possenti}, Andrea and {Puccetti}, Simonetta and {Ramsey}, Brian D. and {Rankin}, John and {Roberts}, Oliver J. and {Romani}, Roger W. and {Sgr{\`o}}, Carmelo and {Slane}, Patrick and {Spandre}, Gloria and {Swartz}, Douglas A. and {Tamagawa}, Toru and {Tavecchio}, Fabrizio and {Taverna}, Roberto and {Tawara}, Yuzuru and {Tennant}, Allyn F. and {Thomas}, Nicholas E. and {Trois}, Alessio and {Tsygankov}, Sergey S. and {Turolla}, Roberto and {Vink}, Jacco and {Weisskopf}, Martin C. and {Xie}, Fei and {IXPE Collaboration}},
        title = "{Tracking the X-Ray Polarization of the Black Hole Transient Swift J1727.8{\textendash}1613 during a State Transition}",
      journal = {\apj},
         year = 2024,
        month = jun,
       volume = {968},
       number = {2},
          eid = {76},
        pages = {76},
          doi = {10.3847/1538-4357/ad3faf},
archivePrefix = {arXiv},
       eprint = {2311.05497},
 primaryClass = {astro-ph.HE},
       adsurl = {https://ui.adsabs.harvard.edu/abs/2024ApJ...968...76I}
}

@ARTICLE{Khabibullin2025,
       author = {{Khabibullin}, Ildar and {Churazov}, Eugene and {Ferrazzoli}, Riccardo and {Kaaret}, Philip and {Kolodziejczak}, Jeffery J. and {Marin}, Fr{\'e}d{\'e}ric and {Sunyaev}, Rashid and {Svoboda}, Jiri and {Vikhlinin}, Alexey and {Barnouin}, Thibault and {Chen}, Chien-Ting and {Costa}, Enrico and {Di Gesu}, Laura and {Di Marco}, Alessandro and {Ehlert}, Steven R. and {Forman}, William and {Kim}, Dawoon E. and {Kraft}, Ralph and {Maksym}, W. Peter and {Matt}, Giorgio and {Poutanen}, Juri and {Soffitta}, Paolo and {Swartz}, Douglas A. and {Agudo}, Ivan and {Antonelli}, Lucio Angelo and {Baldini}, Luca and {Baumgartner}, Wayne H. and {Bellazzini}, Ronaldo and {Bianchi}, Stefano and {Bongiorno}, Stephen D. and {Bonino}, Raffaella and {Brez}, Alessandro and {Bucciantini}, Niccolo and {Capitanio}, Fiamma and {Castellano}, Simone and {Cavazzuti}, Elisabetta and {Ciprini}, Stefano and {De Rosa}, Alessandra and {Del Monte}, Ettore and {Di Lalla}, Niccol{\`o} and {Donnarumma}, Immacolata and {Doroshenko}, Victor and {Dovciak}, Michal and {Enoto}, Teruaki and {Evangelista}, Yuri and {Fabiani}, Sergio and {Garcia}, Javier A. and {Gunji}, Shuichi and {Hayashida}, Kiyoshi and {Heyl}, Jeremy and {Iwakiri}, Wataru and {Jorstad}, Svetlana G. and {Karas}, Vladimir and {Kislat}, Fabian and {Kitaguchi}, Takao and {Krawczynski}, Henric and {La Monaca}, Fabio and {Latronico}, Luca and {Liodakis}, Ioannis and {Maldera}, Simone and {Manfreda}, Alberto and {Marscher}, Alan P. and {Marshall}, Herman L. and {Massaro}, Francesco and {Mitsuishi}, Ikuyuki and {Mizuno}, Tsunefumi and {Muleri}, Fabio and {Negro}, Michela and {Ng}, Chi-Yung and {O'Dell}, Stephen L. and {Omodei}, Nicola and {Oppedisano}, Chiara and {Papitto}, Alessandro and {Pavlov}, George G. and {Peirson}, Abel Lawrence and {Pesce-Rollins}, Melissa and {Petrucci}, Pierre-Olivier and {Pilia}, Maura and {Possenti}, Andrea and {Puccetti}, Simonetta and {Ramsey}, Brian D. and {Rankin}, John and {Ratheesh}, Ajay and {Roberts}, Oliver J. and {Romani}, Roger W. and {Sgro}, Carmelo and {Slane}, Patrick and {Spandre}, Gloria and {Tamagawa}, Toru and {Tavecchio}, Fabrizio and {Taverna}, Roberto and {Tawara}, Yuzuru and {Tennant}, Allyn F. and {Thomas}, Nicholas E. and {Tombesi}, Francesco and {Trois}, Alessio and {Tsygankov}, Sergey S. and {Turolla}, Roberto and {Vink}, Jacco and {Weisskopf}, Martin C. and {Wu}, Kinwah and {Xie}, Fei and {Zane}, Silvia},
        title = "{Polarization of reflected X-ray emission from Sgr A molecular complex: multiple flares, multiple sources?}",
      journal = {arXiv e-prints},
         year = 2025,
        month = aug,
          eid = {arXiv:2508.04862},
        pages = {arXiv:2508.04862},
          doi = {10.48550/arXiv.2508.04862},
archivePrefix = {arXiv},
       eprint = {2508.04862},
 primaryClass = {astro-ph.HE},
       adsurl = {https://ui.adsabs.harvard.edu/abs/2025arXiv250804862K}
}

@ARTICLE{Kislat2015,
       author = {{Kislat}, F. and {Clark}, B. and {Beilicke}, M. and {Krawczynski}, H.},
        title = "{Analyzing the data from X-ray polarimeters with Stokes parameters}",
      journal = {Astroparticle Physics},
         year = 2015,
        month = aug,
       volume = {68},
        pages = {45-51},
          doi = {10.1016/j.astropartphys.2015.02.007},
archivePrefix = {arXiv},
       eprint = {1409.6214},
 primaryClass = {astro-ph.IM},
       adsurl = {https://ui.adsabs.harvard.edu/abs/2015APh....68...45K}
}

@ARTICLE{Kouch2025,
       author = {{Kouch}, Pouya M. and {Liodakis}, Ioannis and {Fenu}, Francesco and {Zhang}, Haocheng and {Boula}, Stella and {Middei}, Riccardo and {Di Gesu}, Laura and {Paraschos}, Georgios F. and {Agudo}, Iv{\'a}n and {Jorstad}, Svetlana G. and {Lindfors}, Elina and {Marscher}, Alan P. and {Krawczynski}, Henric and {Negro}, Michela and {Hu}, Kun and {Kim}, Dawoon E. and {Cavazzuti}, Elisabetta and {Errando}, Manel and {Blinov}, Dmitry and {Gourni}, Anastasia and {Kiehlmann}, Sebastian and {Kourtidis}, Angelos and {Mandarakas}, Nikos and {Triantafyllou}, Nikolaos and {Vervelaki}, Anna and {Borman}, George A. and {Kopatskaya}, Evgenia N. and {Larionova}, Elena G. and {Morozova}, Daria A. and {Savchenko}, Sergey S. and {Vasilyev}, Andrey A. and {Troitskiy}, Ivan S. and {Grishina}, Tatiana S. and {Shishkina}, Ekaterina V. and {Zhovtan}, Alexey V. and {Aceituno}, Francisco Jos{\'e} and {Bonnoli}, Giacomo and {Casanova}, V{\'\i}ctor and {Escudero}, Juan and {Ag{\'\i}s-Gonz{\'a}lez}, Beatriz and {Husillos}, C{\'e}sar and {Otero-Santos}, Jorge and {Piirola}, Vilppu and {Sota}, Alfredo and {Myserlis}, Ioannis and {Gurwell}, Mark and {Keating}, Garrett K. and {Rao}, Ramprasad and {Angelakis}, Emmanouil and {Kraus}, Alexander and {Antonelli}, Lucio Angelo and {Bachetti}, Matteo and {Baldini}, Luca and {Baumgartner}, Wayne H. and {Bellazzini}, Ronaldo and {Bianchi}, Stefano and {Bongiorno}, Stephen D. and {Bonino}, Raffaella and {Brez}, Alessandro and {Bucciantini}, Niccol{\`o} and {Capitanio}, Fiamma and {Castellano}, Simone and {Chen}, Chien-Ting and {Ciprini}, Stefano and {Costa}, Enrico and {De Rosa}, Alessandra and {Del Monte}, Ettore and {Di Lalla}, Niccol{\`o} and {Di Marco}, Alessandro and {Donnarumma}, Immacolata and {Doroshenko}, Victor and {Dov{\v{c}}iak}, Michal and {Ehlert}, Steven R. and {Enoto}, Teruaki and {Evangelista}, Yuri and {Fabiani}, Sergio and {Ferrazzoli}, Riccardo and {Garcia}, Javier A. and {Gunji}, Shuichi and {Hayashida}, Kiyoshi and {Heyl}, Jeremy and {Iwakiri}, Wataru and {Kaaret}, Philip and {Karas}, Vladimir and {Kislat}, Fabian and {Kitaguchi}, Takao and {Kolodziejczak}, Jeffery J. and {La Monaca}, Fabio and {Latronico}, Luca and {Maldera}, Simone and {Manfreda}, Alberto and {Marin}, Fr{\'e}d{\'e}ric and {Marinucci}, Andrea and {Marshall}, Herman L. and {Massaro}, Francesco and {Matt}, Giorgio and {Mitsuishi}, Ikuyuki and {Mizuno}, Tsunefumi and {Muleri}, Fabio and {Ng}, Chi-Yung and {O'Dell}, Stephen L. and {Omodei}, Nicola and {Oppedisano}, Chiara and {Papitto}, Alessandro and {Pavlov}, George G. and {Peirson}, Abel Lawrence and {Perri}, Matteo and {Pesce-Rollins}, Melissa and {Petrucci}, Pierre-Olivier and {Pilia}, Maura and {Possenti}, Andrea and {Poutanen}, Juri and {Puccetti}, Simonetta and {Ramsey}, Brian D. and {Rankin}, John and {Ratheesh}, Ajay and {Roberts}, Oliver J. and {Sgr{\`o}}, Carmelo and {Slane}, Patrick and {Soffitta}, Paolo and {Spandre}, Gloria and {Swartz}, Douglas A. and {Tamagawa}, Toru and {Tavecchio}, Fabrizio and {Taverna}, Roberto and {Tawara}, Yuzuru and {Tennant}, Allyn F. and {Thomas}, Nicholas E. and {Tombesi}, Francesco and {Trois}, Alessio and {Tsygankov}, Sergey S. and {Turolla}, Roberto and {Romani}, Roger W. and {Vink}, Jacco and {Weisskopf}, Martin C. and {Wu}, Kinwah and {Xie}, Fei and {Zane}, Silvia},
        title = "{IXPE observation of the low-synchrotron peaked blazar S4 0954+65 during an optical-X-ray flare}",
      journal = {\aap},
         year = 2025,
        month = mar,
       volume = {695},
          eid = {A99},
        pages = {A99},
          doi = {10.1051/0004-6361/202453127},
archivePrefix = {arXiv},
       eprint = {2411.16868},
 primaryClass = {astro-ph.HE},
       adsurl = {https://ui.adsabs.harvard.edu/abs/2025A&A...695A..99K}
}

@ARTICLE{Krawczynski2022,
       author = {{Krawczynski}, Henric and {Muleri}, Fabio and {Dov{\v{c}}iak}, Michal and {Veledina}, Alexandra and {Rodriguez Cavero}, Nicole and {Svoboda}, Jiri and {Ingram}, Adam and {Matt}, Giorgio and {Garcia}, Javier A. and {Loktev}, Vladislav and {Negro}, Michela and {Poutanen}, Juri and {Kitaguchi}, Takao and {Podgorn{\'y}}, Jakub and {Rankin}, John and {Zhang}, Wenda and {Berdyugin}, Andrei and {Berdyugina}, Svetlana V. and {Bianchi}, Stefano and {Blinov}, Dmitry and {Capitanio}, Fiamma and {Di Lalla}, Niccol{\`o} and {Draghis}, Paul and {Fabiani}, Sergio and {Kagitani}, Masato and {Kravtsov}, Vadim and {Kiehlmann}, Sebastian and {Latronico}, Luca and {Lutovinov}, Alexander A. and {Mandarakas}, Nikos and {Marin}, Fr{\'e}d{\'e}ric and {Marinucci}, Andrea and {Miller}, Jon M. and {Mizuno}, Tsunefumi and {Molkov}, Sergey V. and {Omodei}, Nicola and {Petrucci}, Pierre-Olivier and {Ratheesh}, Ajay and {Sakanoi}, Takeshi and {Semena}, Andrei N. and {Skalidis}, Raphael and {Soffitta}, Paolo and {Tennant}, Allyn F. and {Thalhammer}, Phillipp and {Tombesi}, Francesco and {Weisskopf}, Martin C. and {Wilms}, Joern and {Zhang}, Sixuan and {Agudo}, Iv{\'a}n and {Antonelli}, Lucio A. and {Bachetti}, Matteo and {Baldini}, Luca and {Baumgartner}, Wayne H. and {Bellazzini}, Ronaldo and {Bongiorno}, Stephen D. and {Bonino}, Raffaella and {Brez}, Alessandro and {Bucciantini}, Niccol{\`o} and {Castellano}, Simone and {Cavazzuti}, Elisabetta and {Ciprini}, Stefano and {Costa}, Enrico and {De Rosa}, Alessandra and {Del Monte}, Ettore and {Di Gesu}, Laura and {Di Marco}, Alessandro and {Donnarumma}, Immacolata and {Doroshenko}, Victor and {Ehlert}, Steven R. and {Enoto}, Teruaki and {Evangelista}, Yuri and {Ferrazzoli}, Riccardo and {Gunji}, Shuichi and {Hayashida}, Kiyoshi and {Heyl}, Jeremy and {Iwakiri}, Wataru and {Jorstad}, Svetlana G. and {Karas}, Vladimir and {Kolodziejczak}, Jeffery J. and {La Monaca}, Fabio and {Liodakis}, Ioannis and {Maldera}, Simone and {Manfreda}, Alberto and {Marscher}, Alan P. and {Marshall}, Herman L. and {Mitsuishi}, Ikuyuki and {Ng}, Chi-Yung and {O{\textquoteright}Dell}, Stephen L. and {Oppedisano}, Chiara and {Papitto}, Alessandro and {Pavlov}, George G. and {Peirson}, Abel L. and {Perri}, Matteo and {Pesce-Rollins}, Melissa and {Pilia}, Maura and {Possenti}, Andrea and {Puccetti}, Simonetta and {Ramsey}, Brian D. and {Romani}, Roger W. and {Sgr{\`o}}, Carmelo and {Slane}, Patrick and {Spandre}, Gloria and {Tamagawa}, Toru and {Tavecchio}, Fabrizio and {Taverna}, Roberto and {Tawara}, Yuzuru and {Thomas}, Nicholas E. and {Trois}, Alessio and {Tsygankov}, Sergey and {Turolla}, Roberto and {Vink}, Jacco and {Wu}, Kinwah and {Xie}, Fei and {Zane}, Silvia},
        title = "{Polarized x-rays constrain the disk-jet geometry in the black hole x-ray binary Cygnus X-1}",
      journal = {Science},
         year = 2022,
        month = nov,
       volume = {378},
       number = {6620},
        pages = {650-654},
          doi = {10.1126/science.add5399},
archivePrefix = {arXiv},
       eprint = {2206.09972},
 primaryClass = {astro-ph.HE},
       adsurl = {https://ui.adsabs.harvard.edu/abs/2022Sci...378..650K}
}

@ARTICLE{Kravtsov2025,
       author = {{Kravtsov}, Vadim and {Bocharova}, Anastasiia and {Veledina}, Alexandra and {Poutanen}, Juri and {Hughes}, Andrew K. and {Dov{\v{c}}iak}, Michal and {Egron}, Elise and {Muleri}, Fabio and {Podgorny}, Jakub and {Svoboda}, Ji{\v{r}}i and {Forsblom}, Sofia V. and {Berdyugin}, Andrei V. and {Blinov}, Dmitry and {Bright}, Joe S. and {Carotenuto}, Francesco and {Green}, David A. and {Ingram}, Adam and {Liodakis}, Ioannis and {Mandarakas}, Nikos and {Nitindala}, Anagha P. and {Rhodes}, Lauren and {Trushkin}, Sergei A. and {Tsygankov}, Sergey S. and {Brigitte}, Ma{\"\i}mouna and {Di Marco}, Alessandro and {Iacolina}, Noemi and {Krawczynski}, Henric and {La Monaca}, Fabio and {Loktev}, Vladislav and {Mastroserio}, Guglielmo and {Petrucci}, Pierre-Olivier and {Pilia}, Maura and {Tombesi}, Francesco and {Zdziarski}, Andrzej A.},
        title = "{Variability of X-ray polarization of Cyg X-1}",
      journal = {\aap},
         year = 2025,
        month = sep,
       volume = {701},
          eid = {A115},
        pages = {A115},
          doi = {10.1051/0004-6361/202555411},
archivePrefix = {arXiv},
       eprint = {2505.03942},
 primaryClass = {astro-ph.HE},
       adsurl = {https://ui.adsabs.harvard.edu/abs/2025A&A...701A.115K}
}

@INPROCEEDINGS{LaMonaca2021,
author = {{La Monaca}, F. and {Fabiani}, S. and {Lefevre}, C. and {Morbidini}, A. and {Piazzolla}, R. and {Amici}, F.
and {Attina'}, P. and {Brienza}, D. and {Costa}, E. and {Di Cosimo}, S. and {Di Marco}, A. and {Di Persio}, G.
and {Evangelista}, Y. and {Ferrazzoli}, R. and {Loffredo}, P. and {Muleri}, F. and {Rankin}, J. and {Ratheesh}, R.
and {Rubini}, A. and {Santoli}, F. and {Scalise}, E. and {Soffitta}, P. and {Tobia}, A. and {Trois}, A.
and {Xie}, F. and {Zeiger}, B.
},
title = "{IXPE DU-FM ions-UV filters characterization}",
volume = 11444,
booktitle = {Space Telescopes and Instrumentation 2020: Ultraviolet to Gamma Ray},
editor = {Jan-Willem A. den Herder and Shouleh Nikzad and Kazuhiro Nakazawa},
publisher = {SPIE},
pages = {1029-1039},
year = 2021,
doi = {10.1117/12.2567000},
URL = {https://doi.org/10.1117/12.2567000}
}

@ARTICLE{Lamonaca2024,
       author = {{La Monaca}, Fabio and {Di Marco}, Alessandro and {Poutanen}, Juri and {Bachetti}, Matteo and {Motta}, Sara Elisa and {Papitto}, Alessandro and {Pilia}, Maura and {Xie}, Fei and {Bianchi}, Stefano and {Bobrikova}, Anna and {Costa}, Enrico and {Deng}, Wei and {Ge}, Ming-Yu and {Illiano}, Giulia and {Jia}, Shu-Mei and {Krawczynski}, Henric and {Lai}, Eleonora Veronica and {Liu}, Kuan and {Mastroserio}, Guglielmo and {Muleri}, Fabio and {Rankin}, John and {Soffitta}, Paolo and {Veledina}, Alexandra and {Ambrosino}, Filippo and {Del Santo}, Melania and {Chen}, Wei and {Garcia}, Javier A. and {Kaaret}, Philip and {Russell}, Thomas D. and {Wei}, Wen-Hao and {Zhang}, Shuang-Nan and {Zuo}, Chao and {Arzoumanian}, Zaven and {Cocchi}, Massimo and {Gnarini}, Andrea and {Farinelli}, Ruben and {Gendreau}, Keith and {Ursini}, Francesco and {Weisskopf}, Martin C. and {Zane}, Silvia and {Agudo}, Iv{\'a}n and {Antonelli}, Lucio A. and {Baldini}, Luca and {Baumgartner}, Wayne H. and {Bellazzini}, Ronaldo and {Bongiorno}, Stephen D. and {Bonino}, Raffaella and {Brez}, Alessandro and {Bucciantini}, Niccol{\`o} and {Capitanio}, Fiamma and {Castellano}, Simone and {Cavazzuti}, Elisabetta and {Chen}, Chien-Ting and {Ciprini}, Stefano and {De Rosa}, Alessandra and {Del Monte}, Ettore and {Di Gesu}, Laura and {Di Lalla}, Niccol{\`o} and {Donnarumma}, Immacolata and {Doroshenko}, Victor and {Dov{\v{c}}iak}, Michal and {Ehlert}, Steven R. and {Enoto}, Teruaki and {Evangelista}, Yuri and {Fabiani}, Sergio and {Ferrazzoli}, Riccardo and {Gunji}, Shuichi and {Hayashida}, Kiyoshi and {Heyl}, Jeremy and {Iwakiri}, Wataru and {Jorstad}, Svetlana G. and {Karas}, Vladimir and {Kislat}, Fabian and {Kitaguchi}, Takao and {Kolodziejczak}, Jeffery J. and {Latronico}, Luca and {Liodakis}, Ioannis and {Maldera}, Simone and {Manfreda}, Alberto and {Marin}, Fr{\'e}d{\'e}ric and {Marinucci}, Andrea and {Marscher}, Alan P. and {Marshall}, Herman L. and {Massaro}, Francesco and {Matt}, Giorgio and {Mitsuishi}, Ikuyuki and {Mizuno}, Tsunefumi and {Negro}, Michela and {Ng}, Chi-Yung and {O'Dell}, Stephen L. and {Omodei}, Nicola and {Oppedisano}, Chiara and {Pavlov}, George G. and {Peirson}, Abel L. and {Perri}, Matteo and {Pesce-Rollins}, Melissa and {Petrucci}, Pierre-Olivier and {Possenti}, Andrea and {Puccetti}, Simonetta and {Ramsey}, Brian D. and {Ratheesh}, Ajay and {Roberts}, Oliver J. and {Romani}, Roger W. and {Sgr{\`o}}, Carmelo and {Slane}, Patrick and {Spandre}, Gloria and {Swartz}, Douglas A. and {Tamagawa}, Toru and {Tavecchio}, Fabrizio and {Taverna}, Roberto and {Tawara}, Yuzuru and {Tennant}, Allyn F. and {Thomas}, Nicholas E. and {Tombesi}, Francesco and {Trois}, Alessio and {Tsygankov}, Sergey S. and {Turolla}, Roberto and {Vink}, Jacco and {Wu}, Kinwah and {IXPE Collaboration}},
        title = "{Highly Significant Detection of X-Ray Polarization from the Brightest Accreting Neutron Star Sco X-1}",
      journal = {\apjl},
         year = 2024,
        month = jan,
       volume = {960},
       number = {2},
          eid = {L11},
        pages = {L11},
          doi = {10.3847/2041-8213/ad132d},
archivePrefix = {arXiv},
       eprint = {2311.06359},
 primaryClass = {astro-ph.HE},
       adsurl = {https://ui.adsabs.harvard.edu/abs/2024ApJ...960L..11L}
}

@ARTICLE{Ling2024,
       author = {{Ling}, Yu-Shan and {Xie}, Fei and {Ge}, Ming-Yu and {La Monaca}, Fabio},
        title = "{Polarization Study of Swift J151857.0{\textendash}572147 with IXPE Observation}",
      journal = {Research in Astronomy and Astrophysics},
         year = 2024,
        month = sep,
       volume = {24},
       number = {9},
          eid = {095004},
        pages = {095004},
          doi = {10.1088/1674-4527/ad6edf},
       adsurl = {https://ui.adsabs.harvard.edu/abs/2024RAA....24i5004L}
}

@ARTICLE{Loktev2025,
       author = {{Loktev}, Vladislav and {Forsblom}, Sofia V. and {Tsygankov}, Sergey S. and {Poutanen}, Juri and {Mushtukov}, Alexander A. and {Di Marco}, Alessandro and {Heyl}, Jeremy and {Kelly}, Ruth M.~E. and {La Monaca}, Fabio and {Ng}, Mason and {Ravi}, Swati and {Salganik}, Alexander and {Santangelo}, Andrea and {Suleimanov}, Valery F. and {Zane}, Silvia},
        title = "{Exploring polarization and geometry in the X-ray pulsar 4U 1538{\ensuremath{-}}52}",
      journal = {\aap},
         year = 2025,
        month = may,
       volume = {698},
          eid = {A22},
        pages = {A22},
          doi = {10.1051/0004-6361/202554151},
archivePrefix = {arXiv},
       eprint = {2503.13720},
 primaryClass = {astro-ph.HE},
       adsurl = {https://ui.adsabs.harvard.edu/abs/2025A&A...698A..22L}
}

@ARTICLE{Majumder2024,
       author = {{Majumder}, Seshadri and {Kushwaha}, Ankur and {Das}, Santabrata and {Nandi}, Anuj},
        title = "{First detection of X-ray polarization in thermal state of LMC X-3: spectro-polarimetric study with IXPE}",
      journal = {\mnras},
         year = 2024,
        month = jan,
       volume = {527},
       number = {1},
        pages = {L76-L81},
          doi = {10.1093/mnrasl/slad148},
archivePrefix = {arXiv},
       eprint = {2309.06845},
 primaryClass = {astro-ph.HE},
       adsurl = {https://ui.adsabs.harvard.edu/abs/2024MNRAS.527L..76M}
}

@ARTICLE{Marin2024,
       author = {{Marin}, F. and {Marinucci}, A. and {Laurenti}, M. and {Kim}, D.~E. and {Barnouin}, T. and {Di Marco}, A. and {Ursini}, F. and {Bianchi}, S. and {Ravi}, S. and {Marshall}, H.~L. and {Matt}, G. and {Chen}, C. -T. and {Gianolli}, V.~E. and {Ingram}, A. and {Maksym}, W.~P. and {Panagiotou}, C. and {Podgorny}, J. and {Puccetti}, S. and {Ratheesh}, A. and {Tombesi}, F. and {Agudo}, I. and {Antonelli}, L.~A. and {Bachetti}, M. and {Baldini}, L. and {Baumgartner}, W. and {Bellazzini}, R. and {Bongiorno}, S. and {Bonino}, R. and {Brez}, A. and {Bucciantini}, N. and {Capitanio}, F. and {Castellano}, S. and {Cavazzuti}, E. and {Ciprini}, S. and {Costa}, E. and {De Rosa}, A. and {Del Monte}, E. and {Di Gesu}, L. and {Di Lalla}, N. and {Donnarumma}, I. and {Doroshenko}, V. and {Dovciak}, M. and {Ehlert}, S. and {Enoto}, T. and {Evangelista}, Y. and {Fabiani}, S. and {Ferrazzoli}, R. and {Garcia}, J. and {Gunji}, S. and {Heyl}, J. and {Iwakiri}, W. and {Jorstad}, S. and {Kaaret}, P. and {Karas}, V. and {Kislat}, F. and {Kitaguchi}, T. and {Kolodziejczak}, J. and {Krawczynski}, H. and {La Monaca}, F. and {Latronico}, L. and {Liodakis}, I. and {Madejski}, G. and {Maldera}, S. and {Manfreda}, A. and {Marscher}, A. and {Massaro}, F. and {Mitsuishi}, I. and {Mizuno}, T. and {Muleri}, F. and {Negro}, M. and {Ng}, S. and {O'Dell}, S. and {Omodei}, N. and {Oppedisano}, C. and {Papitto}, A. and {Pavlov}, G. and {Perri}, M. and {Pesce-Rollins}, M. and {Petrucci}, P. -O. and {Pilia}, M. and {Possenti}, A. and {Poutanen}, J. and {Ramsey}, B. and {Rankin}, J. and {Roberts}, O. and {Romani}, R. and {Sgro}, C. and {Slane}, P. and {Soffitta}, P. and {Spandre}, G. and {Swartz}, D. and {Tamagawa}, T. and {Tavecchio}, F. and {Taverna}, R. and {Tawara}, Y. and {Tennant}, A. and {Thomas}, N. and {Trois}, A. and {Tsygankov}, S. and {Turolla}, R. and {Vink}, J. and {Weisskopf}, M. and {Wu}, K. and {Xie}, F. and {Zane}, S.},
        title = "{X-ray polarization measurement of the gold standard of radio-quiet active galactic nuclei : NGC 1068}",
      journal = {arXiv e-prints},
         year = 2024,
        month = mar,
          eid = {arXiv:2403.02061},
        pages = {arXiv:2403.02061},
          doi = {10.48550/arXiv.2403.02061},
archivePrefix = {arXiv},
       eprint = {2403.02061},
 primaryClass = {astro-ph.HE},
       adsurl = {https://ui.adsabs.harvard.edu/abs/2024arXiv240302061M}
}

@ARTICLE{Marra2024,
       author = {{Marra}, L. and {Brigitte}, M. and {Rodriguez Cavero}, N. and {Chun}, S. and {Steiner}, J.~F. and {Dov{\v{c}}iak}, M. and {Nowak}, M. and {Bianchi}, S. and {Capitanio}, F. and {Ingram}, A. and {Matt}, G. and {Muleri}, F. and {Podgorn{\'y}}, J. and {Poutanen}, J. and {Svoboda}, J. and {Taverna}, R. and {Ursini}, F. and {Veledina}, A. and {De Rosa}, A. and {Garc{\'\i}a}, J.~A. and {Lutovinov}, A.~A. and {Mereminskiy}, I.~A. and {Farinelli}, R. and {Gunji}, S. and {Kaaret}, P. and {Kallman}, T. and {Krawczynski}, H. and {Kan}, Y. and {Hu}, K. and {Marinucci}, A. and {Mastroserio}, G. and {Mikus̆incov{\'a}}, R. and {Parra}, M. and {Petrucci}, P. -O. and {Ratheesh}, A. and {Soffitta}, P. and {Tombesi}, F. and {Zane}, S. and {Agudo}, I. and {Antonelli}, L.~A. and {Bachetti}, M. and {Baldini}, L. and {Baumgartner}, W.~H. and {Bellazzini}, R. and {Bongiorno}, S.~D. and {Bonino}, R. and {Brez}, A. and {Bucciantini}, N. and {Castellano}, S. and {Cavazzuti}, E. and {Chen}, C. and {Ciprini}, S. and {Costa}, E. and {Del Monte}, E. and {Di Gesu}, L. and {Di Lalla}, N. and {Di Marco}, A. and {Donnarumma}, I. and {Doroshenko}, V. and {Ehlert}, S.~R. and {Enoto}, T. and {Evangelista}, Y. and {Fabiani}, S. and {Ferrazzoli}, R. and {Hayashida}, K. and {Heyl}, J. and {Iwakiri}, W. and {Jorstad}, S.~G. and {Karas}, V. and {Kislat}, F. and {Kitaguchi}, T. and {Kolodziejczak}, J.~J. and {La Monaca}, F. and {Latronico}, L. and {Liodakis}, I. and {Maldera}, S. and {Manfreda}, A. and {Marin}, F. and {Marscher}, A.~P. and {Marshall}, H.~L. and {Massaro}, F. and {Mitsuishi}, I. and {Mizuno}, T. and {Negro}, M. and {Ng}, C.~Y. and {O'Dell}, S.~L. and {Omodei}, N. and {Oppedisano}, C. and {Papitto}, A. and {Pavlov}, G.~G. and {Peirson}, A.~L. and {Perri}, M. and {Pesce-Rollins}, M. and {Pilia}, M. and {Possenti}, A. and {Puccetti}, S. and {Ramsey}, B.~D. and {Rankin}, J. and {Roberts}, O.~J. and {Romani}, R.~W. and {Sgr{\`o}}, C. and {Slane}, P. and {Spandre}, G. and {Swartz}, D.~A. and {Tamagawa}, T. and {Tavecchio}, F. and {Tawara}, Y. and {Tennant}, A.~F. and {Thomas}, N.~E. and {Trois}, A. and {Tsygankov}, S.~S. and {Turolla}, R. and {Vink}, J. and {Weisskopf}, M.~C. and {Wu}, K. and {Xie}, F.},
        title = "{IXPE observation confirms a high spin in the accreting black hole 4U 1957+115}",
      journal = {\aap},
         year = 2024,
        month = apr,
       volume = {684},
          eid = {A95},
        pages = {A95},
          doi = {10.1051/0004-6361/202348277},
archivePrefix = {arXiv},
       eprint = {2310.11125},
 primaryClass = {astro-ph.HE},
       adsurl = {https://ui.adsabs.harvard.edu/abs/2024A&A...684A..95M}
}

@ARTICLE{Marscher2024,
       author = {{Marscher}, Alan P. and {Di Gesu}, Laura and {Jorstad}, Svetlana G. and {Kim}, Dawoon E. and {Liodakis}, Ioannis and {Middei}, Riccardo and {Tavecchio}, Fabrizio},
        title = "{X-ray Polarization of Blazars and Radio Galaxies Measured by the Imaging X-ray Polarimetry Explorer}",
      journal = {Galaxies},
         year = 2024,
        month = aug,
       volume = {12},
       number = {4},
          eid = {50},
        pages = {50},
          doi = {10.3390/galaxies12040050},
       adsurl = {https://ui.adsabs.harvard.edu/abs/2024Galax..12...50M}
}

@ARTICLE{Marshall2024,
       author = {{Marshall}, Herman L. and {Liodakis}, Ioannis and {Marscher}, Alan P. and {Di Lalla}, Niccol{\`o} and {Jorstad}, Svetlana G. and {Kim}, Dawoon E. and {Middei}, Riccardo and {Negro}, Michela and {Omodei}, Nicola and {Peirson}, Abel L. and {Perri}, Matteo and {Puccetti}, Simonetta and {Laurenti}, Marco and {Agudo}, Iv{\'a}n and {Bonnoli}, Giacomo and {Berdyugin}, Andrei V. and {Cavazzuti}, Elisabetta and {Rodriguez Cavero}, Nicole and {Donnarumma}, Immacolata and {Di Gesu}, Laura and {Jormanainen}, Jenni and {Krawczynski}, Henric and {Lindfors}, Elina and {Madjeski}, Greg and {Marin}, Fr{\'e}d{\'e}ric and {Massaro}, Francesco and {Pacciani}, Luigi and {Poutanen}, Juri and {Tavecchio}, Fabrizio and {Kouch}, Pouya M. and {Aceituno}, Francisco Jos{\'e} and {Bernardos}, Maria I. and {Casanova}, V{\'\i}ctor and {Garc{\'\i}a-Comas}, Maya and {Ag{\'\i}s-Gonz{\'a}lez}, Beatriz and {Husillos}, C{\'e}sar and {Marchini}, Alessandro and {Sota}, Alfredo and {Blinov}, Dmitry and {Bourbah}, Ioakeim G. and {Kielhmann}, Sebastian and {Kontopodis}, Evangelos and {Mandarakas}, Nikos and {Romanopoulos}, Stylianos and {Skalidis}, Raphael and {Vervelaki}, Anna and {Borman}, George A. and {Kopatskaya}, Evgenia N. and {Larionova}, Elena G. and {Morozova}, Daria A. and {Savchenko}, Sergey S. and {Vasilyev}, Andrey A. and {Zhovtan}, Alexey V. and {Casadio}, Carolina and {Escudero}, Juan and {Kramer}, Joana and {Myserlis}, Ioannis and {Trainou}, Efthalia and {Imazawa}, Ryo and {Sasada}, Mahito and {Fukazawa}, Yasushi and {Kawabata}, Koji S. and {Uemura}, Makoto and {Mizuno}, Tsunefumi and {Nakaoka}, Tatsuya and {Akitaya}, Hiroshi and {Masiero}, Joseph R. and {Mawet}, Dimitri and {Panopoulou}, Georgia V. and {Tinyanont}, Samaporn and {Kagitani}, Masato and {Kravtsov}, Vadim and {Sakanoi}, Takeshi and {Dattolo}, Matthew and {Gurwell}, Mark and {Keating}, Garrett and {Rao}, Ramprasad and {Cheong}, Whee Yeon and {Jeong}, Hyeon-Woo and {Kang}, Sincheol and {Kim}, Sang-Hyun and {Lee}, Sang-Sung and {Angelakis}, Emmanouil and {Kraus}, Alexander and {Hales}, Antonio and {Kameno}, Seiji and {Kneissl}, Ruediger and {Messias}, Hugo and {Nagai}, Hiroshi and {Antonelli}, Lucio A. and {Bachetti}, Matteo and {Baldini}, Luca and {Baumgartner}, Wayne H. and {Bellazzini}, Ronaldo and {Bianchi}, Stefano and {Bongiorno}, Stephen D. and {Bonino}, Raffaella and {Brez}, Alessandro and {Bucciantini}, Niccol{\`o} and {Capitanio}, Fiamma and {Castellano}, Simone and {Chen}, Chen-Ting and {Ciprini}, Stefano and {Costa}, Enrico and {De Rosa}, Alessandra and {Del Monte}, Ettore and {Di Marco}, Alessandro and {Doroshenko}, Victor and {Dov{\v{c}}iak}, Michal and {Ehlert}, Steven R. and {Enoto}, Teruaki and {Evangelista}, Yuri and {Fabiani}, Sergio and {Ferrazzoli}, Riccardo and {Garcia}, Javier A. and {Gunji}, Shuichi and {Hayashida}, Kiyoshi and {Heyl}, Jeremy and {Iwakiri}, Wataru and {Kaaret}, Philip and {Karas}, Vladimir and {Kislat}, Fabian and {Kitaguchi}, Takao and {Kolodziejczak}, Jeffery J. and {La Monaca}, Fabio and {Latronico}, Luca and {Maldera}, Simone and {Manfreda}, Alberto and {Marinucci}, Andrea and {Matt}, Giorgio and {Mitsuishi}, Ikuyuki and {Muleri}, Fabio and {Ng}, C.-Y. and {O'Dell}, Stephen L. and {Oppedisano}, Chiara and {Papitto}, Alessandro and {Pavlov}, George G. and {Pesce-Rollins}, Melissa and {Petrucci}, Pierre-Olivier and {Pilia}, Maura and {Possenti}, Andrea and {Ramsey}, Brian D. and {Rankin}, John and {Ratheesh}, Ajay and {Roberts}, Oliver J. and {Romani}, Roger W. and {Sgr{\`o}}, Carmelo and {Slane}, Patrick and {Soffitta}, Paolo and {Spandre}, Gloria and {Swartz}, Douglas A. and {Tamagawa}, Toru and {Taverna}, Roberto and {Tawara}, Yuzuru and {Tennant}, Allyn F. and {Thomas}, Nicholas E. and {Tombesi}, Francesco and {Trois}, Alessio and {Tsygankov}, Sergey S. and {Turolla}, Roberto and {Vink}, Jacco and {Weisskopf}, Martin C. and {Wu}, Kinwah and {Xie}, Fei and {Zane}, Silvia},
        title = "{Observations of Low and Intermediate Spectral Peak Blazars with the Imaging X-Ray Polarimetry Explorer}",
      journal = {\apj},
         year = 2024,
        month = sep,
       volume = {972},
       number = {1},
          eid = {74},
        pages = {74},
          doi = {10.3847/1538-4357/ad5671},
archivePrefix = {arXiv},
       eprint = {2310.11510},
 primaryClass = {astro-ph.HE},
       adsurl = {https://ui.adsabs.harvard.edu/abs/2024ApJ...972...74M}
}

@ARTICLE{Mastroserio2025,
       author = {{Mastroserio}, G. and {De Marco}, B. and {Baglio}, M.~C. and {Carotenuto}, F. and {Fabiani}, S. and {Russell}, T.~D. and {Capitanio}, F. and {Cavecchi}, Y. and {Motta}, S. and {Russell}, D.~M. and {Dov{\v{c}}iak}, M. and {Del Santo}, M. and {Alabarta}, K. and {Ambrifi}, A. and {Campana}, S. and {Casella}, P. and {Covino}, S. and {Illiano}, G. and {Kara}, E. and {Lai}, E.~V. and {Lodato}, G. and {Manca}, A. and {Mariani}, I. and {Marino}, A. and {Miceli}, C. and {Saikia}, P. and {Shaw}, A.~W. and {Svoboda}, J. and {Vincentelli}, F.~M. and {Wang}, J.},
        title = "{X-Ray and Optical Polarization Aligned with the Radio Jet Ejecta in GX 339{\textendash}4}",
      journal = {\apjl},
         year = 2025,
        month = jan,
       volume = {978},
       number = {2},
          eid = {L19},
        pages = {L19},
          doi = {10.3847/2041-8213/ad9913},
archivePrefix = {arXiv},
       eprint = {2408.06856},
 primaryClass = {astro-ph.HE},
       adsurl = {https://ui.adsabs.harvard.edu/abs/2025ApJ...978L..19M}
}

@ARTICLE{Marin2023,
       author = {{Marin}, Fr{\'e}d{\'e}ric and {Churazov}, Eugene and {Khabibullin}, Ildar and {Ferrazzoli}, Riccardo and {Di Gesu}, Laura and {Barnouin}, Thibault and {Di Marco}, Alessandro and {Middei}, Riccardo and {Vikhlinin}, Alexey and {Costa}, Enrico and {Soffitta}, Paolo and {Muleri}, Fabio and {Sunyaev}, Rashid and {Forman}, William and {Kraft}, Ralph and {Bianchi}, Stefano and {Donnarumma}, Immacolata and {Petrucci}, Pierre-Olivier and {Enoto}, Teruaki and {Agudo}, Iv{\'a}n and {Antonelli}, Lucio A. and {Bachetti}, Matteo and {Baldini}, Luca and {Baumgartner}, Wayne H. and {Bellazzini}, Ronaldo and {Bongiorno}, Stephen D. and {Bonino}, Raffaella and {Brez}, Alessandro and {Bucciantini}, Niccol{\`o} and {Capitanio}, Fiamma and {Castellano}, Simone and {Cavazzuti}, Elisabetta and {Chen}, Chien-Ting and {Ciprini}, Stefano and {De Rosa}, Alessandra and {Del Monte}, Ettore and {Di Lalla}, Niccol{\`o} and {Doroshenko}, Victor and {Dov{\v{c}}iak}, Michal and {Ehlert}, Steven R. and {Evangelista}, Yuri and {Fabiani}, Sergio and {Garcia}, Javier A. and {Gunji}, Shuichi and {Hayashida}, Kiyoshi and {Heyl}, Jeremy and {Ingram}, Adam and {Iwakiri}, Wataru and {Jorstad}, Svetlana G. and {Kaaret}, Philip and {Karas}, Vladimir and {Kitaguchi}, Takao and {Kolodziejczak}, Jeffery J. and {Krawczynski}, Henric and {La Monaca}, Fabio and {Latronico}, Luca and {Liodakis}, Ioannis and {Maldera}, Simone and {Manfreda}, Alberto and {Marinucci}, Andrea and {Marscher}, Alan P. and {Marshall}, Herman L. and {Massaro}, Francesco and {Matt}, Giorgio and {Mitsuishi}, Ikuyuki and {Mizuno}, Tsunefumi and {Negro}, Michela and {Ng}, C. -Y. and {O'Dell}, Stephen L. and {Omodei}, Nicola and {Oppedisano}, Chiara and {Papitto}, Alessandro and {Pavlov}, George G. and {Peirson}, Abel L. and {Perri}, Matteo and {Pesce-Rollins}, Melissa and {Pilia}, Maura and {Possenti}, Andrea and {Poutanen}, Juri and {Puccetti}, Simonetta and {Ramsey}, Brian D. and {Rankin}, John and {Ratheesh}, Ajay and {Roberts}, Oliver J. and {Romani}, Roger W. and {Sgr{\`o}}, Carmelo and {Slane}, Patrick and {Spandre}, Gloria and {Swartz}, Doug and {Tamagawa}, Toru and {Tavecchio}, Fabrizio and {Taverna}, Roberto and {Tawara}, Yuzuru and {Tennant}, Allyn F. and {Thomas}, Nicholas E. and {Tombesi}, Francesco and {Trois}, Alessio and {Tsygankov}, Sergey S. and {Turolla}, Roberto and {Vink}, Jacco and {Weisskopf}, Martin C. and {Wu}, Kinwah and {Xie}, Fei and {Zane}, Silvia},
        title = "{X-ray polarization evidence for a 200-year-old flare of Sgr A$^{*}$}",
      journal = {\nat},
         year = 2023,
        month = jul,
       volume = {619},
       number = {7968},
        pages = {41-45},
          doi = {10.1038/s41586-023-06064-x},
archivePrefix = {arXiv},
       eprint = {2304.06967},
 primaryClass = {astro-ph.HE},
       adsurl = {https://ui.adsabs.harvard.edu/abs/2023Natur.619...41M}
}

@ARTICLE{Marinucci2022,
       author = {{Marinucci}, A. and {Muleri}, F. and {Dovciak}, M. and {Bianchi}, S. and {Marin}, F. and {Matt}, G. and {Ursini}, F. and {Middei}, R. and {Marshall}, H.~L. and {Baldini}, L. and {Barnouin}, T. and {Rodriguez}, N. Cavero and {De Rosa}, A. and {Di Gesu}, L. and {Harper}, D. and {Ingram}, A. and {Karas}, V. and {Krawczynski}, H. and {Madejski}, G. and {Panagiotou}, C. and {Petrucci}, P.~O. and {Podgorny}, J. and {Puccetti}, S. and {Tombesi}, F. and {Veledina}, A. and {Zhang}, W. and {Agudo}, I. and {Antonelli}, L.~A. and {Bachetti}, M. and {Baumgartner}, W.~H. and {Bellazzini}, R. and {Bongiorno}, S.~D. and {Bonino}, R. and {Brez}, A. and {Bucciantini}, N. and {Capitanio}, F. and {Castellano}, S. and {Cavazzuti}, E. and {Ciprini}, S. and {Costa}, E. and {Del Monte}, E. and {Di Lalla}, N. and {Di Marco}, A. and {Donnarumma}, I. and {Doroshenko}, V. and {Ehlert}, S.~R. and {Enoto}, T. and {Evangelista}, Y. and {Fabiani}, S. and {Ferrazzoli}, R. and {Garcia}, J.~A. and {Gunji}, S. and {Hayashida}, K. and {Heyl}, J. and {Iwakiri}, W. and {Jorstad}, S.~G. and {Kitaguchi}, T. and {Kolodziejczak}, J.~J. and {La Monaca}, F. and {Latronico}, L. and {Liodakis}, I. and {Maldera}, S. and {Manfreda}, A. and {Marscher}, A.~P. and {Mitsuishi}, I. and {Mizuno}, T. and {Ng}, C. -Y. and {O'Dell}, S.~L. and {Omodei}, N. and {Oppedisano}, C. and {Papitto}, A. and {Pavlov}, G.~G. and {Peirson}, A.~L. and {Perri}, M. and {Pesce-Rollins}, M. and {Pilia}, M. and {Possenti}, A. and {Poutanen}, J. and {Ramsey}, B.~D. and {Rankin}, J. and {Ratheesh}, A. and {Romani}, R.~W. and {Sgr{\v{s}}}, C. and {Slane}, P. and {Soffitta}, P. and {Spandre}, G. and {Tamagawa}, T. and {Tavecchio}, F. and {Taverna}, R. and {Tawara}, Y. and {Tennant}, A.~F. and {Thomas}, N.~E. and {Trois}, A. and {Tsygankov}, S.~S. and {Turolla}, R. and {Vink}, J. and {Weisskopf}, M.~C. and {Wu}, K. and {Xie}, F. and {Zane}, S.},
        title = "{Polarization constraints on the X-ray corona in Seyfert Galaxies: MCG-05-23-16}",
      journal = {\mnras},
         year = 2022,
        month = nov,
       volume = {516},
       number = {4},
        pages = {5907-5913},
          doi = {10.1093/mnras/stac2634},
archivePrefix = {arXiv},
       eprint = {2207.09338},
 primaryClass = {astro-ph.HE},
       adsurl = {https://ui.adsabs.harvard.edu/abs/2022MNRAS.516.5907M}
}

@ARTICLE{Marshall2022,
       author = {{Marshall}, Herman L. and {Ng}, Mason and {Rogantini}, Daniele and {Heyl}, Jeremy and {Tsygankov}, Sergey S. and {Poutanen}, Juri and {Costa}, Enrico and {Zane}, Silvia and {Malacaria}, Christian and {Agudo}, Iv{\'a}n and {Antonelli}, Lucio A. and {Bachetti}, Matteo and {Baldini}, Luca and {Baumgartner}, Wayne H. and {Bellazzini}, Ronaldo and {Bianchi}, Stefano and {Bongiorno}, Stephen D. and {Bonino}, Raffaella and {Brez}, Alessandro and {Bucciantini}, Niccol{\`o} and {Capitanio}, Fiamma and {Castellano}, Simone and {Cavazzuti}, Elisabetta and {Ciprini}, Stefano and {De Rosa}, Alessandra and {Del Monte}, Ettore and {Di Gesu}, Laura and {Di Lalla}, Niccol{\`o} and {Di Marco}, Alessandro and {Donnarumma}, Immacolata and {Doroshenko}, Victor and {Dov{\v{c}}iak}, Michal and {Ehlert}, Steven R. and {Enoto}, Teruaki and {Evangelista}, Yuri and {Fabiani}, Sergio and {Ferrazzoli}, Riccardo and {Garcia}, Javier A. and {Gunji}, Shuichi and {Hayashida}, Kiyoshi and {Iwakiri}, Wataru and {Jorstad}, Svetlana G. and {Karas}, Vladimir and {Kitaguchi}, Takao and {Kolodziejczak}, Jeffery J. and {Krawczynski}, Henric and {La Monaca}, Fabio and {Latronico}, Luca and {Liodakis}, Ioannis and {Maldera}, Simone and {Manfreda}, Alberto and {Marin}, Fr{\'e}d{\'e}ric and {Marinucci}, Andrea and {Marscher}, Alan P. and {Matt}, Giorgio and {Mitsuishi}, Ikuyuki and {Mizuno}, Tsunefumi and {Muleri}, Fabio and {Ng}, C.-Y. and {O'Dell}, Stephen L. and {Omodei}, Nicola and {Oppedisano}, Chiara and {Papitto}, Alessandro and {Pavlov}, George G. and {Peirson}, Abel L. and {Perri}, Matteo and {Pesce-Rollins}, Melissa and {Petrucci}, Pierre-Olivier and {Pilia}, Maura and {Possenti}, Andrea and {Puccetti}, Simonetta and {Ramsey}, Brian D. and {Rankin}, John and {Ratheesh}, Ajay and {Romani}, Roger W. and {Sgr{\`o}}, Carmelo and {Slane}, Patrick and {Soffitta}, Paolo and {Spandre}, Gloria and {Tamagawa}, Toru and {Tavecchio}, Fabrizio and {Taverna}, Roberto and {Tawara}, Yuzuru and {Tennant}, Allyn F. and {Thomas}, Nicholas E. and {Tombesi}, Francesco and {Trois}, Alessio and {Turolla}, Roberto and {Vink}, Jacco and {Weisskopf}, Martin C. and {Wu}, Kinwah and {Xie}, Fei and {IXPE Collaboration} and {Schulz}, Norbert S. and {Chakrabarty}, Deepto},
        title = "{Observations of 4U 1626-67 with the Imaging X-Ray Polarimetry Explorer}",
      journal = {\apj},
         year = 2022,
        month = nov,
       volume = {940},
       number = {1},
          eid = {70},
        pages = {70},
          doi = {10.3847/1538-4357/ac98c2},
archivePrefix = {arXiv},
       eprint = {2210.03194},
 primaryClass = {astro-ph.HE},
       adsurl = {https://ui.adsabs.harvard.edu/abs/2022ApJ...940...70M}
}

@ARTICLE{Meszaros1988,
   author = {{M{\`e}sz{\`a}ros}, P. and {Novick}, R. and {Szentgyorgyi}, A. and {Chanan}, G.~A. and
{Weisskopf}, M.~C.},
    title = "{Astrophysical implications and observational prospects of X-ray polarimetry}",
  journal = {\apj},
     year = 1988,
   volume = 324,
    pages = {1056},
      doi = {10.1086/165962},
   adsurl = {http://adsabs.harvard.edu/abs/1988ApJ...324.1056M}
}

@ARTICLE{Middei2023,
       author = {{Middei}, Riccardo and {Liodakis}, Ioannis and {Perri}, Matteo and {Puccetti}, Simonetta and {Cavazzuti}, Elisabetta and {Di Gesu}, Laura and {Ehlert}, Steven R. and {Madejski}, Grzegorz and {Marscher}, Alan P. and {Marshall}, Herman L. and {Muleri}, Fabio and {Negro}, Michela and {Jorstad}, Svetlana G. and {Ag{\'\i}s-Gonz{\'a}lez}, Beatriz and {Agudo}, Iv{\'a}n and {Bonnoli}, Giacomo and {Bernardos}, Maria I. and {Casanova}, V{\'\i}ctor and {Garc{\'\i}a-Comas}, Maya and {Husillos}, C{\'e}sar and {Marchini}, Alessandro and {Sota}, Alfredo and {Kouch}, Pouya M. and {Lindfors}, Elina and {Borman}, George A. and {Kopatskaya}, Evgenia N. and {Larionova}, Elena G. and {Morozova}, Daria A. and {Savchenko}, Sergey S. and {Vasilyev}, Andrey A. and {Zhovtan}, Alexey V. and {Casadio}, Carolina and {Escudero}, Juan and {Myserlis}, Ioannis and {Hales}, Antonio and {Kameno}, Seiji and {Kneissl}, Ruediger and {Messias}, Hugo and {Nagai}, Hiroshi and {Blinov}, Dmitry and {Bourbah}, Ioakeim G. and {Kiehlmann}, Sebastian and {Kontopodis}, Evangelos and {Mandarakas}, Nikos and {Romanopoulos}, Stylianos and {Skalidis}, Raphael and {Vervelaki}, Anna and {Masiero}, Joseph R. and {Mawet}, Dimitri and {Millar-Blanchaer}, Maxwell A. and {Panopoulou}, Georgia V. and {Tinyanont}, Samaporn and {Berdyugin}, Andrei V. and {Kagitani}, Masato and {Kravtsov}, Vadim and {Sakanoi}, Takeshi and {Imazawa}, Ryo and {Sasada}, Mahito and {Fukazawa}, Yasushi and {Kawabata}, Koji S. and {Uemura}, Makoto and {Mizuno}, Tsunefumi and {Nakaoka}, Tatsuya and {Akitaya}, Hiroshi and {Gurwell}, Mark and {Rao}, Ramprasad and {Di Lalla}, Niccol{\'o} and {Cibrario}, Nicol{\'o} and {Donnarumma}, Immacolata and {Kim}, Dawoon E. and {Omodei}, Nicola and {Pacciani}, Luigi and {Poutanen}, Juri and {Tavecchio}, Fabrizio and {Antonelli}, Lucio A. and {Bachetti}, Matteo and {Baldini}, Luca and {Baumgartner}, Wayne H. and {Bellazzini}, Ronaldo and {Bianchi}, Stefano and {Bongiorno}, Stephen D. and {Bonino}, Raffaella and {Brez}, Alessandro and {Bucciantini}, Niccol{\'o} and {Capitanio}, Fiamma and {Castellano}, Simone and {Ciprini}, Stefano and {Costa}, Enrico and {De Rosa}, Alessandra and {Del Monte}, Ettore and {Di Marco}, Alessandro and {Doroshenko}, Victor and {Dov{\v{c}}iak}, Michal and {Enoto}, Teruaki and {Evangelista}, Yuri and {Fabiani}, Sergio and {Ferrazzoli}, Riccardo and {Garcia}, Javier A. and {Gunji}, Shuichi and {Hayashida}, Kiyoshi and {Heyl}, Jeremy and {Iwakiri}, Wataru and {Karas}, Vladimir and {Kitaguchi}, Takao and {Kolodziejczak}, Jeffery J. and {Krawczynski}, Henric and {La Monaca}, Fabio and {Latronico}, Luca and {Maldera}, Simone and {Manfreda}, Alberto and {Marin}, Fr{\'e}d{\'e}ric and {Marinucci}, Andrea and {Massaro}, Francesco and {Matt}, Giorgio and {Mitsuishi}, Ikuyuki and {Ng}, C. -Y. and {O'Dell}, Stephen L. and {Oppedisano}, Chiara and {Papitto}, Alessandro and {Pavlov}, George G. and {Peirson}, Abel L. and {Pesce-Rollins}, Melissa and {Petrucci}, Pierre-Olivier and {Pilia}, Maura and {Possenti}, Andrea and {Ramsey}, Brian D. and {Rankin}, John and {Ratheesh}, Ajay and {Romani}, Roger W. and {Sgr{\'o}}, Carmelo and {Slane}, Patrick and {Soffitta}, Paolo and {Spandre}, Gloria and {Tamagawa}, Toru and {Taverna}, Roberto and {Tawara}, Yuzuru and {Tennant}, Allyn F. and {Thomas}, Nicholas E. and {Tombesi}, Francesco and {Trois}, Alessio and {Tsygankov}, Sergey and {Turolla}, Roberto and {Vink}, Jacco and {Weisskopf}, Martin C. and {Wu}, Kinwah and {Xie}, Fei and {Zane}, Silvia},
        title = "{X-Ray Polarization Observations of BL Lacertae}",
      journal = {\apjl},
         year = 2023,
        month = jan,
       volume = {942},
       number = {1},
          eid = {L10},
        pages = {L10},
          doi = {10.3847/2041-8213/aca281},
archivePrefix = {arXiv},
       eprint = {2211.13764},
 primaryClass = {astro-ph.HE},
       adsurl = {https://ui.adsabs.harvard.edu/abs/2023ApJ...942L..10M}
}

@ARTICLE{Mondal2024,
       author = {{Mondal}, Santanu and {Chatterjee}, Rwitika and {Agrawal}, Vivek K. and {Nandi}, Anuj},
        title = "{Spectro-polarimetric study to constrain accretion-ejection properties of MCG-5-23-16 using IXPE and NuSTAR observations}",
      journal = {\pasa},
         year = 2024,
        month = oct,
       volume = {41},
          eid = {e072},
        pages = {e072},
          doi = {10.1017/pasa.2024.58},
archivePrefix = {arXiv},
       eprint = {2403.14169},
 primaryClass = {astro-ph.HE},
       adsurl = {https://ui.adsabs.harvard.edu/abs/2024PASA...41...72M}
}

@inproceedings{Muleri2018,
       author = {{Muleri}, Fabio and {Lefevre}, Carlo and {Piazzolla}, Raffaele and {Morbidini}, Alfredo and {Amici}, Fabrizio and {Attina}, Primo and {Centrone}, Mauro and {Del Monte}, Ettore and {Di Cosimo}, Sergio and {Di Persio}, Giuseppe and {Evangelista}, Yuri and {Fabiani}, Sergio and {Ferrazzoli}, Riccardo and {Loffredo}, Pasqualino and {Maiolo}, Luca and {Maita}, Francesco and {Primicino}, Leandra and {Rankin}, John and {Rubini}, Alda and {Santoli}, Francesco and {Soffitta}, Paolo and {Tobia}, Antonino and {Tortosa}, Alessia and {Trois}, Alessio},
        title = "{Calibration of the IXPE instrument}",
    booktitle = {Space Telescopes and Instrumentation 2018: Ultraviolet to Gamma Ray},
         year = 2018,
       editor = {{den Herder}, Jan-Willem A. and {Nikzad}, Shouleh and {Nakazawa}, Kazuhiro},
       series = {Society of Photo-Optical Instrumentation Engineers (SPIE) Conference Series},
       volume = {10699},
        month = jul,
          eid = {106995C},
        pages = {106995C},
          doi = {10.1117/12.2312203},
archivePrefix = {arXiv},
       eprint = {2111.14511},
 primaryClass = {astro-ph.IM},
       adsurl = {https://ui.adsabs.harvard.edu/abs/2018SPIE10699E..5CM}
}

@ARTICLE{Mushtukov2023,
       author = {{Mushtukov}, A.~A. and {Tsygankov}, S.~S. and {Poutanen}, J. and {Doroshenko}, V. and {Salganik}, A. and {Costa}, E. and {Marco}, A. Di and {Heyl}, J. and {Monaca}, F. La and {Lutovinov}, A.~A. and {Mereminsky}, I.~A. and {Papitto}, A. and {Semena}, A.~N. and {Shtykovsky}, A.~E. and {Suleimanov}, V.~F. and {Forsblom}, S.~V. and {Gonz{\'a}lez-Caniulef}, D. and {Malacaria}, C. and {Sunyaev}, R.~A. and {Agudo}, I. and {Antonelli}, L.~A. and {Bachetti}, M. and {Baldini}, L. and {Baumgartner}, W.~H. and {Bellazzini}, R. and {Bianchi}, S. and {Bongiorno}, S.~D. and {Bonino}, R. and {Brez}, A. and {Bucciantini}, N. and {Capitanio}, F. and {Castellano}, S. and {Cavazzuti}, E. and {Chen}, C.-T. and {Ciprini}, S. and {De Rosa}, A. and {Del Monte}, E. and {Gesu}, L. Di and {Lalla}, N. Di and {Donnarumma}, I. and {Dov{\v{c}}iak}, M. and {Ehlert}, S.~R. and {Enoto}, T. and {Evangelista}, Y. and {Fabiani}, S. and {Ferrazzoli}, R. and {Garcia}, J.~A. and {Gunji}, S. and {Hayashida}, K. and {Iwakiri}, W. and {Jorstad}, S.~G. and {Kaaret}, P. and {Karas}, V. and {Kislat}, F. and {Kitaguchi}, T. and {Kolodziejczak}, J.~J. and {Krawczynski}, H. and {Latronico}, L. and {Liodakis}, I. and {Maldera}, S. and {Manfreda}, A. and {Marin}, F. and {Marscher}, A.~P. and {Marshall}, H.~L. and {Massaro}, F. and {Matt}, G. and {Mitsuishi}, I. and {Mizuno}, T. and {Muleri}, F. and {Negro}, M. and {Ng}, C.-Y. and {O'Dell}, S.~L. and {Omodei}, N. and {Oppedisano}, C. and {Pavlov}, G.~G. and {Peirson}, A.~L. and {Perri}, M. and {Pesce-Rollins}, M. and {Petrucci}, P.-O. and {Pilia}, M. and {Possenti}, A. and {Puccetti}, S. and {Ramsey}, B.~D. and {Rankin}, J. and {Ratheesh}, A. and {Roberts}, O.~J. and {Romani}, R.~W. and {Sgr{\`o}}, C. and {Slane}, P. and {Soffitta}, P. and {Spandre}, G. and {Swartz}, D.~A. and {Tamagawa}, T. and {Tavecchio}, F. and {Taverna}, R. and {Tawara}, Y. and {Tennant}, A.~F. and {Thomas}, N.~E. and {Tombesi}, F. and {Trois}, A. and {Turolla}, R. and {Vink}, J. and {Weisskopf}, M.~C. and {Wu}, K. and {Xie}, F. and {Zane}, S.},
        title = "{X-ray polarimetry of X-ray pulsar X Persei: another orthogonal rotator?}",
      journal = {\mnras},
         year = 2023,
        month = sep,
       volume = {524},
       number = {2},
        pages = {2004-2014},
          doi = {10.1093/mnras/stad1961},
archivePrefix = {arXiv},
       eprint = {2303.17325},
 primaryClass = {astro-ph.HE},
       adsurl = {https://ui.adsabs.harvard.edu/abs/2023MNRAS.524.2004M}
}

@ARTICLE{Mushtukov2015,
       author = {{Mushtukov}, Alexander A. and {Suleimanov}, Valery F. and {Tsygankov}, Sergey S. and {Poutanen}, Juri},
        title = "{The critical accretion luminosity for magnetized neutron stars}",
      journal = {\mnras},
         year = 2015,
        month = feb,
       volume = {447},
       number = {2},
        pages = {1847-1856},
          doi = {10.1093/mnras/stu2484},
archivePrefix = {arXiv},
       eprint = {1409.6457},
 primaryClass = {astro-ph.HE},
       adsurl = {https://ui.adsabs.harvard.edu/abs/2015MNRAS.447.1847M}
}

@ARTICLE{Novick1972,
   author = {{Novick}, R. and {Weisskopf}, M.~C. and {Berthelsdorf}, R. and
    {Linke}, R. and {Wolff}, R.~S.},
    title = "{Detection of X-Ray Polarization of the Crab Nebula}",
  journal = {\apjl},
     year = 1972,
   volume = 174,
    pages = {L1},
   adsurl = {http://adsabs.harvard.edu/abs/1972ApJ...174L...1N}
}

@ARTICLE{Pal2023,
       author = {{Pal}, Indrani and {Stalin}, C.~S. and {Chatterjee}, Rwitika and {Agrawal}, Vivek K.},
        title = "{X-ray polarization observations of IC 4329A with IXPE: Constraining the geometry of X-ray corona}",
      journal = {Journal of Astrophysics and Astronomy},
         year = 2023,
        month = dec,
       volume = {44},
       number = {2},
          eid = {87},
        pages = {87},
          doi = {10.1007/s12036-023-09981-5},
archivePrefix = {arXiv},
       eprint = {2305.09365},
 primaryClass = {astro-ph.HE},
       adsurl = {https://ui.adsabs.harvard.edu/abs/2023JApA...44...87P}
}

@ARTICLE{Pal2025,
       author = {{Pal}, I. and {Marchesi}, S. and {Torres-Alb{\`a}}, N. and {Cox}, I. and {Ajello}, M. and {Banerjee}, A. and {Silver}, R. and {Pizzetti}, A. and {Imam}, K.},
        title = "{X-ray polarization observations of NGC 2110 with IXPE}",
      journal = {\aap},
         year = 2025,
        month = may,
       volume = {697},
          eid = {A182},
        pages = {A182},
          doi = {10.1051/0004-6361/202453641},
archivePrefix = {arXiv},
       eprint = {2503.20892},
 primaryClass = {astro-ph.HE},
       adsurl = {https://ui.adsabs.harvard.edu/abs/2025A&A...697A.182P}
}

@ARTICLE{Poutanen2020,
       author = {{Poutanen}, Juri},
        title = "{Relativistic rotating vector model for X-ray millisecond pulsars}",
      journal = {\aap},
         year = 2020,
        month = sep,
       volume = {641},
          eid = {A166},
        pages = {A166},
          doi = {10.1051/0004-6361/202038689},
archivePrefix = {arXiv},
       eprint = {2006.10448},
 primaryClass = {astro-ph.HE},
       adsurl = {https://ui.adsabs.harvard.edu/abs/2020A&A...641A.166P}
}

@ARTICLE{Radhakrishnan1969,
       author = {{Radhakrishnan}, V. and {Cooke}, D.~J.},
        title = "{Magnetic Poles and the Polarization Structure of Pulsar Radiation}",
      journal = {\aplett},
         year = 1969,
        month = jan,
       volume = {3},
        pages = {225},
       adsurl = {https://ui.adsabs.harvard.edu/abs/1969ApL.....3..225R}
}
\bibliographystyle{aasjournal}
\end{document}